\documentstyle[epsfig,aps]{revtex}
\newcommand{\nc}{\newcommand}
\nc{\rnc}{\renewcommand}
\nc{\acs}{\arraycolsep}
\nc{\mc}{\multicolumn}
\nc{\bsk}{\baselineskip}
\nc{\vsp}{\vspace}
\nc{\hsp}{\hspace}
\nc{\stl}{\setlength}
\nc{\stc}{\setcounter}
\nc{\addl}{\addtolength}
\nc{\beq}{\begin{equation}}
\nc{\eeq}{\end{equation}}
\nc{\beqa}{\begin{eqnarray}}
\nc{\eeqa}{\end{eqnarray}}
\nc{\tfrac}[2]{\raisebox{.4ex}{\tiny $\frac{#1}{#2}$}} 
\nc{\romlist}{ \setcounter{num1}{0}%
  \begin{list}{(\roman{num1})}{\usecounter{num1}} }
\nc{\arblist}{ \setcounter{num1}{0}%
  \begin{list}{(\arabic{num1})}{\usecounter{num1}} }
\nc{\alphlist}{ \setcounter{num2}{0}%
  \begin{list}{(\alph{num2})}{\usecounter{num2}} }
\nc{\bullist}{\begin{list}{$\bullet$}{ }}
\nc{\nr}{\\ \hline}
\nc{\hrl}{{\center \stl{\unitlength}{\textwidth} 
 \begin{picture}(1,0)  \put(0,0){\line(1,0){1}}
 \end{picture} \vsp{.001\bsk} }}
\nc{\cents}{{\scriptsize$\mbox{\rm C}\!\!\!\mbox{\raisebox{.2ex}%
{$|$}}\,\,\,\,$}}
\nc{\figsp}[5]{\begin{figure}[#1] \vsp{#2} \caption[#4]{#3} 
\label{#5} \vsp{2\bsk} \end{figure}}
\nc{\fig}{\figsp{tbp}}
\nc{\figb}{\figsp{b}}
\nc{\figh}{\figsp{h}}
\nc{\llist}{\begin{list}{}{} \stl{\labelsep}{.4in}}
\nc{\lit}[2]{
 \item[\raggedright #1]{#2}}
\nc{\lbit}[2]{ 
 \item[\raggedright\bf #1]{#2}}
\nc{\lemit}[2]{ 
 \item[\raggedright\em #1]{#2}}
\nc{\lbemit}[2]{ 
 \item[\raggedright\bf\em #1]{#2}}
\nc{\clst}[1]{\stl{\coltwo}{\textwidth}
\addl{\coltwo}{-#1} \addl{\coltwo}{-5.56ex} \newline
\begin{tabular}{p{#1}p{\coltwo}} \citem{}{}}
 
\nc{\citem}[2]{{\raggedright \bf #1} & #2 \\ }
\nc{\cemitem}[2]{{\raggedright \em #1} & #2 \\ }
\nc{\cbemitem}[2]{{\raggedright \bf \em #1} & #2 \\ }
\nc{\cend}{\citem{}{} \end{tabular} 
\mbox{}
} 
\nc{\SSP}{{\rm \hsp{.4in}}}
\nc{\SSPP}{{\rm \hsp{.2in}}}
\nc{\ds}{\displaystyle}
\nc{\tx}{\textstyle}
\nc{\scst}{\scriptstyle}
\nc{\sscst}{\scriptscriptstyle}
\nc{\prt}{\partial}
\nc{\fr}{\frac}
\nc{\lf}{\left}
\nc{\rt}{\right}
\nc{\la}{\langle}
\nc{\ra}{\rangle}
\nc{\V}{\vec}
\nc{\str}{\stackrel}
\nc{\ovl}{\overline}
\nc{\ul}{\underline}
\nc{\ovb}{\overbrace}
\nc{\ub}{\underbrace}
\nc{\wh}{\widehat}
\nc{\B}{\bar}
\nc{\D}{\dot}
\nc{\C}{\cdot}
\nc{\dd}{\ddot}
\nc{\tl}{\tilde}
\nc{\ha}{\hat}
\nc{\nn}{\nonumber}
\nc{\app}{\approx}
\nc{\al}{\alpha}
\nc{\RA}{\rightarrow}
\nc{\LRA}{\leftrightarrow}
\nc{\SRA}{\SSP\rightarrow\SSP}
\nc{\SSRA}{\SSPP\rightarrow\SSPP}
\nc{\dg}{\dagger}
\nc{\vp}{\varphi}
\nc{\ve}{\varepsilon}
\nc{\Dl}{\Delta}
\nc{\dl}{\delta}
\nc{\gm}{\gamma}
\nc{\Gm}{\Gamma}
\nc{\ep}{\epsilon}
\nc{\sg}{\sigma}
\nc{\Sg}{\Sigma}
\nc{\ua}{\uparrow}
\nc{\da}{\downarrow}
\nc{\lam}{\lambda}
\nc{\eql}[1]{\parbox{#1\textwidth}}
\nc{\eqm}[1]{\makebox[#1\textwidth][l]}
\nc{\enu}[1]{\mbox{\hspace{.4in}(\theequation.#1)}}
\nc{\son}{\\ \\ \ds}
\nc{\stw}{\\ & \\ \ds}		%   Equation formatting   %
\nc{\sth}{\\ & & \\ \ds}
\nc{\sfo}{\\ & & & \\ \ds}
\nc{\sfi}{\\ & & & & \\ \ds}
\nc{\A}{& \ds}
\nc{\bbr}{\lf\{\rule[-1.5ex]{0in}{0.01in}\rt.}
\nc{\hf}{\fr{1}{2}}
\nc{\mhf}{\mbox{\footnotesize$\hf$}}
\nc{\dv}{\/!}
\nc{\dint}{\int\!\!\int}
\nc{\tint}{\int\!\!\dint}               % integrals %
\nc{\qint}{\int\!\!\tint}
\nc{\Pd}[2]{\fr{\prt #1}{\prt #2}}
\nc{\Pdt}[1]{\Pd{#1}{t}}
\nc{\Pdx}[1]{\Pd{#1}{x}}
\nc{\Pdy}[1]{\Pd{#1}{y}}
\nc{\Pdz}[1]{\Pd{#1}{z}}           	%  Derivatives  %           
\nc{\Pdr}[1]{\Pd{#1}{r}}
\nc{\Pds}[1]{\Pd{#1}{s}}
\nc{\Dv}[2]{\fr{d#1}{d#2}}
\nc{\Dvt}[1]{\Dv{#1}{t}}
\nc{\Dvx}[1]{\Dv{#1}{x}}
\nc{\Dvy}[1]{\Dv{#1}{y}}
\nc{\Dvz}[1]{\Dv{#1}{z}}
\nc{\Dvr}[1]{\Dv{#1}{r}}
\nc{\Drs}[1]{\Dv{#1}{s}}
\nc{\inpp}[3]{\la #1| #2| #3\ra}
\nc{\inp}[2]{\inpp{#1}{#2}{#1}}
\nc{\rb}[1]{| #1\ra}
\nc{\lb}[1]{\la#1|}
\nc{\dtpp}[2]{\lb{#1}\rb{#2}}		%  Inner Products  %
\nc{\dtp}[1]{\dtpp{#1}{#1}}
\nc{\otpp}[2]{\rb{#1}\lb{#2}}
\nc{\otp}[1]{\otpp{#1}{#1}}
\rnc{\L}{{\cal L}}                      %  Miscellaneous  %
\nc{\lapp}{\mbox{\raisebox{-.6ex}{$\,\stackrel{\textstyle <}{\sim}\,$}}}
\nc{\gapp}{\mbox{\raisebox{-.6ex}{$\,\stackrel{\textstyle >}{\sim}\,$}}}
\newcounter{num1} \newcounter{num2}  %for \romlist and \alphlist
\newlength{\coltwo}
\nc{\als}{\fr{\al_s(Q^2)}{2\pi}}
\nc{\gpx}{g_1^p(x,Q^2)}
\nc{\gpz}{g_1^p(z,Q^2)}
\nc{\muq}{\lf(\fr{\mu^2}{Q^2}\rt)}
\nc{\xy}{(\fr{x}{y})}
\nc{\ASQ}{\al_s(Q^2)}
\nc{\Li}{{\rm Li}_2}
\nc{\dqx}{\Dl q_i(x,Q^2)} \nc{\dqy}{\Dl q_i(y,Q^2)} 
\nc{\dQ}{\Dl q_i(Q^2)}
\nc{\dgx}{\Dl g(x,Q^2)} \nc{\dgy}{\Dl g(y,Q^2)} 
\nc{\dG}{\Dl g(Q^2)}
\nc{\xq}{(x,Q^2)} \nc{\yq}{(y,Q^2)}
\nc{\Tt}{\tl{t}} \nc{\Ts}{\tl{s}} \nc{\Tu}{\tl{u}}
\nc{\Hs}{\ha{s}} \nc{\Ht}{\ha{t}} \nc{\Hu}{\ha{u}}
\nc{\Hsg}{\hat{\sg}}
\nc{\GeV}{\mbox{\rm GeV}}
\nc{\sS}{\!\not{\!s}}  
\nc{\pS}{\!\not{\!p}}  \nc{\kS}{\!\not{\!k}}
\nc{\poS}{\!\not{\!p}_1}  \nc{\pwS}{\!\not{\!p}_2}
\nc{\ptS}{\!\not{\!p}_3}  \nc{\pfS}{\!\not{\!p}_4}
\nc{\AS}{\!\not{\!\!A}}  \nc{\ASS}{\!\not{\!\!A}^*}
\nc{\BS}{\!\not{\!\!B}}  \nc{\BSS}{\!\not{\!\!B}^*}
\nc{\Tr}{\mbox{\rm Tr}}
\nc{\pT}{p_T}
\nc{\xT}{x_T}
\nc{\AoS}{\!\not{\!\!A}_1}  \nc{\AoSS}{\!\not{\!\!A}_1^*}
\nc{\AwS}{\!\not{\!\!A}_2}  \nc{\AwSS}{\!\not{\!\!A}_2^*}
\nc{\BoS}{\!\not{\!\!B}_1}  \nc{\BoSS}{\!\not{\!\!B}_1^*}
\nc{\BwS}{\!\not{\!\!B}_2}  \nc{\BwSS}{\!\not{\!\!B}_2^*}
\nc{\aS}{\!\not{\!a}}  \nc{\bS}{\!\not{\!b}}
\nc{\aoS}{\!\not{\!a}_1}  \nc{\boS}{\!\not{\!b}_1}
\nc{\awS}{\!\not{\!a}_2}  \nc{\bwS}{\!\not{\!b}_2}
\nc{\anS}{\!\not{\!a}_n}  \nc{\bnS}{\!\not{\!b}_n}
\nc{\anpoS}{\!\not{\!a}_{n+1}}  \nc{\bnpoS}{\!\not{\!b}_{n+1}}
\nc{\anmoS}{\!\not{\!a}_{n-1}}  \nc{\bnmoS}{\!\not{\!b}_{n-1}}
\nc{\refi}[1]{$^{\,\mbox{\scriptsize \ref{#1}}}$}
\nc{\refii}[2]{$^{\,\mbox{\scriptsize \ref{#1},\ref{#2}}}$}
\nc{\refiii}[3]{$^{\,\mbox{\scriptsize \ref{#1},\ref{#2},\ref{#3}}}$}
\nc{\refr}[2]{$^{\,\mbox{\scriptsize \ref{#1}--\ref{#2}}}$}
\nc{\sint}{\int \!\!}
\nc{\qB}{\stackrel{(-)}{q}}
\nc{\Lam}{\Lambda}

\begin{document}

%\stl{\bsk}{1.5\bsk}
%\pagestyle{empty}

\draft

\nc{\stW}{$\sin^2 \theta_W$}
\nc{\afb}{$A_{FB}$}

%\begin{center} \begin{Large} \begin{bf}
\title{Drell-Yan forward-backward and spin asymmetries for arbitrary 
vector boson
production at next-to-leading order}
%\end{bf} \end{Large} \end{center}
%\vglue 0.35cm
%{\begin{center}
\author{B.\ Kamal}
%\end{center}}
%\parbox{6.4in}{%\leftskip=1.0pc
%\begin{center}
\address{Physics Department, Brookhaven National Laboratory, Upton,
 New York 11973, U.S.A.}
%}
%\end{center}
%\begin{center}
%\vglue 1.0cm
%\begin{bf} ABSTRACT \end{bf}
%\end{center}
%%\vglue 1.0cm
%{%\rightskip=1.5pc
% %\leftskip=1.5pc
% %\tenrm\baselineskip=12pt
% \noindent

\date{Submitted October 16, 1997 to Phys.\ Rev.\ D. In press.}

\maketitle

%\widetext

\begin{abstract}
Longitudinally polarized, unpolarized and forward-backward 
mass differential cross sections
for Drell-Yan lepton-pair production by arbitrary vector bosons are 
calculated in next-to-leading order (NLO) QCD. Analytical results are presented
in a form valid for all consistent
$n$-dimensional regularization schemes, with the mass
factorization scheme kept general.  
NLO predictions for all 
Drell-Yan type processes ($W^\pm$, $Z$ and $\gm^*$) at BNL's
relativistic heavy ion collider (RHIC) are made using polarized parton
distributions which fit the recent deep-inelastic scattering data.
These are examined as tools in the determination of the polarized
parton distributions and the unpolarized $\B{u}/\B{d}$ ratio. NLO 
predictions for the forward-backward lepton asymmetry at Fermilab
are made and the precision determination of \stW\  from future runs
is studied. In all the above, the QCD corrections are found to be 
significant. An introductory discussion is given of various theoretical issues,
such as allowable factorization and regularization schemes, and scale
dependences.
\end{abstract}

%\vsp{.15in}
%\noindent

%\narrowtext

%\renewcommand{\thefootnote}{\fnsymbol{footnote}}
%\addtocounter{footnote}{1}
%\footnotetext{  }

%\newpage

%\pagestyle{plain}
%\setcounter{page}{1}

%\vglue .3cm
%\begin{center}\begin{large}\begin{bf}
\section{INTRODUCTION}
%\end{bf}\end{large}\end{center}
%\vglue .3cm

Two major areas of interest within the standard model are the determination
of the polarized parton distributions of the proton and higher precision
determinations of the electroweak mixing angle,
\stW,  as a constraint on the Higgs boson mass and new physics. 
One process useful in exploring both areas is Drell-Yan 
lepton-pair production. We 
shall present a clear picture of the Drell-Yan process at one-loop in
QCD within a general framework which should be both instructive and
useful. We will do so by considering the general interference between
two vector bosons with arbitrary mass, width and couplings, which decay
into a general lepton-antilepton pair (including neutrinos) and are produced
via quark-antiquark fusion (one of the (anti)quarks may of course arise from
an initial state gluon). In this way, we may consider possible new physics
contributions, such as $Z'$ bosons and four-fermion interactions, by 
appropriate choice of couplings, etc\ldots 

The emphasis here will be on presenting complete analytical results in
a form valid for all consistent
$n$-dimensional regularization schemes and within a general mass factorization
framework. In addition, we will consider all possible 
longitudinal polarization states of the initial
hadrons. Mass differential cross sections and asymmetries will be 
presented and the effect of the next-to-leading order (NLO) subprocesses
will be highlighted and explained  pedagogically. We will also
discuss various constraints on allowed regularization and factorization
schemes and describe the origin of the scale dependence
of the one-loop corrected predictions in a general fashion.

The one-loop QCD corrections to the longitudinally polarized Drell-Yan
process have been studied in several papers \cite{Rat}--\cite{Gehr}
using various
regularization prescriptions (for the $\gm_5$) and  factorization 
schemes
while considering the
production by specific bosons. Here, we keep the formalism
completely general. Results are kept in a simple form by considering explicitly
only mass differential cross sections. As an application,
we will study how lepton-pair
production by $\gm^*$, $Z$ and $W^\pm$ bosons can be used to extract
the polarized parton distributions from forthcoming planned polarized
$pp$ collision experiments at BNL's relativistic heavy ion collider
(RHIC), which is scheduled to start running in 1999.
Longitudinally polarized
$pp$ collisions with both beams polarized are expected to begin in 2000
\cite{Design} 
with sufficient muon coverage to perform precision spin studies, in 
particular using $W^\pm$'s.

The QCD corrections are necessary in order to reduce the process dependence
of the parton distributions.
This allows comparison with and use of those
distributions obtained in polarized deep-inelastic scattering (DIS).
This form of global analysis will prove invaluable since,
with the Drell-Yan process, we are  sensitive to the sea-quark
distributions, which are currently
almost totally unconstrained from DIS. We can make
use of the DIS determinations of the polarized valence distributions 
however, since we work consistently at NLO in QCD.
Eventually RHIC will be able to improve those valence determinations due
to the increased flavor separation in $W^\pm$ production.

 Another use of Drell-Yan at RHIC is a precision
determination of the unpolarized $\B{u}/\B{d}$ ratio at 
fairly large $x$ and
at very high energy scales where  other experiments fall short. 
This can provide
information on the limit $x\RA 1$ at lower energy scales, via perturbative
evolution. Similar statements apply to the polarized parton distributions
with regard to the large-$x$ sensitivity. All predictions will 
lie within
the energy range $100\,\, \GeV \leq \sqrt{s_{pp}} \leq 700\,\,\GeV$,
although RHIC, as currently envisioned,
may not be able to run much above $500\,\,\GeV$.

The Drell-Yan process also provides a useful way of obtaining \stW.
One may ask why we would try to measure \stW\  in hadron-hadron collisions
when good measurements exist in $Z$ production from $e^+e^-$ annihilation
at LEP and SLAC. The main reason is that these measurements are 
nearly complete. Future SLAC 
measurements will improve the earlier SLAC value, but a large discrepancy
with the LEP measurement presently exists and should not be expected
to vanish. Thus, what is needed is an independent high precision measure
of \stW. With various high luminosity scenarios planned for Fermilab's
Run II \cite{TeV2000}, such a high precision measurement is indeed possible.
This is because the forward-backward lepton asymmetry (in the 
lepton-pair rest frame) is sensitive to \stW. In a best case
scenario, one could surpass the present (and future)
average from LEP and SLAC combined. In a worst case scenario, a measurement
at the same level of  precision 
as the present SLAC measurement should be attainable.
Various studies \cite{Sirlin,TeV2000}
 have been done to show how such a high precision determination
would significantly constrain the allowed mass region for the standard
model Higgs and provide a probe of new physics.

QCD corrections to the charge asymmetry of lepton pairs produced
in $p\B{p}$ collisions have previously been
investigated \cite{Call}. Unfortunately, the physical observables
considered are not the ones used in determining \stW\  and there does not
appear to be any way to straightforwardly convert them into a useful
form. Very recently, a paper appeared \cite{Baur} performing detailed
numerical studies of the QED and QCD corrections to the forward-backward
Drell-Yan
asymmetry in $p\B{p}$ and $pp$ collisions. There, Monte Carlo methods were
used and similar results were obtained, 
even though a different definition of the
forward-backward asymmetry beyond leading order was used, apparently
in order to minimize the QCD corrections. 
As well, recently, soft gluon resummation effects on the lepton angular
distribution from the decay of $Z$'s produced at the Fermilab Tevatron
were briefly considered in \cite{CP}.
Here, complete analytical results are presented for
the mass differential cross sections and these do not appear elsewhere
to our best knowledge. Also, hadron polarization effects as well 
as regularization
and factorization scheme dependences are presented explicitly, which
is not done elsewhere. One also sees quite clearly the structure of the
QCD corrections and the origin of that structure.

The paper is organized as follows. In Section II we present our 
general formalism and describe the observables being considered.
In Section III we discuss various features of dimensional regularization
and dimensional reduction. In Sections IV--VI we compute the 
(singular) subprocess cross sections. In Section VII we perform the
factorization of the mass singularities and discuss constraints on
allowable regularization and factorization schemes. We also discuss
how to convert subprocess cross sections and parton distributions
from one scheme to another. In Section VIII we present the final 
analytical results and discuss the scale dependence issue in a general
fashion. In Sections IX--XI we present numerical results relevant to
RHIC and examine the sensitivity to the polarized parton distributions
as well as the unpolarized $\B{u}/\B{d}$ ratio. In Section XII we
present the forward-backward Drell-Yan asymmetry relevant for the Fermilab
Tevatron and discuss the extraction of \stW\ from its measurement.
Finally, in Section XIII we present our conclusions and summarize
the work. Throughout, we have tried to avoid giving {\em standard}
discussions of general issues in order to present new and more general 
perspectives which should be useful to non-experts.

%\vglue 1cm
%\begin{center}\begin{large}\begin{bf}
\section{GENERAL PROCESS AND FORMALISM}
%\end{bf}\end{large}\end{center}
%\vglue .3cm

The Drell-Yan process with initial hadrons $A$, $B$ 
of definite chirality is
\beq
\label{e13}
A(P_1,\lam_A) + B(P_2,\lam_B) \RA l(p_3) + \B{l}(p_4) + X,
\eeq
where $\lam_A$, $\lam_B$ denote chiralities, $l\B{l}$ represents a
lepton pair and $X$ is an arbitrary hadronic final state.
We have the following invariant observables,
\beq
S\equiv (P_1+P_2)^2, \SSPP M^2\equiv q^2, \SSPP \tau \equiv M^2/S,
\eeq
where
\beq
q=p_3+p_4.
\eeq
Denoting the $y$-axis as the direction of motion of hadron $A$, we
may also define, in the c.m.\ of $A$ and $B$, the observable
\beq
x_F \equiv \fr{2 q_y}{\sqrt{S}}.
\eeq

The general $2\RA 2$ [$2\RA 3$] parton
subprocess contributing to (\ref{e13})
may be written as
\beqa
\nn
\lefteqn{a(p_1,\lam_1) + b(p_2, \lam_2)} \\
\label{e15}
  && \RA B_i^*(q) + [c(k)]
\RA l(p_3) + \B{l}(p_4) + [c(k)],
\eeqa
where $B_i = \gm,Z,W^\pm$ and $l=l^-,\nu_l$; $\B{l}=l^+, \B{\nu}_l$
(in the standard model).
We must consider the cases $a=q$, $b=\B{q}$, $c=g$; $a=\B{q}$, $b=q$,
$c=g$; $a=\qB$, $b=g$, $c=\qB$; $a=g$, $b=\qB$, $c=\qB$.

Using the parton model relations
\beq
p_1 = x_a P_1,\SSP p_2=x_b P_2,
\eeq
we may define the subprocess invariants,
\beq
s\equiv (p_1+p_2)^2 = x_ax_bS, \SSPP w\equiv \fr{M^2}{s} =
\fr{M^2}{Sx_ax_b} = \fr{\tau}{x_ax_b},
\eeq
where we took all external momenta to be massless, as usual.
The parton momentum distributions are given by
\beq
 F^{i/I}_k (x_i, \mu^2) = x_i  f^{i/I}_k (x_i, \mu^2),
 \SSPP k=u,l,
\eeq
where
\beq
f^{i/I}_u (x_i, \mu^2) = f^{i/I}_{+/+} (x_i, \mu^2) +
f^{i/I}_{-/+} (x_i, \mu^2)
\eeq
and
\beq
f^{i/I}_l (x_i, \mu^2) = f^{i/I}_{+/+} (x_i, \mu^2) -
f^{i/I}_{-/+} (x_i, \mu^2).
\eeq
Here $f^{i/I}_{\lam_i/\lam_I}(x_i,\mu^2)$ is the probability of
finding parton $i$ with chirality $\lam_i$ and momentum fraction
$x_i$ in hadron $I$ having chirality $\lam_I$,
evaluated at renormalization scale $\mu^2$.

Introduce
\beq
z \equiv \cos\theta^*,
\eeq
where $\theta^*$ is the angle between $p_3$ and $P_1$ 
(or, equivalently, $p_1$) in the 
$l\B{l}$ rest frame. Then we may define
\beq
\label{fbdef}
\sg^{F\pm B} \equiv \int_0^1\!\!dz \fr{d\sg}{dz}
\pm \int_{-1}^0\!\!dz \fr{d\sg}{dz}. 
\eeq
Hence $\sg^{F+B}$ has the interpretation as the usual leptonic
integrated cross section. We call $\sg^{F-B}$ the 
{\em forward-backward} cross section.

Let $\sg^{ab}(\lam_a,\lam_b)$ denote the cross section for colliding
partons (or hadrons with some degree of polarization)
$a$, $b$ having chiralities (or beam polarization directions)
$\lam_a$, $\lam_b$. Then there
are only 4 combinations of $\lam_a$, $\lam_b$ leading to well factorized
parton model expressions, as may be straightforwardly verified.
All other observables may be expressed in terms of these. They are
%\widetext
\beqa
\sg^{ab}_{uu} &\equiv& \fr{1}{4}[\sg^{ab}(+,+)+\sg^{ab}(+,-)
+\sg^{ab}(-,+)+\sg^{ab}(-,-)] \\
\label{q13}
\sg^{ab}_{lu} &\equiv& \fr{1}{4}[\sg^{ab}(+,+)+\sg^{ab}(+,-)
-\sg^{ab}(-,+)-\sg^{ab}(-,-)] \\
\label{q14}
\sg^{ab}_{ul} &\equiv& \fr{1}{4}[\sg^{ab}(+,+)-\sg^{ab}(+,-)
+\sg^{ab}(-,+)-\sg^{ab}(-,-)] \\
\sg^{ab}_{ll} &\equiv& \fr{1}{4}[\sg^{ab}(+,+)-\sg^{ab}(+,-)
-\sg^{ab}(-,+)+\sg^{ab}(-,-)].
\eeqa
The notation is straightforward: $\sg_{mn}^{ab}$ denotes the cross section
when $a$ has polarization $m$ and $b$ has polarization $n$. In shorthand,
\beq
u = (\lam=+) + (\lam=-), \SSP l = (\lam=+) - (\lam=-),
\eeq
so that $u$ denotes unpolarized and $l$ denotes longitudinally 
polarized. The factor $1/4$ is required so that $\sg_{uu}$ has
the interpretation as the spin averaged cross section.

The parton model expression for the mass differential
Drell-Yan cross section, for general beam polarization, is
\beqa \label{e20} \nn
\fr{d\sg^{AB, F\pm B}_{mn}}{dM} &=& {\cal S}_m^A {\cal S}_n^B
 \sum_{ab}
\int_{\tau}^1\!\! dx_a \int_{\tau/x_a}^1 dx_b 
f_m^{a/A}(x_a,\mu^2) f_n^{b/B}(x_b,\mu^2) 
\fr{d\ha{\sg}^{ab, F\pm B}_{mn}}{dM} \\
 &=& {\cal S}_m^A {\cal S}_n^B
 \sum_{ab}
\int_{\tau}^1\!\! \fr{dx_a}{x_a} \int_{\tau/x_a}^1 \fr{dw}{w} 
F_m^{a/A}(x_a,\mu^2) F_n^{b/B}(x_b,\mu^2) 
\fr{d\ha{\sg}^{ab, F\pm B}_{mn}}{dM} \\
 &=& {\cal S}_m^A {\cal S}_n^B
 \sum_{ab}
\nn
\int_{-(1-\tau)}^{1-\tau}\!\! dx_{F0} \int_{w_1}^1 \fr{dw}{w}
\fr{1}{x_a+x_b} 
F_m^{a/A}(x_a,\mu^2) F_n^{b/B}(x_b,\mu^2) 
\fr{d\ha{\sg}^{ab, F\pm B}_{mn}}{dM},
\eeqa
where
\beq \label{ea18}
x_{F0} \equiv x_a-x_b, \SSPP w_1 = \fr{\tau}{1-|x_{F0}|}, \SSPP
x_{a,b} = \fr{\sqrt{x_{F0}^2+4\tau/w} \pm x_{F0}}{2}.
\eeq 
%\narrowtext \noindent
$\ha{\sg}^{ab}$ is the subprocess cross section corresponding
to (\ref{e15}) and
\beq
\label{beampol}
{\cal S}^I_u = 1, \SSP {\cal S}^I_l = {\cal P}^I,
\eeq
where ${\cal P}^I$ is the degree of polarization of the beam of 
hadrons $I$. The only assumption made in the derivation of (\ref{e20})
is that the degree of
beam polarization is the same in both polarization directions.
The same factors of beam polarization 
will always enter, regardless of the specific
differential cross section being considered.
The experimental results
(for the $\sg_{mn}$) will have to be divided by the appropriate
factors of beam polarization in constructing asymmetries. 
In general, this is unavoidable since the
beam polarization will vary as a function of time for any given
experiment. In numerical calculations taking evenly spaced integration
points (i.e.\ Simpson's rule) the third form of (\ref{e20}) is
most convenient because $x_a$ and $x_b$ are integrated over symmetrically
while the inner integral is over $w$ and $d\ha{\sg}/dM$ contains
functions singular in $1-w$ (i.e.\ $\dl(1-w)$, $1/(1-w)_+$, \ldots). 

We will consider lepton-pair production by $N$ arbitrary 
vector bosons, $B_i$, having masses $M_i$, widths $\Gm_i$ and
couplings to fermions given by
\beq
\label{coupl}
i c_f^i \gm^{\al} (g_{vf}^i - g_{af}^i \gm_5).
\eeq
Hence we may write
\beq
\fr{d\ha{\sg}^{ab, F\pm B}_{mn}}{dM}
= \sum_{i \geq j}
\fr{d\ha{\sg}^{ab, F\pm B}_{mn,B_iB_j}}{dM},
\eeq
where $\ha{\sg}^{ab, F\pm B}_{mn,B_iB_j}$ is the total interference 
contribution
from $B_i$, $B_j$ (i.e.\ for $i\neq j$ it is the sum over the $i,j$ and
$j,i$ contributions). 
For $M$, $\Gm$, $g_A$ $\RA 0$ we obtain the photon
contribution so that the argumentation is completely general. This
formalism applies to $Z$ and $W^\pm$ production and the 
Drell-Yan production
of any bosons whose coupling to fermions is given by (\ref{coupl}),
provided there are no new contributing diagrams. This includes 
non standard model $Z'$ and ${W^{\pm}}'$ contributions. As well, we can
include four-fermion interactions.
For $W^\pm$ 
production, the $c_f^i$ will depend on both quark flavors at the vertex via
the appropriate CKM matrix element.

The normalization convention of our squared amplitudes is 
easily inferred from the standard leading order parton model
cross section expression,
\beqa  
\nn
\fr{d\sg^{AB}_{mn}}{dMdx_Fdz} &=& 
\fr{1}{\pi M^3} \fr{{\cal S}^A_m{\cal S}^B_n}{x_a+x_b}  \sum_{ab}
F_m^{a/A}(x_a,\mu^2) F_n^{b/B}(x_b,\mu^2) \\
\label{ea22} 
 && \times
\sum_{i \geq j} 
|M|^{2 \,\, ab}_{mn,B_iB_j},
\eeqa
where
\beq
ab = q\B{q},\B{q}q, \SSP x_F = x_{F0}
\eeq
and $|M|^{2 \,\, ab}_{mn,B_iB_j}$ denotes the net interference
between the amplitude involving $B_i$ and that involving $B_j$.
This corresponds to the convention $\sum_{\lam} u(p,\lam)
\B{u}(p,\lam) = \pS/2$ for $p^2=0$, which we maintain for consistency with
\cite{bkprd}.

As far as QCD corrections are concerned, we will only consider
$d\sg/dM$. In imposing exerimental cuts, one restricts the phase space of the
outgoing leptons. Since the QCD corrections effect the hadronic 
sector, the features of the corrections should not be greatly 
influenced by the leptonic cuts.
In the limit of small cuts (i.e.\ good leptonic coverage) the expressions
in this paper become exact.  We will not address further the issue of 
leptonic cuts in what follows.

%\vglue 1cm
%\begin{center}\begin{large}\begin{bf}
\section{\lowercase{$n$}-DIMENSIONAL REGULARIZATION SCHEME PECULIARITIES}
%\end{bf}\end{large}\end{center}
%\vglue .3cm

Since we will need to present the leading order (LO) cross sections in 
$n$ ($\equiv 4-2\ve$) dimensions in the next section, it is appropriate
to discuss the various $n$-dimensional regularization schemes at this
point. We will consider regularization by
dimensional regularization (DREG) \cite{HV,Bol} and by
dimensional reduction (DRED) \cite{Sie}. 

In \cite{bkprd}, the details concerning DREG and DRED were summarized
and will not be repeated here. The important points will be discussed
however. Within DREG, there are two commonly used schemes for dealing
with the $\gm_5$ matrix (and $\ve^{\mu\nu\rho\sg}$ tensor)
which arises in polarized processes. In the
't Hooft-Veltman-Breitenlohner-Maison (HVBM) scheme \cite{HV,BM},
all quantities are mathematically well-defined, but the
non-anticommuting $\gm_5$ leads to physical problems. These necessitate
finite renormalizations of the polarized parton distributions and 
UV counterterms (for non QCD/QED vertices) as will be discussed throughout
this paper. In the anticommuting-$\gm_5$ scheme \cite{Chan}, massless
quark helicity is conserved, but only at the expense of mathematical 
consistency when an odd number of $\gm_5$'s arise in the traces. Also,
one needs to devise prescriptions \cite{cpt} for dealing with the
$\ve^{\mu\nu\rho\sg}$ tensor (arising from polarized gluons), since
nothing other than the HVBM definition 
has been proven to be mathematically consistent
in DREG. 

Even when performing calculations where only an
even number of $\gm_5$'s occur in the traces, one may wonder if the 
result is meaningful in general, using an anticommuting $\gm_5$. 
For the case of longitudinal polarization,
the $\gm_5$ is connected with helicity, and we understand the need for
helicity conservation on physical grounds. For processes such as the
transverse Drell-Yan process, the $\gm_5$ also arises. Here though, there
is no physical motivation for using an anticommuting $\gm_5$ in 
DREG, other than that it simplifies calculations.
More study is required in this area since recently two-loop transversity
splitting functions have been calculated by three groups \cite{Kum,Vog,Haya}
using an
anticommuting $\gm_5$. The transverse case is somewhat more subtle than
the longitudinal case, however, due to the additional axis which enters.
We do not consider transversity further in this work.

So, in general, the anticommuting-$\gm_5$ scheme \cite{Chan} should
only be used in situations where it is physically motivated. And in those
situations, it should always be shown that the same results can be obtained
using the HVBM scheme via finite renormalizations of the parton distributions
or by the addition of UV counterterms. The above renormalizations and
UV counterterms should be unique (in any given gauge) 
for some class of subprocesses.
This is the approach we will
follow in this paper. All results are understood to refer to 
consistent schemes: those schemes which can be related
to the HVBM scheme as described above.  Hence, we will always refer to the
HVBM result as being the DREG result.

More recently, a $\gm_5$ scheme was introduced \cite{Reading}
which reproduces many of the desirable features of the HVBM scheme, such
as the correct ABJ anomaly, and claims mathematical consistency
without violating helicity conservation. This scheme
uses a {\em reading point}
(a specific leftmost $\gm$-matrix in the trace) and maintains an
anticommuting $\gm_5$, which requires non-cyclicity of the traces.
What remains, therefore,
is to show that this scheme leads to process independent
$n$-dimensional splitting functions. This will be discussed below.

Because of the tedious nature of the HVBM scheme, and the other problems
mentioned earlier,
it is more straightforward to
first use DRED to calculate subprocess cross sections, then present them
in a form valid for all consistent
$n$-dimensional regularization 
schemes using the technique of \cite{bkprd}. The only drawback is that
one has to add a UV counterterm to the quark-$\gm(Z,W^\pm)$ vertex when using
DRED. Fortunately, this counterterm is well established and unambiguous.

The DRED result for a physical cross section is (formally speaking)
defined as the result obtained by contracting all tensors and taking
all traces in 4 dimensions, then performing all phase space integrals
in $n$ dimensions (and adding any necessary UV counterterms). When 
considering QCD/QED corrections, one needn't actually take the traces
first. One may work directly at the amplitude level for loop diagrams.
For other theories, however, like supersymmetry, the ill-definedness of
$g_n^{\mu\nu}$ for noninteger $n<4$ leads to an ambiguity such that
the results may depend on the order of operations \cite{Sie2}.
In the HVBM scheme ($n>4$), $g_n^{\mu\nu}$ is defined since there
are an infinite number of integer dimensions with $n>4$.
Hence we may continue $g_n^{\mu\nu}$ to $n>4$ by simply using relations
valid in integer dimensions, but take the continuous limit $n\RA 4$
on the real axis (or complex plane) at the end of the calculation (i.e.\
after all divergences have been cancelled), as if one compressed the 
infinite number of dimensions into a continuum for $n>4$. 
In practice, of course, one continues to noninteger $n$ from the
beginning rather than at the end. As there
are a finite number of dimensions with $n<4$, we cannot continue
$g_n^{\mu\nu}$ into the region $n<4$. Hence, strictly speaking, we
must first perform all contractions and traces in 4 dimensions, then
go to $n$ dimensions to perform the phase space integrals, since only
scalar products of vectors will remain, which are manifestly continuable
to $n$ dimensions. In that way, $g_n^{\mu\nu}$ will never be contracted
into traces containing $\gm_5$ or with $\ve^{\mu\nu\rho\sg}$, which
is where the ambiguity could arise.

In order to present $d\ha{\sg}_{mn}^{ab,F\pm B}/dM$ in a form
valid for all consistent $n$-dimensional schemes, we must first give the
general form of the $n$-dimensional Altarelli-Parisi \cite{AP}
splitting functions, $P_{ij}^n(z)$,
related to the probability of parton $j$ splitting into a 
collinear parton $i$ having momentum fraction $z$, plus an 
arbitrary final state carrying the rest of the momentum.
The reason is that the regularization
scheme dependences arising from mass singularities
are entirely contained in these  functions, as was shown
in \cite{bkprd} (and references therein). Hence, we discuss them here.
We may write
\beq
P_{ij}^n(z,\ve) \equiv P_{ij}^4(z) + \ve P_{ij}^{\ve}(z),
\eeq
where $P_{ij}^4(z)$ is the usual 4-dimensional  splitting function.
In DRED,
\beq
P_{ij}^{n,{\rm DRED}}(z,\ve) = P_{ij}^4(z).
\eeq
We will also use the notation,
\beq
P^n_{ij}(z,\ve) \equiv P_{ij}^{n,<}(z,\ve) + \dl(1-z) P_{ij}^{n,\dl}(\ve),
\eeq
and define
\beq
\Dl_u P_{ij} \equiv P_{ij,++} + P_{ij,-+}, \SSPP
\Dl_l P_{ij} \equiv P_{ij,++} - P_{ij,-+},
\eeq
where the $+,-$ denote the respective chiralities.

In DREG, the unpolarized one-loop
splitting functions are given by \cite{ERT}
%\widetext
\beqa
\label{splitu}
\nn
\Dl_u P_{qq}^4(z) &=&
 C_F \lf[ \fr{2}{(1-z)_+} -1 -z + \fr{3}{2} \dl(1-z) \rt],
\SSPP \Dl_u P_{qq}^{\ve}(z) = C_F\lf[ -(1-z) + \fr{1}{2} \dl(1-z) \rt],
\\
\Dl_u P^4_{qg}(z) &=& \fr{1}{2} (1-2z+2z^2), 
\SSPP \Dl_u P^{\ve}_{qg}(z) = z^2-z,
\\ \nn
\Dl_u P_{gq}^4(z) &=& C_F \lf[ \fr{2}{z} -2 + z\rt], \SSPP
\Dl_u P_{gq}^{\ve}(z) = - C_F z,
\\ \nn
\Dl_u P_{gg}^4(z) &=& 2 N_C \lf[ \fr{1}{(1-z)_+}
+ \fr{1}{z} - 2 + z(1-z)\rt] + \fr{\beta_0}{2} \dl(1-z),
\SSPP \Dl_u P_{gg}^{\ve}(z) = \fr{N_F}{6} \dl(1-z), 
\eeqa
where $\beta_0 = \fr{11}{3} N_C - \fr{2}{3} N_F$ and the usual
convention of $2-2\ve$ ($=n-2$) gluon polarization states was used.
In fact, the latter is somewhat more than a convention, since it is
justified on physical grounds.

Within DREG, we need only consider longitudinally polarized
splitting functions determined in the HVBM scheme since, at
present, it is the only established
mathematically consistent scheme therein, which also leads to process
independent $n$-dimensional splitting functions. 
At one-loop they are given by 
\beqa
\label{splitp}
\nn
\Dl_l P_{qq}^4(z) &=& \Dl_u P_{qq}^4(z),
\SSPP \Dl_l P_{qq}^{\ve}(z) = C_F\lf[ 3(1-z) + \fr{1}{2} \dl(1-z) \rt],
\\
\Dl_l P^4_{qg}(z) &=& z-1/2, 
\SSPP \Dl_l P^{\ve}_{qg}(z) = -(1-z),
\\ \nn
\Dl_l P_{gq}^4(z) &=& C_F (2-z), \SSPP
\Dl_l P_{gq}^{\ve}(z) = 2 C_F (1-z),
\\ \nn
\Dl_l P_{gg}^4(z) &=& 2 N_C \lf[ \fr{1}{(1-z)_+}
- 2 z + 1\rt] + \fr{\beta_0}{2} \dl(1-z),
\SSPP \Dl_l P_{gg}^{\ve}(z) = 4N_C(1-z) + \fr{N_F}{6} \dl(1-z),
\eeqa
%\narrowtext \noindent
as can be inferred from \cite{GorV}.

These $n$-dimensional splitting functions were determined by
investigating the factorization properties of the NLO squared amplitudes
in various collinear limits for various subprocesses. In \cite{GorV}
it was found that one always obtains a Born term multiplied by the 
appropriate $n$-dimensional splitting function in the limit where
two partons are collinear, leading to a pole in a propagator.
It is important that the Born term be exactly the $n$-dimensional 
one appropriate to the regularization scheme in question; in the above 
case, HVBM. We investigated what happens in the scheme of \cite{Reading}
for the $qg$ subprocess of longitudinally polarized Drell-Yan.
For simplicity, we chose the $\gm_5$ coming from the polarized
incoming quark as reading point. We found
that the wrong Born term arose, in the collinear limit where the
outgoing quark is collinear with the initial state gluon. Namely, that of
HVBM rather than that of \cite{Reading}. The basic reason is that the
squared amplitude for the $qg$ subprocess in the scheme of \cite{Reading}
is equivalent to the HVBM result, with our choice of reading point.
Hence, in the collinear limit, the Born term ($q\B{q}$) and $n$-dimensional
splitting function ($\Dl_l P_{qg}^n$) which arise correspond to the
HVBM ones. Unfortunately, the Born term for the $q\B{q}$ subprocess
in the scheme of \cite{Reading} demands an anticommuting $\gm_5$,
hence the result is simply minus the unpolarized one. This is not true
in HVBM, due to the non-anticommuting $\gm_5$. Hence the Born terms
are different in the two schemes and the behaviour of the $qg$ squared
amplitude in the collinear 
limit is unphysical. This physical inconsistency  is clearly unacceptable and 
it will lead to process dependent $n$-dimensional splitting functions.
A different choice of reading point might rectify the situation, but
such an ambiguity is also unacceptable. Unfortunately, the prescription
of \cite{Reading} gives no unambiguous definition of what the 
reading point should be in all cases. Therefore some extension is
in order before it can be applied in practical QCD corrections to
polarized processes. 
We will not discuss the scheme of \cite{Reading} in
further detail.

The only subtlety associated with calculating the $\Dl_l P_{ij}^n$ in
the HVBM scheme is that one must additionally perform an integration
over the $\ha{k}$ momenta (i.e.\ the components between 4 and $n$ dimensions), 
rather than simply taking a collinear limit \cite{bkprd}
as is done in anticommuting-$\gamma_5$ schemes and  in the unpolarized
case. One still obtains process independent splitting functions since the
collinear phase space structure is process independent. Even in some
anticommuting--$\gamma_5$ schemes \cite{cpt}, one must resort to 
Sudakov kinematics in the collinear limit,
making the process nontrivial. In that case, one must drop certain potentially
finite terms of ${\cal O}(\ve)$ which only vanish after integration.

In what follows, we shall present tensors and squared amplitudes
calculated using DRED. Then, using the method of \cite{bkprd}, we present
the necessary one-loop subprocess
cross sections in a form valid for all consistent
$n$-dimensional regularization schemes. 
By ``$n$-dimensional regularization schemes'', we mean schemes in 
which all divergences are regularized by dimensional continuation.
This excludes all forms of off-shell and cutoff regularization schemes
which also involve dimensional continuation to regularize some of the
divergences.
The mass factorization scheme will be kept general.

%\vglue 1cm
%\begin{center}\begin{large}\begin{bf}
\section{LEADING ORDER CROSS SECTIONS}
%\end{bf}\end{large}\end{center}
%\vglue .3cm

Unless otherwise stated, the expressions 
presented here are calculated using DRED.
The unintegrated leptonic tensor $L_{BB'}^{\al\beta}$ is defined as the
product of $L_B^\al$ and $L_{B'}^{\beta^*}$, where $L_B^{\al}$ is the
leading order amplitude consisting of a lepton pair attached to boson $B$ at a
vertex having index $\al$. It is given by 
\beqa
\nn
L_{BB'}^{\al\beta} &=& \fr{\mu^{2\ve}}{4} c_l c_l'
[(g_{vl}g_{vl}' + g_{al}g_{al}') T_{1l}^{\al\beta} \\
\label{e31p}
 & & -(g_{al}g_{vl}' + g_{vl}g_{al}') T_{2l}^{\al\beta}],
\eeqa
with
\beqa
\nn
T_{1l}^{\al\beta} &=& 4(p_3^{\al}p_4^{\beta} + p_4^{\al}p_3^{\beta}
- \fr{M^2}{2}g^{\al\beta}), \\
T_{2l}^{\al\beta} &=& 4i\ve^{\al\beta\mu\nu}p_{3\mu}p_{4\nu},
\eeqa
where the arbitrary mass scale $\mu^{2\ve}$ arises from the
$n$-dimensional coupling $e^2\RA e^2\mu^{2\ve}$. 
Since $\mu$ is arbitrary, the physical predictions should not depend
on it, order by order in $\al_s$. We will show how this is 
explicitly satisfied at one-loop in section VIII.

Now define the integrated leptonic tensor as
\beq
{\cal L}_{F+B}^{\al\beta} = 
\ovb{  \sint \fr{d^{n-1}p_3}{(2\pi)^{n-1}2p_{3,0}}
\fr{\dl[(q-p_3)^2]}{q^2}   }^{\equiv \int_{F+B}\!\! {\cal D} p_3}
 L^{\al\beta}.
\eeq
Here we omitted the $BB'$ indices, and will often do so for compactness.
One finds \cite{bkprd}
\beq
\label{Labfpb}
{\cal L}^{\al\beta}_{F+B} =
\kappa 
\lf[ (1-\ve) \fr{q^\al q^\beta}{q^2} + \fr{g_n^{\al\beta}}{2}
- (3-2\ve) \fr{g^{\al\beta}}{2} \rt],
\eeq
where $\kappa$ now generalizes to
\beq
\kappa = c_l c_l'
\fr{\mu^{2\ve}}{2^{4-2\ve}}
\fr{(q^2)^{-\ve}}{\pi^{2-\ve}}
\fr{\Gm(1-\ve)}{\Gm(1-2\ve)} \fr{(g_{vl}g_{vl}' + g_{al}g_{al}')}
{(3-2\ve)(1-2\ve)}.
\eeq
It is interesting to note that
the corresponding DREG tensor, in an 
anticommuting-$\gm_5$ scheme, is obtained by replacing
$g^{\al\beta}\RA g_n^{\al\beta}$. This gives
\beq
{\cal L}_{F+B,{\rm DREG}}^{\al\beta} = 
\kappa  (1-\ve)
\lf[  \fr{q^\al q^\beta}{q^2} - g_n^{\al\beta} \rt].
\eeq
Of course, $T_{2l}^{\al\beta}$ is not defined in such a scheme,
but it does not contribute here. In all schemes,
the part $\sim q^\al q^\beta$ does not contribute to the cross
section, as follows from gauge invariance. 
${\cal L}_{F+B}^{\al\beta}$ 
is thus effectively a constant tensor. We nonetheless
keep all the terms for completeness.

We may define the forward-backward integrated leptonic tensor
with respect to an arbitrary massless vector, $p$, in a covariant
fashion using
\beqa
\nn
{\cal L}_{F-B(p)}^{\al\beta} &\equiv &
\int_{F+B}\!\! {\cal D} p_3
  [\theta(p\C q/2 - p\C p_3) \\
\label{Labfmbdef}
& & - \theta(p\C p_3 - p\C q/2)] L^{\al\beta} \\
&\equiv& \int_{F-B(p)}\!\! {\cal D} p_3 L^{\al\beta} \\
& = & 
\label{e38p}
- \ovl{\kappa} i \ve^{\al\beta\mu\nu} \fr{p_{\mu}q_{\nu}}
{2 p\C q},
\eeqa
where $\theta$ is the step function and
\beq
\ovl{\kappa} = c_l c_l' 
\fr{\mu^{2\ve}}{2^{5-2\ve}}
\fr{(q^2)^{-\ve}}{\pi^{2-\ve}}
\fr{\Gm(1-\ve)}{\Gm(1-2\ve)} 
F(\ve) (g_{al}g_{vl}' + g_{vl}g_{al}'),
\eeq
with
\beqa
\nn
F(\ve) &\equiv & 4 \fr{\Gm(1-2\ve)}{\Gm^2(1-\ve)} 
  \int_0^{1/2} \!\! dy [y^{-\ve}(1-y)^{1-\ve} \\
 & &
 - y^{1-\ve}(1-y)^{-\ve}]                               
 = 1 + {\cal O}(\ve).
\eeqa

The most general tensor structure of ${\cal L}_{F-B(p)}^{\al\beta}$
may be obtained by noting that only the antisymmetric part of
$L^{\al\beta}$ contributes. Hence ${\cal L}_{F-B(p)}^{\al\beta}$  
must be antisymmetric in $\al, \beta$ and can only depend on the momenta
$p$ and $q$. Then one uses the usual projection methods, as was done
to obtain (\ref{Labfpb}). We will use the notation
\beq
{\cal L}_{F-B}^{\al\beta} \equiv {\cal L}_{F-B(p_1)}^{\al\beta},
\SSP \int_{F-B}\!\! {\cal D} p_3 \equiv \int_{F-B(p_1)}\!\! {\cal D} p_3.
\eeq

It is easy to check that (\ref{Labfmbdef}) leads to the usual
definition of the forward-backward asymmetry, Eq.\ 
(\ref{fbdef}), by working in the rest
frame of $q$. Choices of reference axes other than $p_1$ are of
course possible, but using $p_1$ allows a straightforward, covariant
treatment.
Having performed the forward-backward leptonic
integration analytically greatly simplifies the next-to-leading
order calculation.

We may next define the  subprocess 
hadronic tensor $W_{ab,BB'}^{\al\beta}$
through the subprocess squared Feynman amplitude
contribution from $B,B'$ interference,
\beq
\label{sqampl}
|M|_{ab,BB'}^{2} \equiv D_{BB'}^{-1} L_{\al\beta}^{BB'} 
W_{ab,BB'}^{\al\beta}(2-\dl_{BB'}),
\eeq
where
%\widetext
\beq
D_{BB'}^{-1} \equiv \fr{ (M^2-M_B^2)(M^2-M_{B'}^2)
+ M_B M_{B'} \Gamma_B \Gamma_{B'} }
{ [(M^2-M^2_B)(M^2-M_{B'}^2) + M_B M_{B'} \Gamma_B \Gamma_{B'}]^2
+ [M_B \Gamma_B (M^2-M_{B'}^2) - M_{B'}\Gamma_{B'}(M^2-M_B^2)]^2}.
\eeq

For $a=q,b=\B{q}$, the leading order hadronic tensor is
\beqa
\nn
W_{q\B{q}}^{\al\beta} &=& \fr{\mu^{2\ve}}{2^4N_C} c_q c_q'
\{(g_{vq}g_{vq}' + g_{aq}g_{aq}') [T_{1h}^{\al\beta}(1-\lam_1\lam_2)
+ T_{2h}^{\al\beta}(\lam_1-\lam_2)]  \\
\label{lohad}
 &  & - (g_{aq}g_{vq}' + g_{vq}g_{aq}') [T_{1h}^{\al\beta}(\lam_1 - \lam_2)
+ T_{2h}^{\al\beta}(1 - \lam_1\lam_2)] \},
\eeqa
where
\beq
T_{1h}^{\al\beta} = 4(p_1^{\al}p_2^{\beta} + p_2^{\al}p_1^{\beta}
- \fr{M^2}{2}g^{\al\beta}), \SSPP
T_{2h}^{\al\beta} = - 4i\ve^{\al\beta\mu\nu}p_{1\mu}p_{2\nu}.
\eeq 

Using (\ref{e31p}) and (\ref{lohad}) in (\ref{sqampl}) gives the
LO result  for $a=q,b=\B{q}$,
\beqa
\nn
\lefteqn{ |M|^2_{q\B{q},BB'}(z)  }  \\
\nn &=& (2-\dl_{BB'}) D_{BB'}^{-1}
\fr{\mu^{4\ve}M^4}{2^4N_C} c_l c_l' c_q c_q' 
\{(1-\lam_1\lam_2)[(g_{vl}g_{vl}' + g_{al}g_{al}')
(g_{vq}g_{vq}' + g_{aq}g_{aq}') (1+z^2) \\
\nn
&& +2 (g_{al}g_{vl}' + g_{vl}g_{al}') (g_{aq}g_{vq}' + g_{vq}g_{aq}') z] 
+ (\lam_2 - \lam_1) [(g_{vl}g_{vl}' + g_{al}g_{al}') \\
\label{lomm}
&& \times  (g_{aq}g_{vq}' + g_{vq}g_{aq}') (1+z^2) 
+ 2 (g_{al}g_{vl}' + g_{vl}g_{al}')(g_{vq}g_{vq}' + g_{aq}g_{aq}') z] \}. 
\eeqa
This reproduces the result of \cite{Taxil} which includes the effect
of $Z'$ bosons in the Drell-Yan process. We may express the above in
a covariant fashion using
\beqa
\nn
z &=& (u^2-t^2)/M^2, \SSP 1+z^2 = 2(t^2+u^2)/M^2, \\
t &\equiv& (p_1-p_3)^2 = (p_2-p_4)^2, \SSP
u \equiv (p_2-p_3)^2 = (p_1-p_4)^2.
\eeqa

Defining
\beq
\label{e46p}
  {\cal M}_{ab,BB'}^{F\pm B} \equiv \int_{F\pm B}\!\! {\cal D} p_3
L^{BB'}_{\al\beta} W^{\al\beta}_{ab,BB'}
= {\cal L}^{BB', F\pm B}_{\al\beta} W^{\al\beta}_{ab,BB'},
\eeq
we have, for $a= q,b=\B{q}$ and in leading order,
\beq
{\cal M}^{F+B}_{q\B{q}} = \fr{\mu^{2\ve}M^2}{2^3N_C} c_q c_q' \kappa (2-\ve)
[(g_{vq}g_{vq}' + g_{aq}g_{aq}') (1-\lam_1\lam_2)
+ (g_{aq}g_{vq}' + g_{vq}g_{aq}') (\lam_2 - \lam_1)]
\eeq
and
\beq
{\cal M}^{F-B}_{q\B{q}} = \fr{\mu^{2\ve}M^2}{2^3N_C} c_q c_q' \ovl{\kappa} 
[(g_{aq}g_{vq}' + g_{vq}g_{aq}')  (1-\lam_1\lam_2)
+ (g_{vq}g_{vq}' + g_{aq}g_{aq}')  (\lam_2 - \lam_1)],
\eeq
where we used (\ref{Labfpb}) and (\ref{e38p}), respectively, in 
(\ref{e46p}).
The correctness of this approach (i.e.\ integrating over the leptonic
tensor first, then contracting the leptonic and hadronic tensors)
was verified by contracting first, then performing the leptonic integration.
One obtains exactly the same result, in $n$ dimensions.

This is straightforwardly checked by noting that, in the rest frame of
$q$,
\beq
\int_{F+B}\!\! {\cal D} p_3  \RA \fr{1}{2^{5-2\ve}}
\fr{(q^2)^{-1-\ve}}{\pi^{3-\ve}} \fr{\Gm (1-\ve)}{\Gm (1-2\ve)}
\sint d^2 \omega_3,
\eeq
where
\beq
\sint d^2 \omega_3 = \int_0^\pi \!\! d\theta_1 \sin^{1-2\ve}\theta_1 
\int_0^\pi \!\! d\theta_2 \sin^{-2\ve}\theta_2
\eeq
and the $\theta_i$ represent the first two of the $n-2$
angles of $p_3$ in $n$ 
dimensions \cite{Mar} and $\theta_1$ is taken to be
the angle between $p_3$
and $p_1$ (i.e.\ $\theta^*$). Since $|M|^2$ does not depend on $\theta_2$,
\beq
\sint  d^2 \omega_3 \RA 2\pi \fr{\Gm(1-2\ve)}{\Gm^2(1-\ve)}
\int_0^1 \!\! dy y^{-\ve} (1-y)^{-\ve}, \SSPP y \equiv \fr{1+z}{2}.
\eeq
We simply make the substitutions,
\beq
z=y-(1-y), \SSP 1+z^2 = 2 - 4y(1-y),
\eeq
in (\ref{lomm}), then integrate over $y$.

The $2\RA 2$ phase space is 
\beqa
\lf(\fr{d\ha{\sg}^{ab,F\pm B}_{mn, BB'}}{dM}\rt)_{\rm LO} &=& 
\fr{32 \pi}{M} \dl(1-w) \int_{F\pm B}\!\! {\cal D} p_3
 |M|^{2\,\,ab}_{mn, BB'} \\ 
&=& \label{q55}
\fr{32 \pi}{M} D_{BB'}^{-1} {\cal M}^{ab,F\pm B}_{mn, BB'} (2-\dl_{BB'})
\dl(1-w)
\equiv \chi^{ab,F\pm B}_{mn, BB'}(\ve) \dl(1-w).  
\eeqa
 $\chi$ has the form
\beq
\chi^{ab,F\pm B}_{mn, BB'}(\ve)
= (2-\dl_{BB'}) \fr{M}{2\pi N_C} \fr{c_q c_q' c_l c_l'}{D_{BB'}}
K^{l^+l^-,F\pm B}_{BB'} F^{ab,F\pm B}_{mn,BB'}.
\eeq
Here the $K^{l^+l^-,F\pm B}_{BB'}$ are leptonic factors given by
\beq
K^{l^+l^-,F + B}_{BB'} = \fr{1}{3} (g_{vl}g_{vl}' + g_{al}g_{al}') 
\lf[ \fr{3}{2} \fr{\mu^{4\ve}}{2^{-2\ve}} \fr{\pi^\ve}{q^{2\ve}}
\fr{\Gm(1-\ve)}{\Gm(1-2\ve)} \fr{2-\ve}{(3-2\ve)(1-2\ve)} \rt],
\eeq
\beq
K^{l^+l^-,F - B}_{BB'} =  \fr{1}{4} (g_{al}g_{vl}' + g_{vl}g_{al}') 
\lf[ \fr{\mu^{4\ve}}{2^{-2\ve}} \fr{\pi^\ve}{q^{2\ve}}
\fr{\Gm(1-\ve)}{\Gm(1-2\ve)} F(\ve) \rt].
\eeq
The factors in the square brackets are equal to $1+{\cal O}(\ve)$
and hence may be taken as 1 for calculational purposes, as they
will always factorize exactly in next-to-leading order. Similarly,
the Born term may be computed analogously in any valid $n$-dimensional
regularization scheme. However,
the Born term always factors out of the singular parts, which cancel
(after mass factorization). Hence, the scheme dependence of the Born
term cancels in the limit $\ve \RA 0$.
This will be
demonstrated explicitly in the following sections.

The $F^{ab}$ are hadronic factors whose expressions are
\beq
F^{q\B{q},F + B}_{uu,BB'} =
- F^{q\B{q},F + B}_{ll,BB'} =
F^{\B{q}q,F + B}_{uu,BB'} =
-F^{\B{q}q,F + B}_{ll,BB'} =
 g_{vq}g_{vq}' + g_{aq}g_{aq}',
\eeq
\beq
F^{q\B{q},F + B}_{ul,BB'} =
- F^{q\B{q},F + B}_{lu,BB'} =
F^{\B{q}q,F + B}_{lu,BB'} =
-F^{\B{q}q,F + B}_{ul,BB'} =
 g_{aq}g_{vq}' + g_{vq}g_{aq}',
\eeq
and
\beq
F^{q\B{q},F - B}_{uu,BB'} =
- F^{q\B{q},F - B}_{ll,BB'} =
- F^{\B{q}q,F - B}_{uu,BB'} =
F^{\B{q}q,F - B}_{ll,BB'} =
 g_{aq}g_{vq}' + g_{vq}g_{aq}',
\eeq
\beq
F^{q\B{q},F - B}_{ul,BB'} =
- F^{q\B{q},F - B}_{lu,BB'} =
-F^{\B{q}q,F - B}_{lu,BB'} =
F^{\B{q}q,F - B}_{ul,BB'} =
 g_{vq}g_{vq}' + g_{aq}g_{aq}'.
\eeq
%\narrowtext

%\vglue 1cm
%\begin{center}\begin{large}\begin{bf}
\section{VIRTUAL CORRECTIONS}
%\end{bf}\end{large}\end{center}
%\vglue .3cm

The string of gamma matrices,
\beq
\label{gstring}
\gm^\mu \gm^\rho \gm^\al (g_{vq}-g_{aq}\gm_5) \gm^\sg \gm_\mu,
\eeq
arises in the calculation of the vertex graph (the massless
self-energies vanish in $n$ dimensions). In order to get the correct
form for the virtual corrections, we must satisfy the relation,
\beq
\label{vertrel}
\gm^\mu \gm^\rho \gm^\al (g_{vq}-g_{aq}\gm_5) \gm^\sg \gm_\mu
=
\gm^\mu \gm^\rho \gm^\al  \gm^\sg \gm_\mu (g_{vq}-g_{aq}\gm_5),
\eeq
which holds for an anticommuting $\gm_5$. The relation
(\ref{vertrel}) is necessitated by the fact that the QED Ward identity 
\cite{Ward}
between vertex and self-energy graphs holds for the vector part. 
Therefore, we must be able to anticommute the $\gm_5$ out of the 
vertex correction so that the axial-vector part factors correctly
and the identity is trivially satisfied.
In non-anticommuting-$\gm_5$ schemes,  
this relation would be violated by terms of ${\cal O}(\ve)$ (giving rise
to a finite piece when multiplied by the $1/\ve$
UV singularity) which would have to be removed
by UV a counterterm analogous to that of DRED which corrects the
$\gm$-$q$ vertex. Then, the result would be equivalent to that obtained
using an anticommuting $\gm_5$ in $n$ dimensions, except that, in general,
there could be additional terms of soft origin which must cancel when
the bremsstrahlung contributions are added. If they did not cancel, the 
scheme would not be physically consistent since soft divergences cannot
give rise to scheme dependences in physical cross sections.

It should be noted that we are assuming a vertex having the form
(\ref{coupl}). It was shown in \cite{Nasr} that using a symmetrized
vertex can avoid the spurious UV terms, in the HVBM scheme, for the
vertex $b\RA s+H$. We investigated whether such a procedure would
work here. Unfortunately, the situation is somewhat different than
in $b\RA s+H$. Considering, for simplicity, the case of $W^\pm$ production,
we find that the symmetrization procedure of \cite{Nasr} simply amounts
to replacing the $n$-dimensional $\gamma^\alpha$ by the four-dimensional
one. This clearly does not help remove the spurious UV terms;
a counterterm will still be necessary.
Of course, 
the ambiguity about whether to put the $\gm_5$ on the right-
or left-hand side of the $\gamma^\alpha$ is no longer present.

The string of gamma matrices (\ref{gstring}) gets contracted with a tensor
$I_{\rho\sg}$ arising from the vertex loop integral. This is a
good example to illustrate how the unphysical term arises in DRED.
We will keep the 
argumentation general so that it is valid for both massless and 
massive quarks. Firstly, we have in DRED,
\beq
\gm^\mu \gm^\rho \gm^\al  \gm^\sg \gm_\mu
= -2  \gm^\sg \gm^\al  \gm^\rho.
\eeq
$I_{\rho\sg}$ has the form
\beq
I_{\rho\sg} = g_{\rho\sg}^{n} C_{24}
+ T_{\rho\sg},
\eeq
where $T_{\rho\sg}$ is a tensor which depends on the external momenta
and $C_{24}$  is a scalar coefficient.
Performing the contraction gives
\beqa
\nn
I_{\rho\sg} \gm^\sg \gm^\al  \gm^\rho  &=& 
\gm^\sg \gm^\al \gm^\rho g_{\rho\sg}^n C_{24}
+\gm^\sg \gm^\al \gm^\rho T_{\rho\sg} \\
&=& -2 C_{24} [(1-\ve)\gm^\al + \gm_{\ve}^\al]
+ \gm^\sg \gm^\al \gm^\rho T_{\rho\sg},
\eeqa
where we used the relation, valid for $n<4$,
\beq
 \gm^\sg \gm^\al  \gm^\rho g_{\rho\sg}^n =
 -2 (1-\ve) \gm^\al - 2 \gm_\ve^\al.
\eeq
It is precisely the term $\sim \gm_\ve^\al$ which is removed by the
DRED counterterm. The remaining terms give the correct result, i.e.\
that required to satisfy the QED Ward identity between vertex and
self-energy graphs. That identity is most straightforwardly checked
by retaining the quark mass, in order to avoid mass singularities.
Having satisfied (\ref{vertrel}), the virtual corrections are 
analogous to those of the $\gm$-$q$ vertex, since the Born term
simply factors out.

Using the Born term of the specific scheme,
the net virtual contribution in a form valid for
all consistent $n$-dimensional regularization schemes is
($ab = q\B{q},\B{q}q$)
\beqa
\nn
 \lf(\fr{d\ha{\sg}^{ab,F\pm B}_{mn, BB'}}{dM}\rt)_{\rm V} &=& 
  \chi^{ab,F\pm B}_{mn,BB'}(\ve) \dl(1-w) C_F \fr{\al_s}{2\pi}
C(\ve) \\
\label{e50}
&& \times
 \lf[-\fr{2}{\ve^2}  - 7 + \fr{2\pi^2}{3} -\fr{2}{\ve}
\fr{P_{qq}^{n,\dl}}{C_F}  \rt],
\eeqa 
plus possible extra soft terms which would cancel with  opposite
ones in the soft bremsstrahlung, as discussed above.
Here we defined
\beq
C(\ve) \equiv
\lf(\fr{4\pi\mu^2}{M^2}\rt)^\ve \fr{\Gm(1-\ve)}{\Gm(1-2\ve)}.
\eeq
The result (\ref{e50}) follows directly from that given in \cite{bkprd}.

%\vglue 1cm
%\begin{center}\begin{large}\begin{bf}
\section{BREMSSTRAHLUNG CORRECTIONS}
%\end{bf}\end{large}\end{center}
%\vglue .3cm

The leptonic integrated $2\RA 3$ particle squared amplitudes have
the form
\beq
\label{e69}
{\cal M}^{ab,F\pm B}_{mn, 2\RA 3}
= [c_F] \lf( \fr{2\pi\al_s\mu^{2\ve}}{M^2} \rt) 
\tilde{{\cal M}}^{ab,F\pm B}_{mn, 2\RA 2}
F^{ab,F\pm B}_{mn, 2\RA 3},
\eeq
where the factor $c_F$ is present only for $ab = q\B{q},\B{q}q$.
The $\tilde{{\cal M}}$ are given by
\beq
\label{e70}
\tilde{{\cal M}}^{q\B{q},F\pm B}_{mn, 2\RA 2} =
\tilde{{\cal M}}^{qg,F\pm B}_{mn, 2\RA 2} =
\tilde{{\cal M}}^{g\B{q},F\pm B}_{mn, 2\RA 2} =
{\cal M}^{q\B{q},F\pm B}_{mn, 2\RA 2} 
\eeq
and
\beq
\label{e71}
\tilde{{\cal M}}^{\B{q}q,F\pm B}_{mn, 2\RA 2} =
\tilde{{\cal M}}^{\B{q}g,F\pm B}_{mn, 2\RA 2} =
\tilde{{\cal M}}^{gq,F\pm B}_{mn, 2\RA 2} =
{\cal M}^{\B{q}q,F\pm B}_{mn, 2\RA 2}. 
\eeq
For the $qg$ subprocess in $W^{\pm}$ production, the r.h.s.\ of
(\ref{e70}),  (\ref{e71})
implicitly contain a sum over the various quark flavors
into which the gluon may split to form the $W^{\pm}$.

For the $q\B{q}$ subprocess, we have
\beq 
\label{e75p}
F^{q\B{q},F+ B}_{mn} =
F^{\B{q}q,F+B}_{mn} =
\fr{s^2(1+w^2)}{p_1\C k \, p_2\C k} -8,
\eeq
\beq
\label{e76p}
F^{q\B{q},F-B}_{mn} =
F^{\B{q}q,F-B}_{mn} =
\fr{s^2(1+w^2)}{p_1\C k \, p_2\C k} 
-4 \fr{s(1+w)}{p_1\C q}.
\eeq

For the $qg$ subprocess, we have
\beq \label{ea75}
F^{\qB g/g\qB, F+ B}_{mn} = 2\lf[ 2w+2\fr{p_{2/1}\C k}{s}
+ s \fr{\Dl_{n/m}P^4_{qg}(w)}{p_{2/1}\C k}\rt],
\eeq
\beq
F^{\qB g, F-B}_{mn} = 2\lf[ s\fr{ h_{mn}^{\qB g}(w)}{p_1\C q}
+ 2w + 2 \fr{p_2\C k}{s}
+ s        \fr{\Dl_{n}P^4_{qg}(w)}{p_2\C k}\rt],
\eeq
where
\beq
h^{\qB g}_{uu/lu} = -1+w, \SSP
h^{\qB g}_{ll/ul} = 1-w
\eeq
and
\beq
F^{g \qB , F-B}_{mn} = 2\lf[ s\fr{h_{mn}^{g \qB}(w)}{p_1\C q}
- 2w - 2 \fr{p_1\C k}{s}
+ s\fr{\Dl_{m}P^4_{qg}(w)}{p_1\C k}\rt],
\eeq
with
\beq
\label{e81p}
h^{g \qB}_{uu/ul} = 2w^2, \SSP
h^{g \qB}_{ll/lu} = 2w.
\eeq

Various checks on the above results are possible at this point.
From crossing symmetry, it follows that
%\widetext
\beq
{\cal M}^{qg,F\pm B}_{uu}
= - \fr{1}{2C_F} {\cal M}^{q\B{q},F\pm B}_{uu} (p_2 \LRA -k), \SSP
{\cal M}^{\B{q} g,F\pm B}_{uu}
= - \fr{1}{2C_F} {\cal M}^{\B{q}q,F\pm B}_{uu} (p_2 \LRA -k)
\eeq
and
\beq
\label{q82}
{\cal M}^{gq,F+B}_{uu}
= -\fr{1}{2C_F} {\cal M}^{\B{q}q,F+B}_{uu} (p_1 \LRA -k), \SSP
{\cal M}^{g \B{q},F+B}_{uu}
= -\fr{1}{2C_F} {\cal M}^{q\B{q},F+B}_{uu} (p_1 \LRA -k).
\eeq
The above correspondences were explicitly verified.
(\ref{q82}) holds only for the $F+B$ case, since the $F-B$ integration
depends on $p_1$ and hence destroys the crossing symmetry.

Also, the $q\LRA \B{q}$ interchanges were 
checked using CPT invariance. The only
difference between the $F-B$ and $F+B$ cases is that in the $F-B$
case, one picks up an extra minus under CPT, since 
CPT implies $l^+\LRA l^-$
and the leptonic integration is antisymmetric.

The $2\RA 3$ particle bremsstrahlung phase space is given by
\beq
\lf(\fr{d\ha{\sg}^{ab,F\pm B}_{mn, BB'}}{dM}\rt)_{\rm Br} = 
\fr{2^{1+2\ve}}{\pi^{1-\ve}} M^{1-2\ve} w^{1+\ve}
\fr{(1-w)^{1-2\ve}}{\Gm(1-\ve)} \int_0^1\!\! dy y^{-\ve}
(1-y)^{-\ve} \int_{F\pm B}\!\! {\cal D} p_3
 |M|^{2\,\,ab}_{mn, BB'},
\eeq
where
\beq
 y=\fr{1+\cos\theta}{2}
\eeq
and $\theta$ is the angle between
$p_1$ and $k$ in the $p_1$, $p_2$ c.m..

Define 
\beq \label{ea84}
I^{ab,F\pm B}_{mn} \equiv \fr{(1-w)^{1-2\ve}}{4}
\fr{\Gm(1-2\ve)}{\Gm^2(1-\ve)}    
\int_0^1\!\! dy y^{-\ve} (1-y)^{-\ve}
F^{ab,F\pm B}_{mn}.
\eeq
Then, noting (\ref{e69}), (\ref{q55}),
\beq \label{ea85}
\lf(\fr{d\ha{\sg}^{ab,F\pm B}_{mn, BB'}}{dM}\rt)_{\rm Br} =
 [c_F] \fr{\al_s}{2\pi} \tl{\chi}^{ab,F\pm B}_{mn,BB'}(\ve)
w^{1+\ve} C(\ve) I^{ab,F\pm B}_{mn},
\eeq
where the $\tl{\chi}^{ab,F\pm B}_{mn}$ are defined analogously
to the $\tl{\cal M}^{ab,F\pm B}_{mn}$ (see (\ref{e70}), (\ref{e71})).
The resulting integrations are rather straightforward. Using the
approach of \cite{bkprd} we may cast the $I^{ab,F\pm B}_{mn}$ in
a form valid for all consistent
$n$-dimensional regularization schemes, making
use of the $n$-dimensional splitting functions $\Dl_k P_{ij}^n$.
As for the virtual graphs, we omit a possible extra soft piece which
would cancel in the virtual plus bremsstrahlung sum and use the Born
term appropriate to the regularization scheme in question.

We obtain, for the $q\B{q}$ subprocess,
\beq
\label{e49}
I^{q\B{q},F\pm B}_{mn}
= 
\fr{2}{\ve^2}\dl(1-w) 
-  \fr{1}{\ve}\fr{\Dl_{mn} P_{qq}^{n,<}(w,\ve)}{C_F} 
+  8\lf(\fr{\ln(1-w)}{1-w}\rt)_+
-4(1+w)\ln(1-w) + k^{q\B{q},F\pm B}_{mn}(w),
\eeq
with
\beq
\label{kqqb}
k^{q\B{q},F+B}_{mn}(w) = -2(1-w), \SSP
k^{q\B{q},F-B}_{mn}(w) = 2(1+w)\ln w
\eeq
and
\beq
\Dl_{mn} P_{ij} \equiv \Dl_m P_{ij} + \Dl_n P_{ij}.
\eeq

For the $qg$ subprocess, we obtain
\beq
\label{iqg}
I^{ab,F\pm B}_{mn} = -\fr{1}{\ve}
\Dl^{ab}_{n/m} P^n_{qg}(w) + 2 \ln(1-w)
\Dl^{ab}_{n/m} P^4_{qg}(w) + k^{ab,F\pm B}_{mn}(w),
\eeq
%\narrowtext \noindent
with
\beqa
\label{kqgfirst}
k^{\qB g,F+B}_{mn}(w) &=&
k^{g \qB,F+B}_{mn}(w) = \fr{(1-w)}{4}(1+3w), \\
k^{\qB g,F-B}_{uu,lu}(w) &=&
\fr{(1-w)}{4}(1+3w) + (1-w)\ln w, \\
k^{\qB g,F-B}_{ll,ul}(w) &=&
\fr{(1-w)}{4}(1+3w) - (1-w)\ln w, \\
k^{g \qB,F-B}_{uu,ul}(w) &=&
-\fr{(1-w)}{4}(1+3w) - 2w^2 \ln w, \\
\label{kqglast}
k^{g \qB,F-B}_{ll,lu}(w) & = &
-\fr{(1-w)}{4}(1+3w) - 2w \ln w
\eeqa
and
\beq
\Dl_{n/m}^{ab} P_{qg} \equiv \lf\{ \begin{array}{ll}
\Dl_n P_{qg} & : b=g \\
\Dl_m P_{qg} & : a=g
\end{array}
\rt. \,\,.
\eeq

%\vglue 1cm
%\begin{center}\begin{large}\begin{bf}
\section{FACTORIZATION OF MASS SINGULARITIES}
%\end{bf}\end{large}\end{center}
%\vglue .3cm

The term $\sim 1/\ve^2$ in (\ref{e49}) represents a soft divergence
(and simultaneous mass singularity)
and is cancelled by an opposite term in (\ref{e50}). The remaining
terms $\sim 1/\ve$ in (\ref{e50}), (\ref{e49}), (\ref{iqg}) represent
mass singularities and do not cancel. Hence, they must be removed by
expressing the bare parton distributions (and fragmentation functions
in processes with final state mass singularities)
in terms of the renormalized ones \cite{Ellis}. Thus, our parton
model expression (\ref{e20}) is understood to be initially written
in terms of the unrenormalized parton distributions, whose relation
to the renormalized ones is given by
\beqa
\label{e41}
\nn
 f_k^{0,i/A}(x,\mu^2) &=&  f_k^{r,i/A}(x,\mu^2) \\
\nn
 & & +  \fr{c(\ve)}{\ve} \sum_j \fr{\al_j}{2\pi}\int_x^1\!\! \fr{dy}{y}
  f_k^{r,j/A}(y,\mu^2) 
 \\ &  & \times
[\Dl_k P_{ij}^4(x/y) + \ve \Dl_k T_{ij}(x/y)],
\eeqa
where
\beq
\label{e42}
\fr{c(\ve)}{\ve} \equiv \fr{1}{\ve} (4\pi)^{\ve}
\fr{\Gm(1-\ve)}{\Gm(1-2\ve)}
= \fr{1}{\ve} - \gm_{\rm E} + \ln 4\pi
+ {\cal O}(\ve)
\eeq
and $\al_j=\al_s$, unless $j$ (=$A$) = $\gm$, in which case
$\al_j = \al$. 
After making the above substitution in (\ref{e20}), we recover the
same form, except that now the subprocess cross section is finite and
the parton distributions which enter are the renormalized (finite) ones.
This will be demonstrated explicitly later.
Note that photons, like hadrons, have internal partonic
structure. Hence (\ref{e41}) is not restricted to hadronic initial 
states.
We do not introduce the any additional mass scale
here associated with the mass singularities as is commonly done.
In this form,
the above definition is consistent with the usual $\ovl{\rm MS}$
definition of renormalization when $\Dl_k T_{ij}=0$, as will be 
discussed shortly.

The fragmentation functions ${\cal D}^{A/i}$, which represent
the probability for quark $i$ to split into a collinear hadron 
(or photon) $A$, with momentum fraction $z$, have the following
renormalization:
\beqa
\label{e43}
\nn
  {\cal D}_k^{0,A/i}(z,\mu^2)
&=& {\cal D}_k^{r,A/i}(z,\mu^2) \\
\nn & &
+  \fr{c(\ve)}{\ve} \sum_j \fr{\al_j}{2\pi}\int_z^1\!\! \fr{dy}{y}
 {\cal D}_k^{r,A/j}(y,\mu^2)
\\ & & \times 
[\Dl_k P_{ji}^4(z/y) + \ve \Dl_k T_{ji}(z/y)].
\eeqa

All of the freedom in the factorization scheme is parametrized
by the subtraction terms $T_{ij}(x/y)$. In theory, the $T_{ij}$
could depend explicitly on $x$ as well.
Although we will discuss constraints on the 
allowed $T_{ij}$ in some
detail later in this section, it is worthwhile to briefly summarize
the most commonly used schemes here. In Table \ref{Tab1} 
we list various schemes
of interest. For the ${\rm MS}$ (as opposed to $\ovl{\rm MS}$) version
of any of these schemes, simply add $(\gamma_E-\ln 4\pi)\Dl_k P_{ij}^4 $
to the corresponding $\Dl_k T_{ij}$.

The $\ve$ modified minimal subtraction scheme ($\ovl{\rm MS}_{\ve}$)
was introduced in \cite{bkprd}, the `polarized' modified minimal 
subtraction scheme ($\ovl{\rm MS}_p$) was introduced in \cite{GorV}
and the `helicity conserving' $\ovl{\rm MS}$ scheme ($\ovl{\rm MS}_{HC}$),
as we shall refer to it, 
corresponds to the scheme used in determining the two-loop polarized
splitting functions in \cite{VanN}--\cite{VogL2}, 
although it was inaccurately referred to
as $\ovl{\rm MS}$ there and in many other papers making use of it.
This inaccuracy appears to have been acknowledged in \cite{VogL2}.

In HVBM regularization, the $\ovl{\rm MS}$ scheme leads to helicity
nonconservation of massless fermions. Hence it is not generally used.
The other three schemes are helicity conserving,
at least in the direct sense. A potential problem in the 
$\ovl{\rm MS}_{HC}$ scheme will be discussed later in this section.
For the 
$\ovl{\rm MS}_{HC}$ and $\ovl{\rm MS}_p$ schemes, one must use the
$\ovl{\rm MS}$ scheme as the corresponding unpolarized factorization
scheme when comparing polarized parton distributions to unpolarized
ones. This is implicit in the definition of those two schemes, which
are defined only for polarized parton distributions. This is an 
advantage since unpolarized parton distributions determined in the
$\ovl{\rm MS}$ scheme are widely available, in contrast to the
$\ovl{\rm MS}_{\ve}$ scheme.
On the other hand, it is not clear
what effect these two schemes
could have on various positivity constraints which
require polarized quantities to be smaller in magnitude than the 
corresponding unpolarized ones, as will be discussed briefly at the end
of this section. Nonetheless, presently available
NLO polarized parton distributions are predominantly determined in the 
$\ovl{\rm MS}_{HC}$ scheme, so we will use it in making physical 
predictions.

The $\ovl{\rm MS}_{\ve}$ scheme is the only of the three helicity conserving
schemes with the following properties. {\bf (a)} It treats the polarized
and unpolarized parton distributions analogously. {\bf (b)} It gives
regularization scheme independent analytical results and parton
distributions (equivalent to those of DRED $\ovl{\rm MS}$).
{\bf (c)} It satisfies the supersymmetric identity 
\beq
\Dl_k P_{\tl{g}\tl{g}} + \Dl_k P_{g\tl{g}} = 
\Dl_k P_{\tl{g}g} + \Dl_k P_{gg},
\SSP \mbox{$\tl{g} =$ gluino},
\eeq
for the unpolarized \cite{Flor,Kun} and polarized 
\cite{VanN,VogL} two-loop spacelike splitting functions as well as the
timelike \cite{VogL2} splitting functions (both polarized and
unpolarized). 
This is due to the equivalence
with DRED, which in turn implies the applicability of four-dimensional
calculational techniques. The timelike splitting functions are those which
enter in (\ref{e43}). They only begin to differ from the spacelike
ones at the two-loop level.

Substituting (\ref{e41}) in (\ref{e20}) and convoluting with
the $n$-dimensional
Born term of the regularization scheme being used
generates several {\em factorization}
counterterms. Those associated with the $q\B{q}$ subprocess 
($ab=q\B{q},\B{q}q$) have the form
%\widetext
\beq
\lf(\fr{d\ha{\sg}^{ab,F\pm B}_{mn, BB'}}{dM}\rt)_{\rm ct} 
\label{e48}
= \fr{1}{\ve} \chi^{ab,F\pm B}_{mn,BB'}(\ve)  \fr{\al_s}{2\pi}
w^{1+\ve} C(\ve)
\lf(\fr{s}{\mu^2}\rt)^{\ve}  
[\Dl_{mn} P_{qq}^4(w) + \ve \Dl_{mn} T_{qq}(w)]
.
\eeq
The remaining terms are associated with the $qg$ subprocess 
($ab=\qB g, g\qB$) and have the form
\beq
\label{e53}
\lf(\fr{d\ha{\sg}^{ab,F\pm B}_{mn, BB'}}{dM}\rt)_{\rm ct} 
= \fr{1}{\ve} \tl{\chi}^{ab,F\pm B}_{mn,BB'}(\ve)  \fr{\al_s}{2\pi}
w^{1+\ve} C(\ve)
\lf(\fr{s}{\mu^2}\rt)^{\ve} 
[\Dl_{n/m}^{ab} P_{qg}^4(w) + \ve \Dl_{n/m}^{ab} T_{qg}(w)]
.
\eeq

From \cite{bkprd} we may infer the relation between parton distributions
of any regularization and factorization scheme, denoted 1, with any
other, denoted 2:
\beqa \nn 
 f_{k,1}^{i/A} (x) 
&=&  f_{k,2}^{i/A}(x)
+  \sum_j \fr{\al_j}{2\pi}\int_x^1\!\! \fr{dy}{y}  
 f_{k,2}^{j/A}(y) 
\{
[\Dl_k P_{1,ij}^\ve(x/y) - \Dl_k T_{1,ij}(x/y)]
\\ \label{fia}
 && -
[\Dl_k P_{2,ij}^\ve(x/y) - \Dl_k T_{2,ij}(x/y)]
\}
+ {\cal O} (\al_j\al_s).
\eeqa
Similarly, for the fragmentation functions,
\beqa \nn 
 {\cal D}_{k,1}^{A/i}(z) 
&=&  {\cal D}_{k,2}^{A/i}(z)
+  \sum_j \fr{\al_j}{2\pi}\int_z^1\!\! \fr{dy}{y}  
 {\cal D}_{k,2}^{A/j}(y) 
\{
[\Dl_k P_{1,ji}^\ve(z/y) - \Dl_k T_{1,ji}(z/y)]
\\ \label{dia}
 && -
[\Dl_k P_{2,ji}^\ve(z/y) - \Dl_k T_{2,ji}(z/y)]
\}
+ {\cal O} (\al_j \al_s).
\eeqa

Going from
scheme 1 to scheme 2 simply amounts to expressing the
parton distributions and fragmentation
functions of scheme 1 in terms of those of scheme 2, as
was shown in \cite{bkprd}. Also, we see that working in any consistent
$n$-dimensional
regularization scheme and any factorization scheme is equivalent to
working in any other consistent $n$-dimensional regularization scheme 
with the appropriate factorization scheme. Thus, 
according to our definition of consistency given
in section III, a consistent $n$-dimensional regularization
scheme must (a) be mathematically
consistent, for the class of subprocesses under consideration,
and (b) have a process independent set of
splitting functions within that class, in order that it may be
related to the HVBM scheme as described in section III
(assuming DREG and HVBM themselves do not fail for some process, unlikely
as it may seem).
The DRED scheme satisfies both criteria for one-loop QCD
corrections to  polarized processes.  

From \cite{bkprd} we may also infer that the conversion term which must
be added to the subprocess
cross section calculated in scheme 1 to obtain the
subprocess cross section in scheme 2 is obtained via
\beqa
\label{q105}
\nn
\ha{\sg}_{conv} = \sum_l \ha{\sg}_{ct,l}\{
[\Dl_{k} P_{ij}^4(w_l) + \ve \Dl_{k} T_{1,ij}(w_l)] 
&\RA & \ve [ \Dl_{k} T_{2,ij}(w_l) - \Dl_k P_{2,ij}^{\ve}(w_l)]   \\
 &-&
\ve [ \Dl_{k} T_{1,ij}(w_l) - \Dl_k P_{1,ij}^{\ve}(w_l)]
\},
\eeqa
%\narrowtext \noindent
where $w_l$ is the argument of the splitting functions in the 
counterterm $\ha{\sg}_{ct,l}$.
The conversion terms are generated by making the above substitution
in all factorization counterterms for both initial and final
state mass singularities in general. 
As well, one must express the couplings of scheme 1 in terms of those
in scheme 2 for processes where coupling constant renormalization
enters. Then, the above form is valid for all one-loop QCD corrections,
not just Drell-Yan and DIS. Also, it is straightforward to verify that
making the substitutions (\ref{fia}), (\ref{dia}) in any one-loop
cross section calculated in scheme 1, leads to the conversion term
(\ref{q105}).

Parton distributions determined using different
processes, but the same scheme, will differ by 
${\cal O}(\al_s^{n+1})$ when ${\cal O}(\al_s^{n})$ corrected
cross sections are used in the fits. 
Since a basic assumption of the
parton model is that the all-orders parton distributions will be
process independent, the process dependence of the parton distributions
can be seen to arise from the neglected higher orders. The physical
parton distributions are in fact the all-orders parton distributions
since, in reality, it is the all-orders 
subprocess cross section that they are 
convoluted with. Hence, if the all-orders parton distributions were
not process independent, there would exist no process independent
parton distributions.  By all orders, we mean sufficently high
order in $\al_s$ that higher orders may be neglected for all practical
purposes.
The process dependence is therefore
easy to understand since some processes have large QCD corrections
(i.e.\ Drell-Yan) while others have relatively small corrections
(i.e.\ DIS). Hence the extent to which the ${\cal O}(\al_s^{n})$
and ${\cal O}(\al_s^{n+1})$ parton distributions differ,
in any particular scheme, will depend
on the process from which they are determined. So, for low $n$
(i.e.\ $n=0$), processes with significantly different QCD corrections
will yield somewhat different parton distributions, but the ones with
smaller QCD corrections will lead to parton distributions which are 
closer to the all-orders ones. 
One expects, on the other hand, that the ratio of the polarized to
unpolarized parton distributions will not have a large process 
dependence since, even for Drell-Yan like processes, spin asymmetries
tend to exhibit perturbative stability as we shall demonstrate in
sections IX -- XI.

Equation (\ref{fia}) is very powerful and restrictive. This is because
it has implications for various conservation rule motivated sum
rules. According to the parton model,  the number
of valence (constituent) quarks, $q_v$, in a hadron, $A$, is given by
\beq
\label{q106}
N^{q_v/A} = \int_0^1 \!\! dy  f_u^{q_v/A}(y) = {\rm const}.
\eeq
More specifically,
\beqa
\nn
N^{d_v/p} &=& N^{d/p}-N^{\B{d}/p} = 1, \\
\label{nquarks}
N^{u_v/p} &=& N^{u/p}-N^{\B{u}/p} = 2.
\eeqa
The fact that this must hold for all energy scales, $\mu$, is
an expression of charge/probability conservation. Since $N^{q/A}$
is just the first moment of $f_u^{q/A}(y)$ and since the $k^{\rm th}$
 moment
of a convolution is the product of the $k^{\rm th}$  moments of
the functions being convoluted, it is easy to see
what happens to the above conserved quantity as one goes from one scheme
to another.

Firstly, we note that the gluonic contribution of (\ref{fia}) to the relation
between quark distributions of different schemes cancels for the 
valence distributions.
Suppose we wish to see what happens when going from regularization
scheme 2 to regularization scheme 1.  Using (\ref{fia}),
we get, for valence quarks,
\beqa
\nn
N^{q_v/A}_1 = N^{q_v/A}_2 \{1 &+& \fr{\al_s}{2\pi}
[(\Dl_u \ovl{P}^\ve_{1,qq} - \Dl_u \ovl{T}_{1,qq}) \\
\label{n1n2}
&-& (\Dl_u \ovl{P}^\ve_{2,qq} - \Dl_u \ovl{T}_{2,qq})]\},
\eeqa
where $\ovl{g}$ denotes the first moment of
$g(y)$, for some function $g$. Suppose scheme 1 is DRED, for which
$\Dl_u P^\ve_{DRED,qq}=0$, and scheme 2 is DREG, for which
\beq
\Dl_u \ovl{P}^\ve_{DREG,qq}=0,
\eeq
as can be seen from (\ref{splitu}).  Assuming the same factorization
scheme, i.e.\ $\Dl_u T_{2,ij} = \Dl_u T_{1,ij}$, we obtain
\beq
N^{q_v/A}_{DRED} =
N^{q_v/A}_{DREG}, 
\eeq
so that the two regularization
schemes are mutually physically consistent regarding
charge conservation. The requirement,
\beq  
\label{q109p}
\Dl_u \ovl{P}^n_{qq}=0, 
\eeq
follows from the fact that the 
${\cal O}(\al_s)$ corrections to charge conservation must be zero
in $n$ dimensions. By corrections, we mean that one can think of
a zeroth order piece to $\Dl_u P_{qq}(z)$, $\Dl_u P_{gg}(z)$ which is
proportional to $\dl(1-z)$ and denotes the no interaction scenario.
Then, it is clear that the usual leading order splitting functions
denote ${\cal O}(\al_s)$ corrections to the non-interacting case.

Let us now consider momentum conservation. Let $P^{i/A}$ denote the total
momentum fraction carried by parton $i$. It is just the second moment
of $f_u^{i/A}$:
\beq
P^{i/A} = \int_0^1 \!\! dy y  f_u^{i/A} (y).
\eeq
Of course, momentum conservation must be applied to the sum
over all partons. One can easily verify that if,
in some factorization scheme, the DRED parton
distributions satisfy the momentum conservation rule
\beq
\label{mquarks}
\sum_i P^{i/A} = 1,
\eeq
then the DREG parton distributions,
in that same factorization scheme, will also satisfy it, and 
vice-versa. The basic reason is that the requirement, which DREG
satisfies, that
\beqa
\nn
\lefteqn{\int_0^1 \!\! dz z [\Dl_u P^n_{qq}(z) + \Dl_u P^n_{gq}(z)]} \\
\label{intpij}
 & & = \int_0^1 \!\! dz z [\Dl_u P^n_{gg}(z) + 2 N_F \Dl_u
P^n_{qg}(z)] = 0,
\eeqa
(which follows from the fact that the ${\cal O}(\al_s)$ corrections
to momentum conservation must be zero, in $n$ dimensions) 
leads to cancellations in the
sum over partons, so that the net difference in total momentum 
between schemes is zero.
This is easily verified from Eq.\ (\ref{splitu}).
The individual quark and gluon momentum fractions will depend on the
regularization (and factorization) scheme
as well as the energy scale, $\mu$,
 to ${\cal O}(\al_s)$,
but a change in the quark momentum fraction is cancelled by an
exactly opposite change in the gluon momentum fraction.
Thus, the total momentum is the same in both schemes, and is 
independent of $\mu$, as follows from DGLAP evolution \cite{DGL,AP}.

All this discussion tells us that DRED and DREG are mutually physically
consistent in general, in the unpolarized case. 
One can convert unpolarized parton densities
from one regularization
scheme to the other and the conserved quantities retain 
their values.
The real problem arises when one
converts from one factorization scheme to another.
It is safe to go from $\ovl{\rm MS}$ to ${\rm MS}$ since,
in the ${\rm MS}$ scheme, all
the $\Dl_k T_{ij}$ are proportional to $\Dl_k P^4_{ij}(\gm_E - \ln 4\pi)$, 
and hence do not change the conserved quantities.
Also, since DRED is equivalent to the 
$\ovl{\rm MS}_\ve$ scheme, we see that $\ovl{\rm MS}_\ve$
and $\ovl{\rm MS}$ are mutually physically consistent as well,
for unpolarized processes. In general though, for arbitrary
$\Dl_k T_{ij}$, there are no analogous constraints to
(\ref{q109p}) and (\ref{intpij}), at least without invoking constraints
from higher order evolution. 
This means that the aforementioned conserved quantities
will differ in general to ${\cal O}(\al_s)$ when going
from one arbitrarily chosen factorization scheme to another.
Since it would be quite a formidable task to measure so precisely
the parton distributions that one could definitely verify all
conserved quantities to high accuracy,
how do we know which schemes are correct a-priori? This is even more 
pertinent to the case of the polarized parton distributions, where the
data is definitely lacking.

We may gain some insight into this issue by studying the scheme
dependence of the polarized (nonsinglet) distributions.
The net spin, $S^{q/A}$, carried by quarks of flavor $q$, in
hadron $A$, and the corresponding total quark spin, 
$\fr{1}{2} \Dl \Sg_A$, are given by
\beq
\label{squarks}
S^{q/A} = \fr{1}{2} \int_0^1 \!\! dy  f_l^{q/A}(y), \SSPP
\fr{1}{2} \Dl \Sg_A = \sum_q S^{q/A},
\eeq
respectively.
The spin carried by the valence quarks
is conserved under one-loop evolution
due to chirality conservation. More formally,
$\Dl_l \ovl{P}_{qq}^4 = 0$, 
substituted in the expression for the first moment of $f_l^{q/A}$,
evolved using the DGLAP equations gives $S^{q_v/A} = {\rm const}$,
under one-loop evolution, as was the case for $N^{q_v/A}$. 
For a discussion of two-loop valence evolution effects, which in
general violate the above conservation, the reader is referred to
\cite{Nonsing}. The main point of interest concerning two-loop 
evolution is that chirality conservation is relevant to nonsinglet
combinations of the form $f_l^q+f_l^{\B{q}}-f_l^{q'}-f_l^{\B{q}'}$, and
linear combinations thereof. The two interesting ones $\Dl A_3$, 
$\Dl A_8$ satisfy
\beq
\label{q113}
\Dl \ovl{A}_{3,8} = {\rm const},
\eeq
where
\beqa
\nn
\Dl A_3 &=& f_l^u+f_l^{\B{u}}-f_l^d-f_l^{\B{d}} \\
\Dl A_8 &=& f_l^u+f_l^{\B{u}}+f_l^d+f_l^{\B{d}}
-2(f_l^s+f_l^{\B{s}}).
\eeqa

Similar
argumentation to that 
used in relating $N_1^{q_v/A}$ to $N_2^{q_v/A}$ 
leads to
\beqa
\label{ed120}
\nn
\Dl \ovl{A}_{3,8}^1 = \Dl \ovl{A}_{3,8}^2 \{1 &+& \fr{\al_s}{2\pi} 
[(\Dl_l \ovl{P}^\ve_{1,qq} - \Dl_l \ovl{T}_{1,qq}) \\
\label{q113p}
&-& (\Dl_l \ovl{P}^\ve_{2,qq} - \Dl_l \ovl{T}_{2,qq})]\},
\eeqa
From Table \ref{Tab1}
and Eq.\ (\ref{splitp}) we get
\beqa
\nn
\Dl \ovl{A}_{3,8}^{DRED(\ovl{\rm MS})} &=&
\Dl \ovl{A}_{3,8}^{HV(\ovl{\rm MS}_{\ve})} =
\Dl \ovl{A}_{3,8}^{HV(\ovl{\rm MS}_{HC})}  \\
&=& \Dl \ovl{A}_{3,8}^{HV(\ovl{\rm MS}_{p})} \neq
\Dl \ovl{A}_{3,8}^{HV(\ovl{\rm MS})}.
\eeqa
Thus, the $\ovl{\rm MS}$ scheme of DREG(HVBM) gives  different
values for $\Dl \ovl{A}_3$, $\Dl \ovl{A}_8$ than all the 
`chirality conserving' schemes. This is an explicit example of the
factorization scheme ambiguity.

The real implication of this seeming nonuniqueness problem
is that certain factorization schemes may be ruled out from the 
beginning since they are manifestly unphysical. One can determine
which schemes are unphysical by imposing the fact that the 
unrenormalized parton distributions (\ref{e41}) themselves satisfy the
conservation rules (\ref{q106}), (\ref{mquarks}) and (\ref{q113}) 
and yield the correct value for the corresponding
conserved quantities (except as will be mentioned below).
The reason for this is that if a particular regularization preserves
the necessary symmetries: unitarity, translation invariance
and chiral symmetry, then the unrenormalized parton distributions
in that regularization will obey the corresponding conservation rules:
conservation of
charge/probability, momentum,  and massless quark
chirality.
Hence, the connection between renormalized and unrenormalized
parton distributions must not destroy those symmetries
in the renormalized parton distributions. If, on 
the other hand, a regularization violates a particular symmetry,
a finite $\Dl_k T_{ij}$ will have to be introduced to insure that
the renormalized parton distributions obey the corresponding
conservation rule, since the unrenormalized parton distributions 
themselves will manifestly violate it.

Thus, using similar rationale to that used in going from one
regularization scheme to another, we arrive at the physical 
consistency constraints on allowable factorization schemes,
\beq \label{intt}
\int_0^1 \!\! dz \Dl_u T_{qq}(z) = 0,
\eeq
\beqa 
\nn
\lefteqn{\int_0^1 \!\! dz z [\Dl_u T_{qq}(z) + \Dl_u T_{gq}(z)]} \\
\label{inttij}
 && = \int_0^1 \!\! dz z  [\Dl_u T_{gg}(z) + 2 N_F \Dl_u
T_{qg}(z)] = 0
\eeqa
and
\beq
\label{dltdlp}
\Dl_l T_{qq}(z) - \Dl_l P^{\ve}_{qq}(z)
=
\Dl_u T_{qq}(z) - \Dl_u P^{\ve}_{qq}(z).
\eeq
The reason we had to subtract $\Dl_k P_{qq}^\ve$ in (\ref{dltdlp}) is that
for regularization schemes like HVBM which violate
chiral symmetry ($\Dl_l P_{qq}^\ve \neq \Dl_u P_{qq}^\ve$),
the unrenormalized parton distributions themselves manifestly
violate  helicity conservation. One must always subtract
the helicity nonconserving part in $\Dl_l P_{qq}^\ve$ in order
to restore this conservation. This is why HVBM($\ovl{\rm MS}$)
disagrees with the other schemes and is manifestly unphysical.
Under two-loop evolution, helicity non-conservation in 
HVBM($\ovl{\rm MS}$) manifests itself via direct violation of
(\ref{q113}). This may be seen by taking scheme 1 to be 
HVBM($\ovl{\rm MS}$) and scheme 2 to be one of the `chirality
conserving' schemes, then differentiating (\ref{ed120}) with
respect to $\ln \mu^2$. Comparing with the DGLAP equation for
$\Dl \ovl{A}^1_{3,8}$ one finds precisely the same nonvanishing
result for $d \Dl \ovl{A}^1_{3,8} / d\ln \mu^2$ as follows from
the two-loop analysis \cite{VanN,VogL}.

The condition (\ref{dltdlp}) is somewhat stronger than the 
conservation rule (\ref{q113}). It also insures that
\beq
\label{helflip}
\ha{\sg}(\mbox{helicity flip}) = 0,
\eeq
where $\ha{\sg}(\mbox{helicity flip})$ generically denotes
any doubly polarized (in the initial state)
subprocess cross section where a massless quark flips helicity,
with the helicity at both ends of the quark line fixed.

The relation between the total quark spin in scheme 1 and that in 
scheme 2 is obtained by taking the first moment of 
(\ref{fia}) and summing over quarks (including antiquarks),
%\widetext
\beqa \nn 
\fr{1}{2} \Dl \Sg_1 
=\fr{1}{2}  \Dl \Sg_2  
+  \fr{\al_s}{4\pi} \biggl\{ && \sum_q   
 \B{f}_l^q  [(\Dl_l \ovl{P}_{1,qq}^\ve - \Dl_l \ovl{T}_{1,qq}) 
- (\Dl_l \ovl{P}_{2,qq}^\ve - \Dl_l \ovl{T}_{2,qq})] 
\\ \label{ec126p}
&& + 2N_f \B{f}_l^g  [(\Dl_l \ovl{P}_{1,qg}^\ve - \Dl_l \ovl{T}_{1,qg})
- (\Dl_l \ovl{P}_{2,qg}^\ve - \Dl_l \ovl{T}_{2,qg})]
\biggr\}.
\eeqa
Since DRED preserves chiral symmetry, it is natural to use it as a 
reference scheme.
If we also desire that the total spin of the quarks 
be the same as in DRED($\ovl{\rm MS}$), 
then $\Dl_l \ovl{P}_{qg}^{\ve}
\neq 0$ in HVBM regularization leads to violation of this in the
$\ovl{\rm MS}_{HC}$ scheme. We must impose
\beq
\label{totspin}
\int_0^1 dz [ \Dl_l T_{qg}(z) -  \Dl_l P_{qg}^{\ve}(z)] = 0
\eeq
to restore equivalence. Hence, of the factorization
schemes listed in Table \ref{Tab1},  only 
$\ovl{\rm MS}_{\ve}$ and $\ovl{\rm MS}_p$ give the value of total
quark spin equal to that in DRED($\ovl{\rm MS}$).

One may ask if there is some deeper physical reason why the 
$\ovl{\rm MS}_{HC}$ scheme gives a different value for the total
quark spin. We notice, using (\ref{splitu}) and (\ref{splitp}),
that in the HVBM scheme
\beq
\label{pqgn}
P^n_{qg, ++} (z) \neq P^n_{qg, -+} (1-z).
\eeq
This means that when
a polarized gluon splits into a collinear $q\B{q}$ pair, the $q$ and $\B{q}$
do not necessarily have opposite chiralities, as required by
quark chirality conservation. The above result is possible since,
if there is an initial state 
gluon, (\ref{helflip}) does not apply.

The $\ovl{\rm MS}_{\ve}$ and
$\ovl{\rm MS}_{p}$ schemes correct (\ref{pqgn}), while
$\ovl{\rm MS}_{HC}$ does not. More precisely, strictly speaking,
chirality conservation imposes the indirect constraint,
\beqa \label{ee129}
\nn
\lefteqn{[\Dl_u T_{qg}(z)+\Dl_l T_{qg}(z)] -
[\Dl_u P^{\ve}_{qg}(z)+\Dl_l P^{\ve}_{qg}(z)]} \\
& & \nn  =
[\Dl_u T_{qg}(1-z)-\Dl_l T_{qg}(1-z)]   - 
[\Dl_u P^{\ve}_{qg}(1-z)-\Dl_l P^{\ve}_{qg}(1-z)] \\
& & \Rightarrow  \Dl_l T_{qg}(z) - \Dl_l P^{\ve}_{qg}(z) 
  =
 - [\Dl_l T_{qg}(1-z) - \Dl_l P^{\ve}_{qg}(1-z)],
\eeqa
which both $\ovl{\rm MS}_{\ve}$ and
$\ovl{\rm MS}_{p}$ satisfy, but $\ovl{\rm MS}_{HC}$ does not. 
The reason, as usual, is that it is the difference,
$\Dl_k P^{\ve}_{ij}(z)-\Dl_k T_{ij}(z)$, which enters in physical
cross sections. Thus, the subtraction term, $\Dl_k T_{ij}(z)$, effectively
modifies the splitting function, $\Dl_k P^{\ve}_{ij}(z)$.
The unpolarized parts in (\ref{ee129}) cancelled due to the unpolarized
constraint, $\Dl_u T_{qg}(z) - \Dl_u P^{\ve}_{qg}(z) 
 = \Dl_u T_{qg}(1-z) - \Dl_u P^{\ve}_{qg}(1-z)$, which is satisfied
in all the schemes considered here.
Satisfaction of (\ref{ee129}) implies satisfaction of
(\ref{totspin}), while the reverse is not true.
An interesting analysis which arrived, from a different perspective,
at a conclusion similar to
(\ref{totspin}), in the context of DIS, may be found in 
\cite{Delg}. How strictly one wishes to impose chirality conservation is 
perhaps a matter of taste, however.

Under one-loop evolution, the total quark spin is conserved. Under
two-loop evolution, conservation or non-conservation depends on the 
choice of factorization scheme.
Analogously to the nonsinglet case, we see that if 
$\Dl \Sg_2$ is conserved and $\Dl \Sg_1 \neq \Dl \Sg_2$, then 
$\Dl \Sg_1$ will not be conserved under two-loop evolution. The scale
dependence of $\Dl \Sg_1$ may be obtained by differentiating 
(\ref{ec126p}) directly. The $\ovl{\rm MS}_{\ve}$ and 
$\ovl{\rm MS}_p$ schemes lead to conservation of $\Dl \Sg$, therefore
$\ovl{\rm MS}_{HC}$ does not. Alternatively worded, $\Dl \Sg$ is not
conserved in the $\ovl{\rm MS}_{HC}$ scheme and the difference in
the schemes exactly compensates this non-conservation such that $\Dl \Sg$
is conserved in the $\ovl{\rm MS}_{\ve}$ and 
$\ovl{\rm MS}_p$ schemes. The evolution of $\Dl \Sg$ in the 
$\ovl{\rm MS}_{HC}$ scheme obtained via differentiation of (\ref{ec126p}) 
agrees with the two-loop analysis \cite{VanN,VogL} and can be seen to
arise from the non-vanishing of $\Dl_l \ovl{P}^{\ve}_{qg}$. Under 
two-loop evolution, it is the non-vanishing of the first moment
of $\Dl_l P_{qq}$ (from the pure singlet part) which leads to the
non-conservation of $\Dl \Sg$, however. Hence the non-vanishing at
two-loops of $\Dl_l \ovl{P}_{qq}$ is seen to be a consequence of the 
non-vanishing at one-loop of $\Dl_l \ovl{P}^{\ve}_{qg}$, while
$\Dl_l \ovl{P}_{qg}$ remains zero at two loops, in all three schemes.

At any rate, we see that satisfaction of (\ref{totspin}) leads to 
conservation of total quark spin. 
The DGLAP evolution equations will be elaborated on in the next section 
in connection with scale dependences.
It is worth noting that analogous 
argumentation can be used to relate the spin carried by the gluons
in different schemes. Since such
differences do not enter in the observables considered here, at
one loop, we will not discuss them. Also, direct differentiation
of (\ref{fia}) gives the correct scheme transformations for all
two-loop splitting functions, generalized to include the regularization
scheme dependence (i.e.\ $T_{ij}\RA T_{ij}-P^{\ve}_{ij}$).

So far, we have not discussed the interesting requirement of 
positivity of the parton distributions,
\beq \label{positivity}
f_{+/+}^{i/I}(x,\mu^2) \geq 0, \SSP 
f_{-/+}^{i/I}(x,\mu^2) \geq 0.
\eeq
The only guarantee of positivity applies to the physical
cross section itself, as measured by experiment.
Individual subprocess cross sections need not satisfy
positivity due to the interconnection of the subprocesses arising
from the renormalization of the parton distributions discussed
earlier in this section.
To the extent that leading order cross sections dominate
and that the parton distributions are process independent
at leading order, positivity
should not be a problem. On the other hand, for processes with
large radiative corrections, some factorization schemes could
lead to parton distributions in violation of (\ref{positivity}). This issue
has not been thoroughly studied at this point, but it will
become important as more experiments involving polarized hadrons
are undertaken.

%\vglue 1cm
%\begin{center}\begin{large}\begin{bf}
\section{ANALYTICAL RESULTS TO ONE-LOOP ORDER}
%\end{bf}\end{large}\end{center}
%\vglue .3cm

We may now present the analytical results for 
$ d\ha{\sg}^{ab,F\pm B}_{mn,BB'}/dM$ in a form valid for all 
consistent $n$-dimensional
regularization schemes and for a general factorization scheme,  starting 
with the $q\B{q}$ subprocess.
Combining (\ref{q55}), (\ref{e50}), (\ref{ea85}), 
 (\ref{e49}) and  (\ref{e48}), we obtain the total result 
($ab = q\B{q},\B{q}q$), 
\beqa
\label{qqans}
\nn
\lf(\fr{d\ha{\sg}^{ab,F\pm B}_{mn,BB'}}{dM}\rt)_{\rm NLO} 
&=& \chi^{ab,F\pm B}_{mn,BB'}(0)
\Biggl( \dl(1-w) +  C_F \fr{\al_s}{2\pi}w
\Biggl\{\lf( \fr{2\pi^2}{3}-7 \rt) \dl(1-w)  \Biggr.\Biggr.\\ 
\nn
& &  +8\lf(\fr{\ln(1-w)}{1-w}\rt)_+
+ \fr{\Dl_{mn}P_{qq}^4(w)}{C_F} \ln\fr{M^2}{w\mu^2} 
 - 4(1+w)\ln(1-w) 
\\  && 
 +\fr{1}{C_F}[\Dl_{mn} T_{qq}(w) - \Dl_{mn} P_{qq}^\ve(w)]
     + k^{q\B{q},F\pm B}_{mn}(w)  
%  \!\!\!\!\!\!  \lf.\lf. \begin{array}{c} \mbox{} 
%   \\ \mbox{} \end{array} \rt\}\rt),
\Biggl. \Biggl.  \Biggr\} \Biggr),
\eeqa
where the $k^{q\B{q},F\pm B}_{mn}$ are given in (\ref{kqqb}).

We now present the result for the $qg$ subprocess. There is no
${\cal O}(1)$ term (in $\al_s$). 
Combining (\ref{ea85}), (\ref{iqg}) and (\ref{e53}), 
we obtain the total result ($ab = \qB g,g\qB$),
\beqa
\label{qgans}
\nn
\lf(\fr{d\ha{\sg}^{ab,F\pm B}_{mn,BB'}}{dM}\rt)_{\rm NLO} 
&=& \tl{\chi}^{ab,F\pm B}_{mn,BB'}(0)
\fr{\al_s}{2\pi} w
\Biggl\{
\Dl_{n/m}^{ab} P_{qg}^4 (w)\lf[\ln\fr{M^2}{w\mu^2} +2\ln(1-w)\rt]  \Biggr. \\
  &&  
+[\Dl_{n/m}^{ab} T_{qg}(w) - \Dl_{n/m}^{ab} P_{qg}^\ve(w)]
     + k^{ab,F\pm B}_{mn}(w)  
%  \!\!\!\!\!\!  \lf. \begin{array}{c} \mbox{} 
%   \\ \mbox{} \end{array} \rt\},
 \Biggl. \Biggr\},
\eeqa
where the $k^{ab,F\pm B}_{mn}(w)$ are given by (\ref{kqgfirst}) -- 
(\ref{kqglast}).

We see at this point that the results of \cite{bkprd} (for the
$uu$ and $ll$, $F+B$, vector coupling case with massless $B=B'$)
are contained within the results presented here. Those, in turn,
were found to be consistent with various existing results calculated
in certain specific schemes, for both the $uu$ \cite{AEM,Hamb} and
$ll$ \cite{Rat,Weber92} cases. Single-spin $W^\pm$ production in 
hadron-hadron collisions has been studied in \cite{Weber93}.
Considering the subprocess $ab \RA W^\pm c$, where $a$ is longitudinally
polarized and $b$ is unpolarized, we see that for $ab=q\B{q},\B{q}q$ and
for $a=g,b=\qB$, our expressions reproduce those given in Appendix B
of \cite{Weber93}. For $a=\qB,b=g$, it seems at first sight that there is
a discrepancy. In order to reproduce the result of \cite{Weber93}, one
needs $\Dl_u P^\ve_{qg} = -1/2$, in disagreement with (\ref{splitu}).
The reason for this disagreement is that \cite{Weber93} averages over
2, rather than $2-2\ve$, gluon polarization states, as can be inferred from
their Eq.\ (10). That convention does indeed lead to 
$\Dl_u P^\ve_{qg} = -1/2$. Unfortunately, all of the presently 
available unpolarized parton distributions are determined using the
convention of $2-2\ve$ gluon polarization states. Consequently, the 
expressions of \cite{Weber93}, for the $qg$ subprocess, should be
appropriately converted using the technique of \cite{bkprd}, which
amounts to Eq.\ (\ref{q105}) of this paper, before being used in
numerical studies demanding complete consistency at NLO.

At this point, we may investigate the scale dependence of the
cross section. Define
\beq
\sg \equiv \fr{d\sg_{mn}^{AB,F\pm B}}{dM}
\eeq
and let $\tl{\sg}$ denote the contribution to $\sg$ which depends
on $\mu^2$ at ${\cal O}(\al_s)$ coming from $a=q$, $b=\B{q}$;
$a=q$, $b=g$ and $a=g$, $b=\B{q}$ for one quark flavor
and for production by one bosonic interference channel. 
These subprocesses are connected via the renormalization of the parton
distributions, but are disconnected from the remaining subprocesses
in that sense. We have (taking ${\cal S}^A_m={\cal S}^B_n=1$)
\beqa \label{sgmu}
\tl{\sg} &=& \chi_{mn}^{q\B{q},F\pm B}(0) 
\lf\{ \int_{\tau}^1 \!\! dx_{a,b}
\int_{\tau/x_{a,b}}^1 \!\!  dx_{b,a} f_m^q(x_a,\mu^2)
f_n^{\B{q}}(x_b,\mu^2) \dl(1-\fr{\tau}{x_a x_b}) \rt. \\ 
\nn
&&  -\fr{\al_s}{2 \pi} \int_{\tau}^1 dx_b  \fr{\tau}{x_b} 
f_n^{\B{q}}(x_b,\mu^2) \int_{\tau/x_b}^1
\!\! \fr{dx_a}{x_a} [ f_m^{q}(x_a,\mu^2) \Dl_m P^4_{qq}(w) \ln\mu^2 
+  f_m^{g}(x_a,\mu^2) \Dl_m P^4_{qg}(w) \ln\mu^2] 
 \\  
\nn
&& \lf.  -\fr{\al_s}{2 \pi} \int_{\tau}^1 dx_a \fr{\tau}{x_a}
f_m^{q}(x_a,\mu^2) \int_{\tau/x_a}^1
\!\! \fr{dx_b}{x_b} [ f_n^{\B{q}}(x_b,\mu^2) \Dl_n P^4_{qq}(w) \ln\mu^2 
+  f_n^{g}(x_b,\mu^2) \Dl_n P^4_{qg}(w) \ln\mu^2] 
 \rt\}, 
\eeqa
where 
\beq
f_k^j(x_{a,b},\mu^2) \equiv f_k^{j/A,B}(x_{a,b},\mu^2).
\eeq
In DIS, one has an analogous expression, except without the convolution
with the $f_n^{\B{q}}$ term.
The independence of $\sg$ on $\mu$ to ${\cal O}(\al_s)$ implies
\beq
\label{e128}
\fr{d \tl{\sg}}{d\ln\mu^2} = {\cal O}(\al_s^2).
\eeq
Here we note
\beq
\label{q133}
\fr{d\al_s}{d\ln\mu^2} =  - \fr{\beta_0}{4\pi} \alpha_s^2
= {\cal O} (\al_s^2).
\eeq
Taking into account the fact that the evolution of $f_m^{q/A}$ is
independent of $f_n^{\B{q}/B}$ and $f_n^{g/B}$, we may arbitrarily set,
for any fixed $\mu^2$,
$f_n^g(x_b,\mu^2)=0$ and $f_n^{\B{q}}(x_b,\mu^2)=\dl(1-x_b)$, after
performing the differentiation in (\ref{e128}). Of course, $f_n^{g/B}$ and
$f_n^{\B{q}/B}$ still evolve with $\mu^2$. We then use (\ref{e128}) with
$m=n$, $f_n^g(x_a,\mu^2)=f_n^g(x_b,\mu^2)=0$ and   
$f_n^{q}(x_a,\mu^2)=\dl(1-x_a)$, $f_n^{\B{q}}(x_b,\mu^2)=\dl(1-x_b)$
to solve for and eliminate $df_n^{\B{q}}(x_b,\mu^2)/d\ln\mu^2$. The result for
$df_n^{\B{q}}(x_b,\mu^2)/d\ln\mu^2$ is simply 
$\fr{\al_s}{2\pi} \Dl_n P^4_{qq}(x_b)$.

We thus obtain
\beq
\label{e129}
\fr{d f_k^{q}(x,\mu^2)}{d \ln\mu^2} = \fr{\al_s(\mu^2)}{2\pi}
\int_x^1\,\, \fr{dy}{y} [ f_k^{q}(y,\mu^2) \Dl_k P^4_{qq}(x/y)
+ f_k^{g}(y,\mu^2) \Dl_k P^4_{qg}(x/y) ] + {\cal O}(\al_s^2),
\eeq
for $k=u,l$.
This is the DGLAP equation \cite{DGL,AP} with one-loop splitting
functions. Therefore, one needs at least 
one-loop evolution to
guarantee $\mu^2$ independence to ${\cal O}(\al_s)$.
Gluonic evolution only enters at the ${\cal O}(\al_s^2)$ level.

We may write, using the Taylor expansion for fixed $x$,
\beqa
\label{tayl}
\nn
f_k^{q}(x,\mu^2) &=&   f_k^{q}(x,\mu_0^2) + 
\fr{\al_s(\mu_0^2)}{2\pi} \ln\lf(\fr{\mu^2}{\mu_0^2}\rt) c_{1,k}^q(x,\mu_0^2)
+ \lf[\fr{\al_s(\mu_0^2)}{2\pi} \ln\lf(\fr{\mu^2}{\mu_0^2}\rt)\rt]^2 
c_{2,k}^q(x,\mu_0^2) 
+ \cdots \\ \nn
&\equiv&  
f_k^{q}(x,\mu_0^2) + 
\fr{\al_s(\mu_0^2)}{2\pi} \ln\lf(\fr{\mu^2}{\mu_0^2}\rt) 
 c_{1,k}^q(x,\mu_0^2) \\
&& +\lf[\fr{\al_s(\mu_0^2)}{2\pi} \ln\lf(\fr{\mu^2}{\mu_0^2}\rt) \rt]^2 
\tl{f}_k^q(x,\mu^2,\mu_0^2),
\eeqa
where
\beq
c_{1,k}^q(x,\mu_0^2) = 
\int_x^1 \fr{dy}{y} [ f_k^{q}(y,\mu_0^2) \Dl_k P^4_{qq}(x/y)
+ f_k^{g}(y,\mu_0^2) \Dl_k P^4_{qg}(x/y) ]. 
\eeq
One can write an analogous expression for $f_k^{g}(x,\mu^2)$
using the DGLAP evolution equation for $f_k^{g}(x,\mu^2)$, yielding
\beq
\label{c1kg}
c_{1,k}^g(x,\mu_0^2) = 
\int_x^1  \fr{dy}{y} [ \sum_{i=1}^{2N_f}f_k^{q_i}(y,\mu_0^2) 
\Dl_k P^4_{gq}(x/y)
+ f_k^{g}(y,\mu_0^2) \Dl_k P^4_{gg}(x/y) ]. 
\eeq
%\narrowtext \noindent
Then one could 
express $\sg$ in terms of the $\tl{f}$ and parametrize explicitly
the ${\cal O}(\al_s^2)$ $\mu^2$ dependence with respect to some
fixed scale $\mu_0^2$.
Suppose $\mu_0^2$ is a typical scale of the process (i.e.\
$M^2$), then there will not be any large logarithms and 
$\sg(\mu_0^2)$ will have reasonable ${\cal O}(\al_s)$ corrections
when $\mu^2 = \mu_0^2 \gg \Lam^2$. Having identified $\mu_0^2$
as being a scale which leads to reasonable ${\cal O}(\al_s)$
corrections, we now ask: what happens when we vary $\mu^2$
away from $\mu_0^2$?
Using (\ref{tayl}) in the parton model expression, we see that there
will be $\mu^2$ dependent terms will be proportional to
\beq
\sg(\mu^2) - \sg(\mu_0^2) \sim 
\lf[\fr{\al_s}{2\pi} \ln(\mu^2/\mu_0^2)\rt]^2
\sim \lf[ \fr{\ln(\mu^2/\mu_0^2)}{\ln(\mu^2/\Lam^2)}\rt]^2
\eeq
times an overall factor having roughly the same magnitude as
the Born term. For typical values of $\mu$ and $\al_s$,
$\fr{\al_s}{2\pi} \ln(\mu^2/\mu_0^2)$ is a small number so that the 
$\mu^2$ dependence is rather suppressed so long as
$\mu^2$ is not too far from $\mu_0^2$.
These terms will only cancel when two-loop corrections to the 
subprocess cross section are included. There are also terms with
lower powers of $\ln(\mu^2/\mu_0^2)$ (i.e.\ from (\ref{q133}))
which require two-loop corrections for cancellation as well.

The next question is: how much better are we doing than a Born
level calculation? In the Born level result, the $\mu^2$
dependence comes entirely from the evolution of the distributions.
Hence, from (\ref{sgmu}), (\ref{tayl}),
\beq
\sg_{\rm Born}(\mu^2) -
\sg_{\rm Born}(\mu_0^2) \sim
\fr{\al_s}{2\pi} \ln(\mu^2/\mu_0^2),
\eeq
again times an overall factor having roughly the same magnitude
as the Born term. We see explicitly that inclusion of one-loop
corrections reduces the $\mu^2$ dependence by an extra factor of 
$\fr{\al_s}{2\pi} \ln(\mu^2/\mu_0^2)$ for reasonable values of
$\mu^2$. The above arguments apply to all one loop calculations
in hadronic processes. In general, though, there is also a contribution
from the evolution of the gluon distribution at ${\cal O}(\al_s)$, 
relative to the Born term. This is because, in general, the gluon
distribution may enter in the Born term. Then, imposing 
the analogue of (\ref{e128})
would determine $c_{1,k}^g$ as given by (\ref{c1kg}).

One could also convolute the Born cross sections with parton distributions
having no evolution (i.e.\ determined at some fixed scale
$\B{\mu}^2$), which would
correspond to an ``off-shell'' renormalization of the parton distributions
in higher order.
But then the physical predictions would be wrong
by terms $\sim \fr{\al_s}{2\pi} \ln(\mu_0^2/\B{\mu}^2)$, 
since the one loop corrections would contain such terms (analogous
to the terms $\sim   \fr{\al_s}{2\pi} \ln(M^2/\mu^2)$ present in
(\ref{qqans}), (\ref{qgans})). So, even though there is no
renormalization
scale dependence, the error will be larger for $\mu_0\gg\B{\mu}$ or
$\mu_0\ll \B{\mu}$. One expects an analogous tradeoff when using
one-loop evolution instead of two-loop evolution in NLO
calculations.

Actually, this argumentation is somewhat naiive in the sense that
we only considered explicitly
the  $\sim [\fr{\al_s}{2\pi} \ln(\mu^2/\mu_0^2)]^2$
behavior. For processes
such as the Drell-Yan process, which receive large $\pi^2$ 
corrections, there will be residual $\mu^2$-dependent effects
arising from the $\mu^2$-dependent part of the parton distributions
multiplied by the large correction terms in the cross section which
will not cancel at the one-loop level. Thus, there will be an
additional $\mu^2$ dependence of the cross section proportional
to
\beqa
\sg^{\pi^2}(\mu^2) - \sg^{\pi^2}(\mu_0^2) &\sim &
\fr{\al_s}{2\pi} \pi^2 
\lf[\fr{\al_s}{2\pi} \ln(\mu^2/\mu_0^2)\rt] \\
&\sim& \fr{\pi}{2} \al_s 
\lf[\fr{\al_s}{2\pi} \ln(\mu^2/\mu_0^2)\rt],
\eeqa
which is still suppressed relative to the $\mu^2$ dependence of the
Born term for typical energy scales. Of course, there may be other
sources of large corrections not strictly proportional 
to $\pi^2$ (see the discussion near the beginning of section XI),
so our $\pi^2$ factor is really a generic representation of large
correction terms in general. The rule then is that the larger the 
one-loop corrections, the greater the $\mu^2$ dependence. Inclusion
of two-loop corrections (and evolution) will reduce this factor.
We will not discuss two-loop effects in any detail here. 
Also, we will not perform detailed numerical studies of scale dependence, as
this has been done extensively in the literature. The above 
discussion was included simply to clarify the origin of the scale
dependence and to show how it is reduced by the inclusion of QCD 
corrections in general.

%\vglue 1cm
%\begin{center}\begin{large}\begin{bf}
\section{DRELL-YAN IN LOW ENERGY RUNNING AT RHIC}
%\end{bf}\end{large}\end{center}
%\vglue .3cm

Throughout, for unpolarized cross sections, we use subprocess cross
sections determined in the $\ovl{\rm MS}$ scheme of DREG. They are
convoluted with the unpolarized MRSG set \cite{MRSG}
of parton distributions. 
Varying the choice of unpolarized parton distributions
produces negligibly small changes in the spin dependent asymmetries
in general. Hence, we only consider the effect of varying the 
polarized parton distributions even though, strictly speaking, one should
use specific unpolarized sets with the various polarized sets.
We use the GSA and GSC sets \cite{GS}
as well as the GRSV 
(standard NLO  $\ovl{\rm MS}$) set \cite{GRSV} and compare the
corresponding predicted asymmetries. Since all these sets are determined
in the $\ovl{\rm MS}_{HC}$ scheme, using HVBM regularization, we use
subprocess cross sections determined in that scheme for doubly
polarized predictions. Clearly then, for single-spin cross sections,
we use the $\ovl{\rm MS}_{HC}$($\ovl{\rm MS}$) scheme for the
renormalization of the 
(un)polarized parton distributions which enter (i.e.\ for the $T_{ij}$).
Consequently, the $\Dl_k P_{ij}^\ve$ are always those of DREG/HVBM, given in
Eq's (\ref{splitu}) and (\ref{splitp}). 
Throughout, we use a two-loop evolved $\al_s$, with 5 flavors.
We also use $\mu=M$ in all numerical calculations.

We will always present both the LO and NLO results for the asymmetries
given. We use NLO parton distributions, polarized and unpolarized, in 
all asymmetries however. The reason, as discussed in section VII, is
that the LO parton distributions possess a process dependence of order
$\alpha_s$, while the NLO ones are process dependent only at order
$\alpha_s^2$. Therefore, it is in general not meaningful to use
LO parton distributions in making predictions for processes with
different structure
from the one where they were determined; in the polarized
case, DIS. In this way, our LO and NLO asymmetries are closer to the
actual first two terms in the ``all-orders'' expansion of the asymmetry
in $\alpha_s$ and the difference in the LO and NLO predictions 
meaningfully
measures the effect of the HOC. Finally, if one uses the LO polarized
distributions determined from DIS, one would also have to use LO
unpolarized distributions determined entirely from DIS as well, for
consistency.

In order to understand the qualitative features of the numerical results,
in $pp$ collisions, 
it is instructive to examine which values of $x$ are being probed in the
parton distributions for a given $M$. We will consider the leading order
contributions for simplicity. From (\ref{e20}), we see that $d\sg/dM$ involves
an integration over $x_F$ ($=x_{F0}$), equivalent to a single integration
over $x_a$ (or $x_b$), both constrained by 
(\ref{ea18}) with $w=1$. $d\sg/dMdx_F$
can be obtained from (\ref{ea22}), though, and involves no integration.
Numerically, it is found to generally 
peak at $x_F\simeq 0$. This implies that dominant
contributions come from the region $x_a\simeq x_b\simeq \sqrt{\tau}$.
The peak is not very sharp though, so this is somewhat of an 
approximation. Nonetheless, this feature holds for the energies 
considered here and becomes more pronounced with decreasing $\tau$.
The basic reason is that at $x_a=x_b=\sqrt{\tau}$ there is an equal
contribution coming from the $q\B{q}$ and $\B{q}q$ subprocesses.
As one moves away from $x_F=0$, one of the contributions becomes smaller
since the $\B{q}$ distribution is being evaluated at larger $x$.
This is somewhat compensated by the other term getting larger, especially
at larger $\tau$. Still, the net effect is a smaller contribution as
$x_F$ moves away from 0. As we go the lower $\tau$, the fact that the
peak is more pronounced is not so surprising due to the $1/(x_a+x_b)$
factor in $d\sg/dMdx_F$, which is not present in the rapidity differential
cross section. The rapidity differential cross sections generally peak
at or near zero rapidity ($x_a=x_b$) as well, although the peak may
be rather broad. $x_F$ is more physical though, since it is simply
$x_a-x_b$ at leading order (i.e.\ it is linear in $x_a$ and $x_b$ unlike
the rapidity). Hence, in what follows, including after QCD corrections,
we will assume that the qualitative features of the numerical results can
be roughly described by taking $x_a\simeq x_b\simeq \sqrt{\tau}$. Of
course, all figures are obtained using the exact expressions only, since
the approximation does not give quantitatively accurate predictions.
We will also use the shorthand notation,
\beq
q_{i},g \equiv f_u^{q_i,g/p}(x,\mu^2), \SSP 
\Dl q_i,\Dl g  \equiv f_l^{q_i,g/p}(x,\mu^2),
\eeq
with $x\simeq \sqrt{\tau}$. Also,
\beq
\sg_{mn} \equiv d\sg_{mn}^{pp,F+B}/dM,
\eeq
where $\sg_{mn}^{F+B}$ was defined in (\ref{fbdef}) as being the 
usual leptonic integrated cross section.

Consider the double-spin asymmetry
\beq
A_{ll} \equiv \fr{1}{{\cal P}^A {\cal P}^B} \fr{\sg_{ll}}{\sg_{uu}}.
\eeq
With the above mentioned assumptions, $A_{ll}$ is roughly given
by
\beq
A_{ll} \simeq - \fr{(4/9)\Dl u \Dl \B{u} + (1/9)\Dl d \Dl \B{d} +
(1/9) \Dl s \Dl \B{s} }{(4/9) u \B{u} + (1/9) d \B{d} +
(1/9) s  \B{s} }.
\eeq
Due to the extra factor of 4 and the relative largeness of the $u_v$ 
and $\Dl u_v$ distributions, 
the cross section and asymmetry are mostly dominated by
the up quarks.

The polarized valence distributions are reasonably well constrained
over a large range of $x$ from polarized DIS. The polarized sea quark
distributions, however, are only moderately constrained at smaller
$x$, where their contribution in DIS is non-negligible. At  
moderate and large $x$,
there is no constraint experimentally on the polarized sea quark 
distributions, as they are vanishingly small compared to the valence ones.

Most models, however, assume the behaviour
\beq
\label{x0x1lim}
|\Dl q_v|/q_v \stackrel{x\RA 1}{\RA} 1, \SSP
|\Dl q_v|/q_v \stackrel{x\RA 0}{\RA} 0,
\eeq
which appears to be consistent with the DIS data. The result is that the 
asymmetries, and the distinction between the various polarized sets,
will be largest at large $\tau$ (i.e.\ large $x$) where the cross sections
are small. Conversely, where the cross sections
are large (small $\tau$) we expect small asymmetries if the polarized
sea quark distributions are small at low $x$, as all the sets assume.

\begin{figure}
\centerline{\epsfig{file=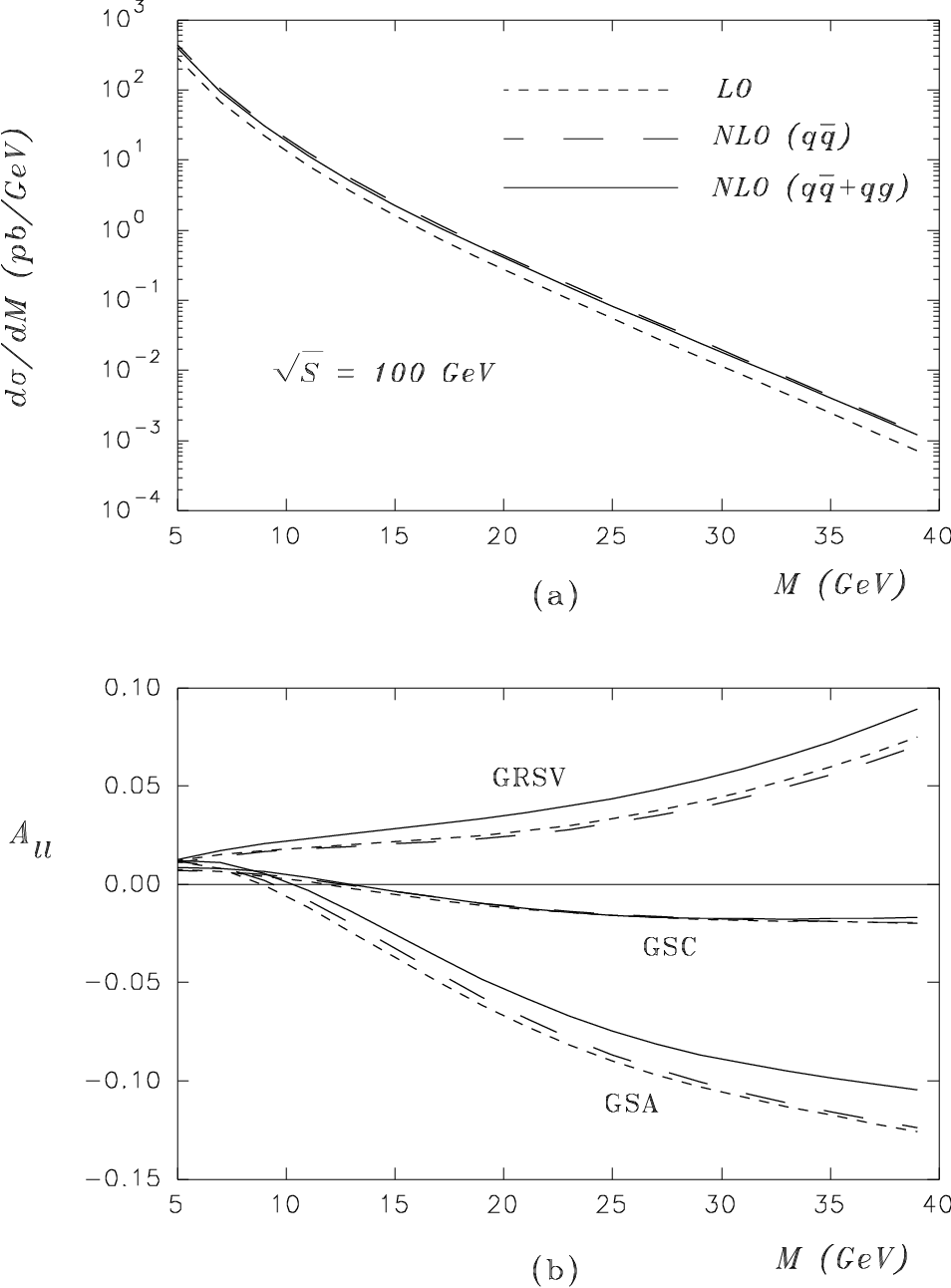,height=11cm,width=8cm}}
\vspace{10pt}
\caption{(a) The cross section, $d\sg/dM$, versus $M$, for 
$l^+l^-$ production in $pp$ collisions at $\sqrt{S}=100\,\,\GeV$;
(b) corresponding double-spin asymmetry, $A_{ll}$, for various sets of
polarized parton distributions. Details given in text.
}
\label{Fig1}
\end{figure}

  Fig.\ \ref{Fig1}(a) gives the (virtual photon dominated) unpolarized
Drell-Yan cross section,
$d\sg/dM$, at $\sqrt{S} = 100\,\,\GeV$, versus $M$ for $pp$ collisions
relevant to RHIC. 
Here, we include only one type of lepton pair (i.e.\ $\mu^+\mu^-$).
Throughout, we present
the leading order, next-to-leading order with $q\B{q}$ contributions 
only, and full next-to-leading order predictions for all cross sections
and asymmetries. This means that, for the asymmetries, the numerators
and denominators are treated in the same way with regard to which corrections
are included.
Here and throughout, we note that the $q\B{q}$ corrections
are positive and large. The $qg$ subprocess makes
a small negative contribution though. This highlights the fact that one
cannot think of the $qg$ subprocess as being physically separate
from the $q\B{q}$ subprocess. They are both related via the renormalization
of the quark distributions. We will now investigate this issue thoroughly.

The negativity of the $qg$ contribution also holds in the $\ovl{\rm MS}_\ve$
scheme. This will (slightly) simplify the process of understanding the
 origin of the negative $qg$ contribution. The analytical result for 
the subprocess cross section is given in (\ref{qgans}).
Since, when using  $\ovl{\rm MS}_\ve$ factorization, 
the scheme dependent part of the subprocess cross section is
zero and since $k_{mn}^{qg,F+B}(w) > 0$ and since $\mu^2=M^2$ here, the
only negative contribution comes from the term
$\sim \Dl_u P_{qg}^4(w) \ln (1-w)$. 
From (\ref{ea75}), (\ref{ea84}), (\ref{ea85}), (\ref{iqg}) we see
that this term arose from $(1-w)^{-2\ve} \Dl_u P^4_{qg}(w)/\ve$
and is therefore of collinear origin. This means that 
such terms could be factorized via
the renormalization of the quark distributions, but would most likely
lead to unphysical distributions violating some of the conservation
rules mentioned in Sec.\ VII. So, we see explicitly
that the issue of the negativity of the $qg$ subprocess is intimately
related to the issue of the renormalization of the quark distribution,
which in turn connects the $q\B{q}$ and $qg$ subprocesses.

The fact that the net $qg$ contribution is negative tells us something
about which regions of the $w$ integration are giving dominant 
contributions. In order that the $qg$ contribution be negative, we 
see from (\ref{qgans}) that we must satisfy
\beq
\label{ineq1}
\Dl_u P_{qg}^4(w) [2\ln(1-w) - \ln w] + \fr{(1-w)}{4} (1+3w) < 0
\eeq
(again working in the $\ovl{\rm MS}_\ve$ scheme for simplicity).
This is satisfied for
\beq
\label{ans1}
w>.572022\,\,.
\eeq
Since the $w$ integration includes $w=1$, we expect large contributions
from the term $\sim \ln(1-w)$ from $w$ near 1. In light of (\ref{ans1}),
we can see clearly why the corrections are negative.

The expected yearly integrated  RHIC $pp$
luminosity at $\sqrt{S}= 100\,\,\GeV$
is ${\cal L} = 160 \,\,{\rm pb}^{-1}$ (assuming it is linear in 
$\sqrt{S}$ down to this energy). 
This means that the statistics
will be quite poor beyond $M=20\,\,\GeV$, concerning asymmetry measurements,
whose error goes like (see (\ref{beampol}))
%\widetext
\beq
\label{aerr}
\Dl A_{mn} = \sqrt{\fr{1-A_{mn}^2}{N}} \fr{1}{{\cal S}_m^A{\cal S}_n^B}
\simeq \fr{1}{\sqrt{N}} \fr{1}{{\cal S}_m^A{\cal S}_n^B}, \SSPP
 A_{mn} \equiv \fr{\sg_{mn}}{\sg_{uu}} \fr{1}{{\cal S}_m^A{\cal S}_n^B},
\eeq
%\narrowtext \noindent
where the approximate equality holds
for $A_{mn}$ not too large. Here $N$ is the number of events.
Assuming 1 $\GeV$ binning,
this means that for $M\simeq 5
\,\,\GeV$, $\Dl A_{ll} \simeq .4 \% /{\cal P}^A{\cal P}^B$. 
Unfortunately, this is a rather low
mass scale where the parton model may not work well. 
For $M\simeq 10\,\,\GeV$, $\Dl A_{ll}\simeq 1.8\% /{\cal P}^A{\cal P}^B$. 
At low $M$ ($\lesssim 10 \,\,\GeV$), one also has
to be careful of resonance backgrounds.
Experimental cuts will make the errors slightly
larger. On the other hand, with two independent experiments (at the
PHENIX and STAR detectors), the combined asymmetry
measurements should have roughly the errors given here.
A more detailed error analysis is beyond the scope of this paper.

Fig.\ \ref{Fig1}(b) gives the corresponding $A_{ll}$. The behaviour is exactly
as expected from the previous arguments. It the low mass region, where
the statistics are good, $A_{ll}$ is of the order $1$--$2$\%. There is
also little distinction between the various sets. This is a result
of the common assumptions
\beq
|\Dl \B{u}|/ \B{u} \ll 1, \SSPP \Dl \B{u} < 0 \SSPP
(\mbox{at small $x$}).
\eeq
There is certainly more freedom in $\Dl \B{q}$ than is manifest
among the various sets. The DIS constraints on $\Dl \B{q}$ are 
rather weak and it will be interesting and important to see whether
or not the low mass Drell-Yan asymmetry is as small (and positive) as it is
predicted to be.

It is also of interest to study the effect of the QCD corrections
on $A_{ll}$. From Fig.\  \ref{Fig1}(b), we 
observe good stability in the $q\B{q}$ contribution in general.
This is not unexpected, taking into account helicity conservation.
For the GRSV and GSA sets, $\Dl g$ is positive and sizable. For
the GSC set, $\Dl g$ is positive at small $x$, but negative 
and small at large $x$. Hence, the gluonic contribution is small and
uninteresting in general for the GSC set. Therefore, we study throughout
the gluonic contributions (and features in general) only for the
GRSV and GSA sets. We notice that the gluonic contribution to $A_{ll}$
is always positive. The gluonic contributions to $\sg_{ll}$
and $\sg_{uu}$ are dominated by the $ug$ subprocess. Since both
$\Dl u$ and $\Dl g$ are positive, the essential difference between
the sign of the gluonic contribution to $\sg_{ll}$ and $\sg_{uu}$
is that arising from the sign of the respective subprocess cross
sections. The most important difference between $\ha{\sg}_{ll}^{qg}$ and
$\ha{\sg}_{uu}^{qg}$ is the difference in the overall factors
(which are proportional to the respective Born terms, see 
(\ref{e69})--(\ref{e71})).
There is a relative minus in the overall factors for $\ha{\sg}_{uu}^{qg}$
and $\ha{\sg}_{ll}^{qg}$.

More precisely, for the correction, $\ha{\sg}_{ll}^{qg}$, to be
positive, we must satisfy (\ref{ineq1}) with $\Dl_u P^4_{qg}(w)$
replaced by $\Dl_l P^4_{qg}(w)$. It is satisfied for
\beq
w< .208233, \SSP w> .652396\,\,.
\eeq
Using the same logic as before, dominant contributions will come from
near $w=1$, where the corrections are positive. Hence the corrections
to $\sg_{ll}$ will be positive. The $qg$ corrections to $\sg_{ll}$
are relatively larger then those to $\sg_{uu}$ since, for typical $x$
and $\mu^2$,
\beq
\label{glineq}
|\Dl g / \Dl \B{q}| >  g/\B{q},
\eeq
in the sets considered.
Hence the $qg$ subprocess typically
weighs in more heavily versus the
$q\B{q}$ subprocess in $\sg_{ll}$ than in $\sg_{uu}$. Thus we can
clearly understand the sizable and positive $qg$ corrections to
$A_{ll}$. This rationale also applies to the corrections to 
$A_{ll}$ in $Z$- and $W^\pm$-boson production.

\begin{figure}
\centerline{\epsfig{file=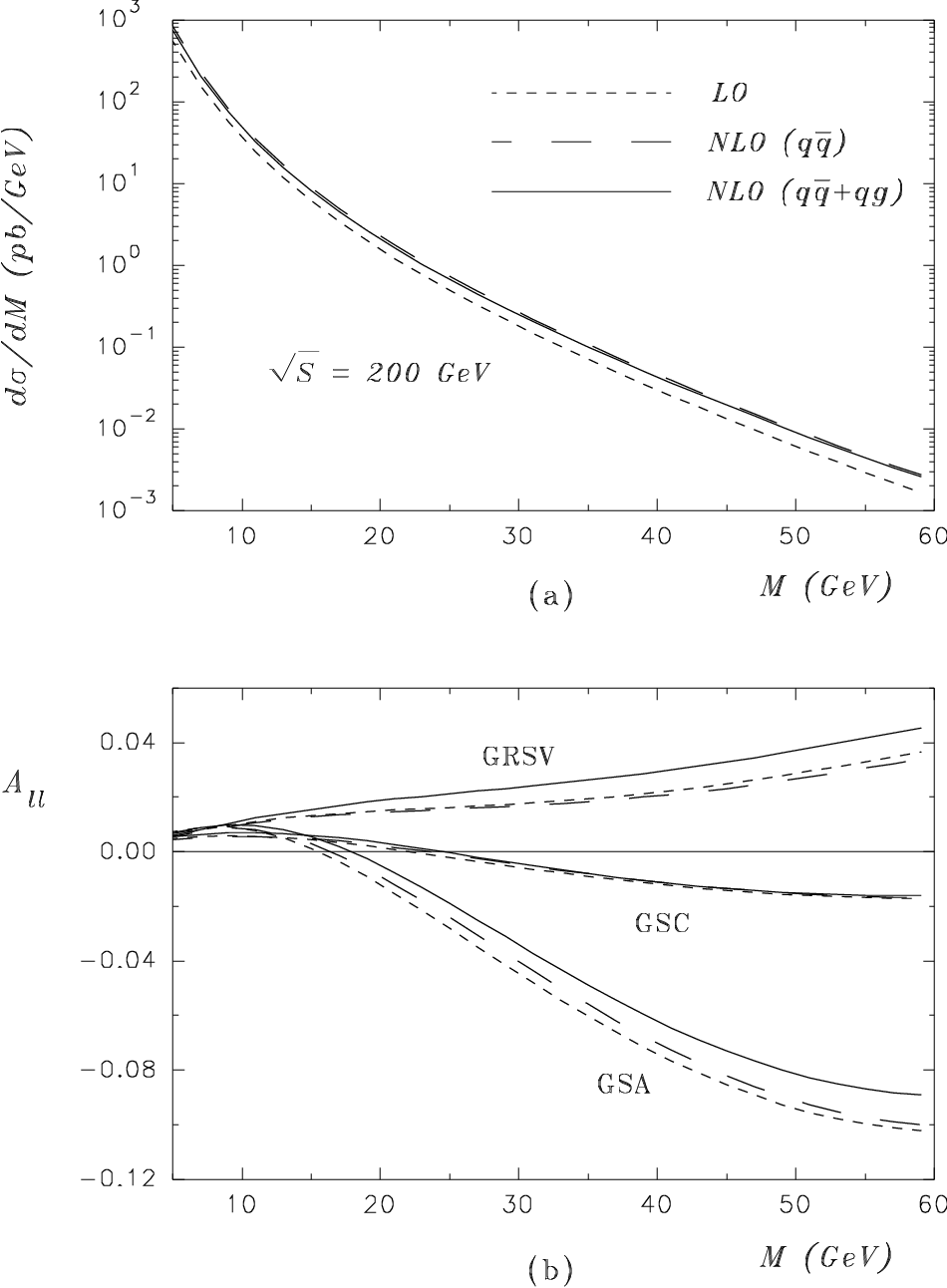,height=11cm,width=8cm}}
\vspace{10pt}
\caption{ As Fig.\ \ref{Fig1}, except at $\sqrt{S}=200\,\,\GeV$.
}
\label{Fig2}
\end{figure}

Figures \ref{Fig2}(a) and \ref{Fig2}(b) show the corresponding cross 
sections and 
asymmetries at $\sqrt{S} = 200 \,\,\GeV$. All the features are the 
same. The luminosity is doubled and the unpolarized 
cross sections are larger for the same $M$ (due to the smaller
$x$, where the sea quark distributions are large).
The asymmetries are somewhat smaller though. Hence the statistical
significance ($\Dl A_{ll}/A_{ll}$) is comparable. 
The main difference is that we
probe lower $x$ than at $\sqrt{S}=100\,\,\GeV$.
Running at more than one energy is definitely an advantage in that
one covers a larger range of $x$ and $\mu^2$, compared with only running
at one energy. This also allows some degree of cross checking via 
perturbative evolution and thus allows detection of various types of 
systematic errors. The net result is lower overall errors on the polarized
parton distributions so determined.

%\vglue 1cm
%\begin{center}\begin{large}\begin{bf}
\section{$Z$-BOSON PRODUCTION AT RHIC}
%\end{bf}\end{large}\end{center}
%\vglue .3cm

$Z$-boson production at RHIC is quite useful since it allows us to measure
the polarized parton distributions at relatively large $x$. Also,
the parton distributions enter as a different linear combination in
the $Z$ double-spin asymmetries than for production by virtual 
photons. This is helpful for disentangling the various contributions.
One can also make use of parity violation to consider additional
asymmetries which vanish for virtual photon production. The major 
limitation is the event rate. Certain asymmetries will be able to
somewhat overcome this limitation, however.

\begin{figure}
\centerline{\epsfig{file=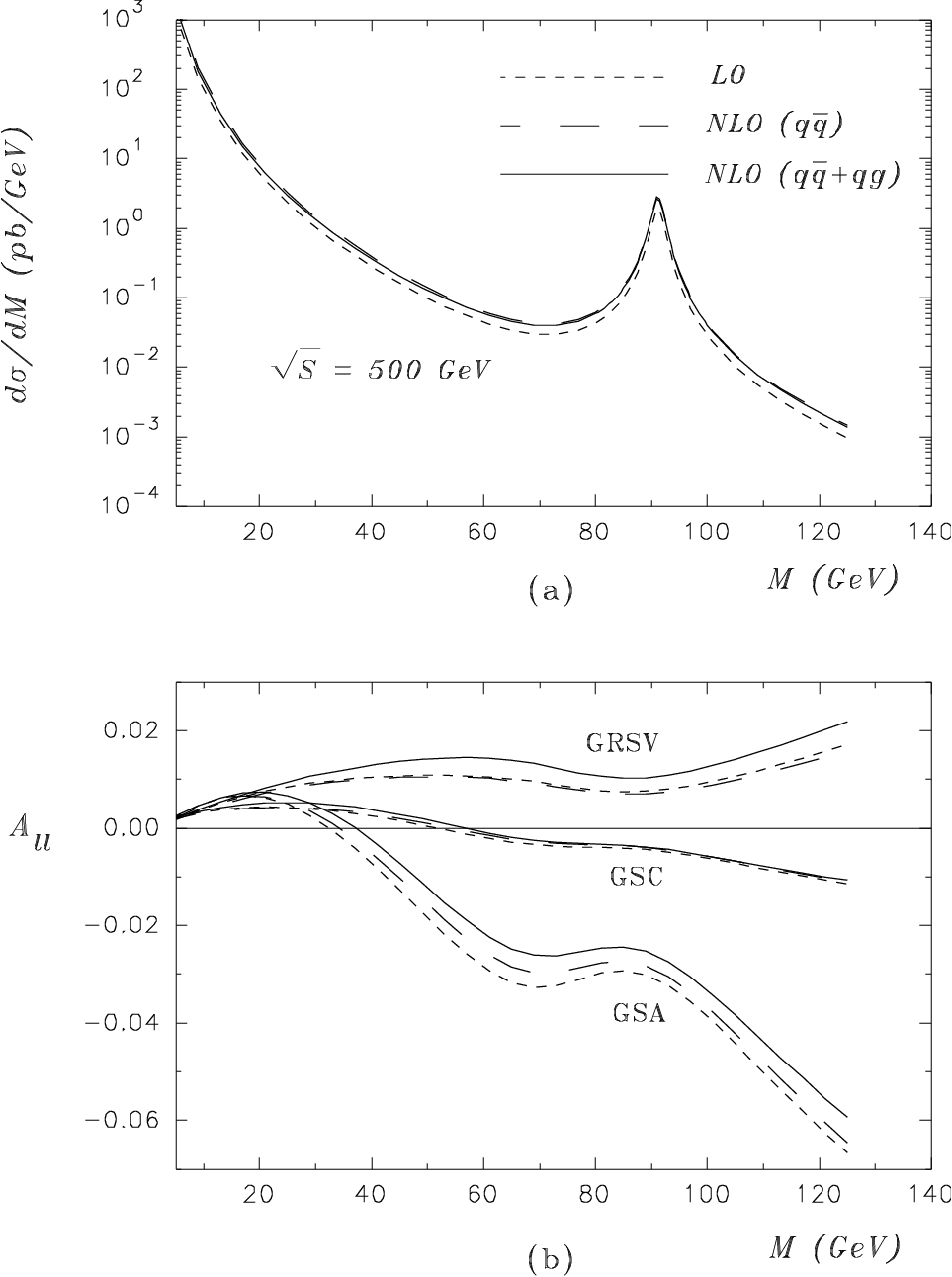,height=11cm,width=8cm}}
\vspace{10pt}
\caption{(a) The cross section, $d\sg/dM$, versus $M$, for 
$l^+l^-$ production in $pp$ collisions at $\sqrt{S}=500\,\,\GeV$;
(b) corresponding double-spin asymmetry, $A_{ll}$, for various sets of
polarized parton distributions. Lines as in Fig.\ \ref{Fig1}.
}
\label{Fig3}
\end{figure}

Fig.\ \ref{Fig3}(a) shows the unpolarized cross section, $d\sg/dM$
for $5\leq M \leq 125 \,\,\GeV$ at $\sqrt{S}=500\,\,\GeV$.
Again, we consider production of only one type of lepton pair
(i.e.\ $\mu^+\mu^-$).
The QCD corrections behave anaolgously to the virtual photon case.
Using the expected yearly integrated luminosity of 800 ${\rm pb}^{-1}$
and integrating the cross section between $80\leq M \leq 100 \,\,
\GeV$ gives approximately 8,000 $\mu^+\mu^-$ pairs. Hence we can
measure asymmetries with an uncertainty (see (\ref{beampol}),
(\ref{aerr}))
\beq
\Dl A_{mn} \simeq \fr{1-1.5\%}{{\cal S}_m^A{\cal S}_n^B},
\eeq
depending on the experimental cuts. Higher energy running would increase
the $Z$ event rate, but somewhat lower the asymmetries, as will become
clear. 

The double-spin asymmetry in the $Z$-pole region goes rougly like,
\beq
A_{ll} \simeq - \fr{\sum_{q} (g_{aq}^2+g_{vq}^2)\Dl q \Dl \B{q}}
{\sum_{q} (g_{aq}^2+g_{vq}^2) q  \B{q}},
\eeq
evaluated near $x\simeq \sqrt{\tau} = .18$ . Putting in 
the appropriate couplings gives
\beq
A_{ll} \simeq - \fr{.29\Dl u \Dl \B{u} + .37 (\Dl d \Dl \B{d}
+ \Dl s \Dl \B{s}) }{.29 u \B{u} + .37 ( d  \B{d} +  s  \B{s}) } \,.
\eeq

In general, for $Z$ production, the GSA and GRSV sets correspond to 
two extreme solutions, with the GSC set lying somewhere in between.
So, as in the last section, we will try to understand qualitatively
only the GSA and GRSV predictions, since one can make a clear prediction
for those sets.

All sets assume an SU(3) symmetric polarized sea and take a negative down 
valence and a positive up valence distribution. For the GRSV set, the sea 
quark distributions are negative everywhere, while for the GSA set
they are positive at intermediate and large $x$ and become negative
only at small $x$. Thus, for the GSA set, $\Dl \B{q} > 0$ in the
$x$ range of interest, $x \simeq .18$ .

There will be some cancellation between the up- and down-quark
contributions, but since
\beq
\label{q156}
\fr{\Dl u_v(.18)}{-\Dl d_v(.18)} \simeq 2.5-4,
\eeq
the $u$ contribution will still be bigger. Due to the smallness
of the polarized sea quark distributions, $\Dl q\simeq \Dl q_v$.
Hence, we expect
\beq
\label{sgnZll}
A_{ll} > 0: {\rm GRSV}, \SSPP A_{ll} < 0: {\rm GSA}
\eeq
in the $Z$-pole region.

Fig.\ \ref{Fig3}(b) presents $A_{ll}$ for $5 \leq M \leq 125 \,\, \GeV$.
The higher order corrections (HOC)
have the same behaviour as for virtual photon production,
for exactly the same reasons. As well, the sign of $A_{ll}$
predicted in (\ref{sgnZll}) is verified. There is roughly a 4\%
variation in $A_{ll}$ between the two extreme cases. This means that
one can rule out one case or the other, but not much more. Hence
we must examine other asymmetries in order to do better. It would also
be interesting to look at $A_{ll}$ in the low mass region, where the 
virtual photons dominate, due to the large event rate. As mentioned 
in the previous section, $A_{ll}$ may turn out to be larger than 
expected in that region.

Define the single-spin asymmetry as
\beq
A_{l} \equiv A_{ul} =
\fr{1}{{\cal P}^B} \fr{\sg_{ul}}{\sg_{uu}} = A_{lu} =
\fr{1}{{\cal P}^A} \fr{\sg_{lu}}{\sg_{uu}} .
\eeq
There is an overall minus relative to the definition often used 
\cite{Sof}, owing to (\ref{q13}), (\ref{q14}). This asymmetry
is nonzero due to the parity violating $Z$ vertices. It
is approximately given by
\beqa
\nn
A_l  &\simeq & \fr{\sum_q [2 g_{aq} g_{vq}(q\Dl\B{q}-\B{q}\Dl q)]}
{\sum_q [(g_{aq}^2+g_{vq}^2) (2q\B{q}) ]  }
\\ &=&
\nn
\fr{g_{au}g_{vu} u_v\Dl \B{u} + g_{ad}g_{vd} d_v \Dl \B{d}}
{(g_{au}^2+g_{vu}^2) u\B{u} + (g_{ad}^2+g_{vd}^2) (d\B{d}+s\B{s})} \\
\nn & &
- \fr{g_{au}g_{vu} \B{u}\Dl u_v + g_{ad}g_{vd} \B{d} \Dl d_v}
{(g_{au}^2+g_{vu}^2) u\B{u} + (g_{ad}^2+g_{vd}^2) (d\B{d}+s\B{s})}
\\ \label{q159}
&\simeq &
\fr{.1 u_v\Dl \B{u} + .17 d_v \Dl \B{d}}
{.29 u\B{u} + .37 (d\B{d}+s\B{s})}
- \fr{.1 \B{u}\Dl u_v + .17 \B{d} \Dl d_v}
{.29 u\B{u} + .37 (d\B{d}+s\B{s})}\,,
\eeqa
where we took $(\Dl) q(x) = (\Dl) \B{q}(x)$ as is done in all the
sets considered. Noting that
\beq
\fr{\B{d}(.18)}{\B{u}(.18)} \simeq 2 
\eeq
and taking into account (\ref{q156}), 
we observe a rather large cancellation between the $\B{u}\Dl u_v$
and $\B{d} \Dl d_v$ contributions in the second term
of (\ref{q159}). The net
effect is that the first and second terms become comparable in 
magnitude. Hence $A_l$ is directly sensitive to $\Dl \B{u}$ and
$\Dl \B{d}$. Also, the $\Dl \B{u}$ and $\Dl \B{d}$ contributions are
comparable, unlike in virtual photon production. The second term is 
reasonably well constrained from DIS, while the first term is almost
completely unconstrained (except in maximum magnitude). From the 
respective signs of the sea quark distributions, we expect $A_l$ to
be most positive for the GSA set and least so for the GRSV set.

\begin{figure}
\centerline{\epsfig{file=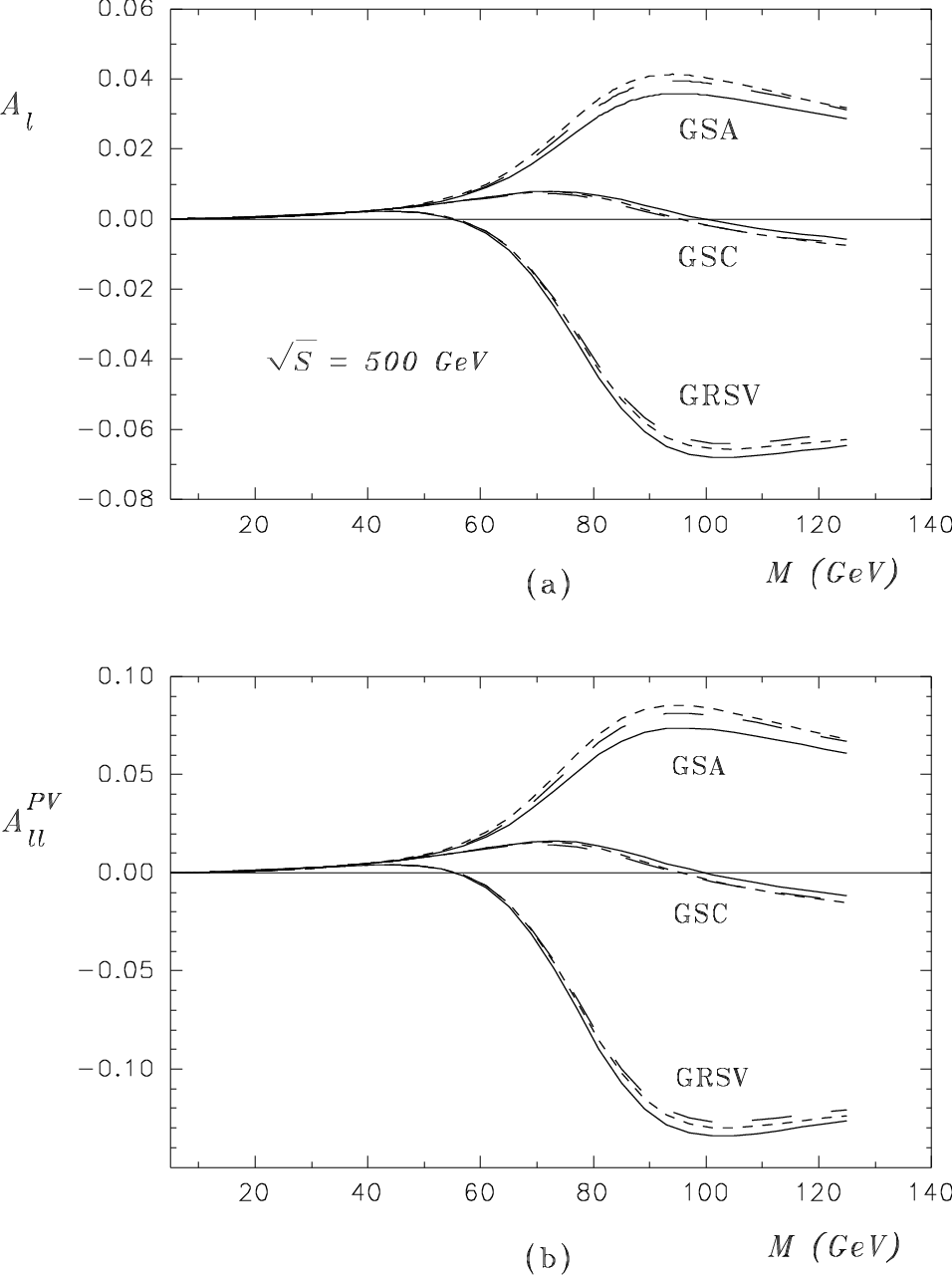,height=11cm,width=8cm}}
\vspace{10pt}
\caption{(a) The single-spin asymmetry, $A_l$, versus $M$, for
$l^+l^-$ production in $pp$ collisions at $\sqrt{S}=500\,\,\GeV$;
(b) corresponding two-spin parity violating asymmetry, $A_{ll}^{PV}$.
Lines as in Fig.\ \ref{Fig1}.
}
\label{Fig4}
\end{figure}

Fig.\ \ref{Fig4}(a) shows $A_l$ for $5\leq M \leq 125 \,\, \GeV$. We observe
the predicted behaviour. At small $M$, $A_l$ vanishes as expected
since the parity violating $Z$ contribution also vanishes. The peak
is just above the $Z$-pole and there is a well measurable separation
between the various sets. So, the sensitivity to the sea quark
distributions has improved as compared to $A_{ll}$.

One can take advantage of having two polarized beams
to improve the statistical significance of the parity violating
asymmetry. Define the two-spin parity violating asymmetry as,
%\widetext
\beqa 
\label{e161p}
A_{ll}^{PV} &\equiv& \lf. \fr{\sg(+,+)-\sg(-,-)}{\sg(+,+)+\sg(-,-)}
\rt|_{{\cal P}^{A,B}=1}
= \fr{2(\sg_{ul}/{\cal P}^B+\sg_{lu}/{\cal P}^A)}
{2(\sg_{uu}+\sg_{ll}/{\cal P}^A{\cal P}^B)}
\\ &\simeq& \fr{2}{{\cal P}} \fr{\sg_{ul}}{\sg_{uu}},
\SSPP \mbox{for}\,\,\, {\cal P}^A\simeq {\cal P}^B\equiv {\cal P}, 
\eeqa
%\narrowtext \noindent
where the last approximate equality holds since $\sg_{ll}\ll \sg_{uu}$
as $\sg_{ll}$ involves two polarized parton distributions and is
thus relatively rather suppressed.
We see explicitly that
$A_{ll}^{PV}$ is proportional to $1/{\cal P}$
(relative to experiment), not $1/{\cal P}^2$ as is
often assumed. In order to get an idea of the statistical error on 
$A_{ll}^{PV}$, we will make the simplifying assumption that, experimentally,
${\cal P}^A = {\cal P}^B = 1$.
Then, the total number of events is given by
\beqa \nn
N_0 &=& {\cal L} \fr{\sg(+,+)+\sg(-,-)}{4}
= {\cal L} \fr{\sg_{uu}+\sg_{ll}}{2}
\\ &\simeq & {\cal L} \fr{\sg_{uu}}{2},
\eeqa
assuming equal running in all four polarization modes. Hence, the number
of events is cut in half, but the asymmetry is doubled. So,
\beq
\fr{\Dl A_{ll}^{PV}}{A_{ll}^{PV}} \simeq \fr{1}{\sqrt{2}}
\fr{\Dl A_l}{A_l}.
\eeq

This means that we gain roughly a factor of $\sqrt{2}$ in precision by looking
at $A_{ll}^{PV}$. One could argue, of course, that there are two $A_l$;
$A_{ul}$ and $A_{lu}$. Then, one could combine
them to improve the errors. Theoretically, the two asymmetries are 
equal, but experimentally they only approach each other in the limit
of infinite events. The problem is that these are not
independent measurements.
Hence, one cannot simply add the $\Dl A_l$ in quadrature. 
If we define $\ovl{A}_l$ as being the experimental average of the
two $A_l$, a proper treatment of the errors yields
\beqa
\nn
\fr{\Dl \ovl{A}_l}{\ovl{A}_l} &=& \fr{1}{\sqrt{N_0}}
\lf[ \fr{1-A_{ll}^{PV\,2}}{A_{ll}^{PV\,2}} + \fr{A_{ll}^{PV}-\ovl{A}_l}
{A_{ll}^{PV}}\rt]^{1/2} \\
&\simeq& \fr{1}{\sqrt{N_0}}
\lf[ \fr{1-A_{ll}^{PV\,2}}{A_{ll}^{PV\,2}} + \fr{1}{2} \rt]^{1/2}.
\eeqa
In the limit of small $A_{ll}^{PV}$, $\Dl\ovl{A}_l/\ovl{A}_l \RA
\Dl A_{ll}^{PV}/A_{ll}^{PV}$, but in general   
$\Dl\ovl{A}_l/\ovl{A}_l > \Dl A_{ll}^{PV}/A_{ll}^{PV}$.
Statistically, one therefore does best
with $A_{ll}^{PV}$ when two polarized beams are available.
Although, $\ovl{A}_l$ is not much worse for typical experiments,
$Z$ production in particular, where the fractional errors should be
virtually identical in $\ovl{A}_l$ and in $A_{ll}^{PV}$. Of course, the
above analysis is only strictly valid in the limit of full beam polarization.
Inclusion of partial beam polarization will not change our conclusions,
however, since we can always `pretend' that the beams are fully polarized.
Then we are measuring the polarized parton distributions multiplied by
the corresponding beam polarizations, rather than just the distributions
themselves. 
When considering $x_F$ (or rapidity) differential cross
sections, $A_l$ is the quantity of interest \cite{Sof} since it is more
directly related to the ratio of the polarized to unpolarized 
parton distributions.

Fig.\ \ref{Fig4}(b) gives $A_{ll}^{PV}$ for $5\leq M\leq 125\,\, \GeV$. It has
exactly the expected behaviour. Since $\Dl A_{ll}^{PV}\simeq
(1.5\sim 2\%)/{\cal P}$, we can clearly distinguish between the 
possible solutions,
and hence are quite sensitive to both $\Dl \B{u}$ and $\Dl \B{d}$
in the region $x\simeq .18$\,.

%\vglue 1cm
%\begin{center}\begin{large}\begin{bf}
\section{$W^\pm$-BOSON PRODUCTION AT RHIC}
%\end{bf}\end{large}\end{center}
%\vglue .3cm

It is important to determine the intermediate-
and large-$x$ behaviour of the antiquark
distributions at high $\mu^2$ since this (along with the large-$x$
gluonic behaviour) influences the behaviour of the antiquark distributions
in the limit $x\RA 1$ at lower $\mu^2$, relevant to deep-inelastic
scattering. This is because when one evolves from a low energy scale
to a high energy scale, the large-$x$ behaviour influences
the evolution at lower $x$, and vice-versa. Using $W^\pm$ production
at RHIC, we can gain insight
into the  $x\RA 1$ behaviour of the antiquark distributions, where other
experiments (such as deep-inelastic scattering) will have 
little or no sensitivity.  
This statement applies equally to the polarized and unpolarized
antiquark distributions in the proton.

\begin{figure}
\centerline{\epsfig{file=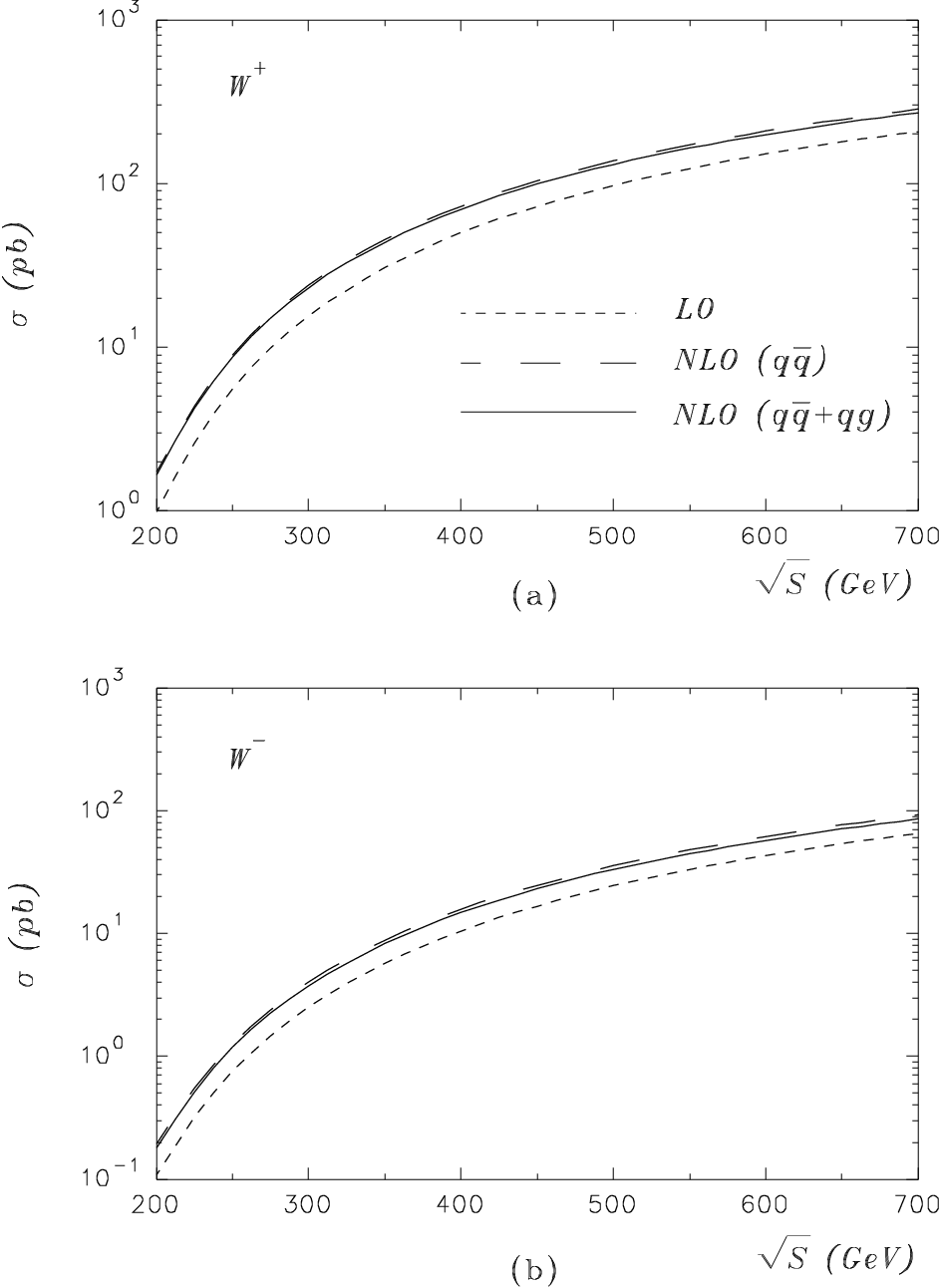,height=11cm,width=8cm}}
\vspace{10pt}
\caption{(a) The total cross section for $W^+(\RA l^+\nu_l)$ production,
versus $\sqrt{S}$, in $pp$ collisions; (b) corresponding cross section
for $W^-(\RA l^-\B{\nu}_l)$ production. Lines as in Fig.\ \ref{Fig1}.
}
\label{Fig5}
\end{figure}

$W^\pm$ production at RHIC
is ideally suited for this purpose because of the
high event rate, resulting from RHIC's high luminosity,
 and because of the flavor specificity of the cross sections. In 
Figures \ref{Fig5}(a) and \ref{Fig5}(b), 
respectively, we plot the $W^+\RA l^+\nu_l$ 
and $W^-\RA l^-\B{\nu}_l$ production
cross sections for $200 \leq \sqrt{S} \leq 700 \,\, \GeV$
and for decay into one type of lepton (i.e.\ muons). Since
one cannot measure $M$ on an event by event basis, we have integrated
over it. The HOC have the same structure as for $\gm^*$ and $Z$ production.
At this point though, it is instructive to look at the variation of the
QCD corrections with $\sqrt{S}$. We understand the origin of the 
negative $qg$ contribution, from the discussion in section IX. The fact
that the 
relative magnitude of the $qg$ contribution increases with $\sqrt{S}$
is understood to be a consequence of the increasing phase space in the
inital state, reflected in a larger integration region over $x_a$ and
$x_b$ (or $w$). What we are
really observing is the effect of varying $\tau$, since $M\simeq M_W$
for all $\sqrt{S}$.

One normally thinks of the large QCD corrections as arising 
mostly from the
term $\sim \pi^2$ in the $q\B{q}$ subprocess, which comes from the
virtual corrections. The rationale 
is that all the other terms are of order one and will tend to largely cancel
among themselves. Clearly, things are somewhat more subtle than that
though, since the term $\sim -7$ in (\ref{e50}) makes the finite virtual
corrections small and negative. Also, the corrections to the $q\B{q}$
subprocess
clearly increase with decreasing $\sqrt{S}$ (i.e.\ increasing $\tau$).
In order for the net correction to be positive, there must be a large,
positive contribution arising from the $q\B{q}$ bremsstrahlung. Such
a large term is indeed present in (\ref{qqans}), namely, the term
$\sim -4(1+w)\ln(1-w)$ which gives a large, positive contribution from
the integration region near $w=1$. There is also a smaller term
$\sim 4\ln^2(1-w_1)\dl(1-w)$ arising from the `+'-distribution term.
As $\tau$ increases, these terms make a larger contribution relative to
the other, potentially negative, bremsstrahlung contributions. The other
term arising from the `+'-distribution could give rise to sizable positive
corrections as well, but its exact relative magnitude depends on the 
details of the parton distributions. In the above picture, the behaviour
of the corrections is well understood.

Considering the decay channel
\beq
W^\pm \RA \mu^\pm \stackrel{(-)}{\nu_{\mu}},
\eeq
for $\sqrt{S} = 500 \,\, \GeV$, one predicts roughly 105,000 $W^+(\mu^+)$
and 27,000 $W^-(\mu^-)$ events. Not taking into account the details of
the cuts, this corresponds roughly to an error on (double- and
single-spin) asymmetries of 
\beq
\Dl A^{W^+}_{mn} \simeq \fr{.3\%}{{\cal S}^A_m{\cal S}^B_n}, 
\SSP \Dl A^{W^-}_{mn} \simeq \fr{.6\%}{{\cal S}^A_m{\cal S}^B_n} \,\, .
\eeq
With such large rates at $\sqrt{S} = 500\,\, \GeV$, it is not unreasonable
to consider going to lower $\sqrt{S}$ as a way of probing 
larger $x$. In fact, the whole energy region,
\beq
250 \leq \sqrt{S} \leq 700 \,\, \GeV,
\eeq
is interesting. Experimentally, the higher energies may be difficult
to access. For the case of greatest experimental relevance 
(500 $\GeV$), we are most sensitive to $x\simeq .16$.

The double-spin asymmetries adopt a simple form
\beq
A_{ll}^{W^+} \simeq -\fr{\Dl u \Dl \B{d}}{u \B{d}}, \SSPP
A_{ll}^{W^-} \simeq -\fr{\Dl \B{u} \Dl d}{\B{u} d},
\eeq
so that $A_{ll}^{W^+}$ is sensitive to $\Dl \B{d}$ and $A_{ll}^{W^-}$
is sensitive to $\Dl \B{u}$. From the signs of the polarized antiquark
distributions for the respective sets, we expect
\beqa \nn
{\rm GSA}: & & A_{ll}^{W^+} < 0, \SSPP A_{ll}^{W^-} > 0, \\
{\rm GRSV}: & & A_{ll}^{W^+} > 0, \SSPP A_{ll}^{W^-} < 0.
\eeqa

\begin{figure}
\centerline{\epsfig{file=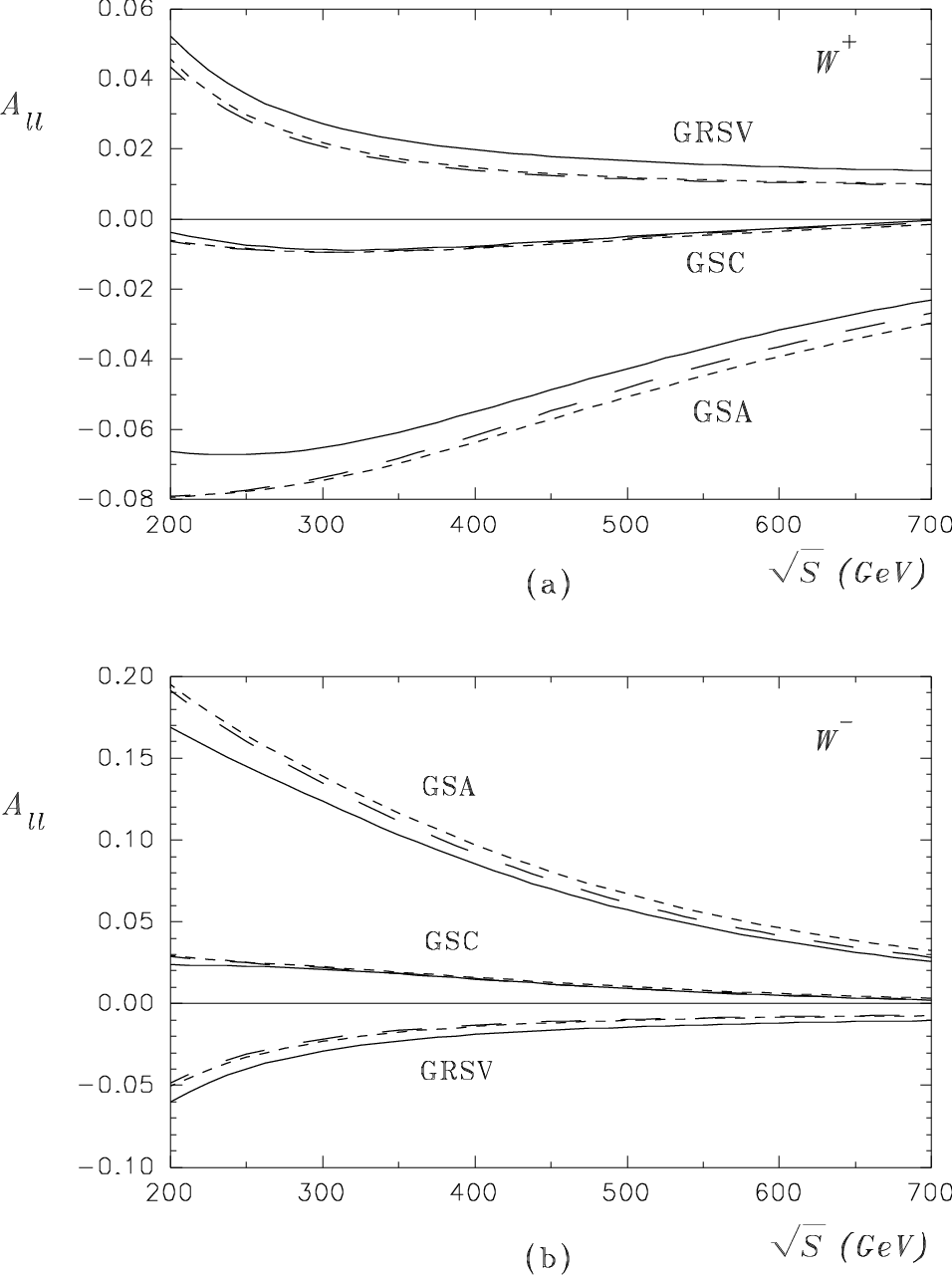,height=11cm,width=8cm}}
\vspace{10pt}
\caption{(a) The double-spin asymmmetry, $A_{ll}$, for $W^+(\RA l^+\nu_l)$
production in $pp$ collisions, versus $\sqrt{S}$; (b) corresponding
asymmetry for $W^-(\RA l^-\B{\nu}_l)$ production. Lines as in Fig.\ \ref{Fig1}.
}
\label{Fig6}
\end{figure}

$A_{ll}^{W^+}$ and $A_{ll}^{W^-}$ are shown in Figures \ref{Fig6}(a) and
\ref{Fig6}(b), respectively. The signs are as expected. We also see that 
$|A_{ll}|$ increases as $\sqrt{S}$ decreases (i.e.\ $x$ increases),
which is a consequence of (\ref{x0x1lim}). This is a general trend
which makes low energy measurements feasible. With high precision
$A_{ll}$ measurements possible at $\sqrt{S}=500\,\,\GeV$, there is
no problem in disentangling $\Dl \B{u}$ and $\Dl \B{d}$ near
$x=.16$.

For $A_{ll}^{W^+}$, the HOC's have the same structure as $A_{ll}^{\gm^*,Z}$,
for the same reasons. For $A_{ll}^{W^-}$, the $qg$ corrections have
opposite sign since it is now the $dg$ subprocess which enters. 
Another interesting feature of this and other asymmetries in $W^\pm$
production is that the effect of the HOC is comparable to
the expected uncertainty in their measurement. Hence, a complete
analysis must make use of the QCD corrections in order not to waste
the good statistics.

The $W^+$ single-spin asymmetry is roughly
\beq
A_l^{W^+} \simeq \fr{u\Dl \B{d}}{2 u \B{d}} - \fr{\B{d} \Dl u}
{2 u \B{d}} = \fr{\Dl \B{d}}{2\B{d}} - \fr{\Dl u}{2 u} < 0.
\eeq
The second term is dominant over the first, but the first term 
does allow distinction between sets. We expect, based on the signs
of the various $\Dl \B{d}$,
\beq
|A_l^{W^+}({\rm GRSV})| > |A_l^{W^+}({\rm GSA})|.
\eeq

The $W^-$ single spin asymmetry is approximately
\beq
A_l^{W^-} \simeq \fr{d\Dl \B{u}}{2 d \B{u}} - \fr{\B{u} \Dl d}
{2 d \B{u}} = \fr{\Dl \B{u}}{2\B{u}} - \fr{\Dl d}{2 d} > 0.
\eeq
Now the two terms are more comparable in magnitude, since $\B{u}<\B{d}$
at large $x$ and all the sets assume $\Dl\B{u}=\Dl\B{d}$. 
Thus, the magnitudes of $\Dl \B{u}$ and $\Dl d$ are 
important as is the sign of $\Dl \B{u}$. In this case, we expect
\beq
A_l^{W^-}({\rm GSA}) > A_l^{W^-}({\rm GRSV}),
\eeq
since there is a cancellation in the GRSV case. In the GSA case, there
is an enhancement rather than a cancellation and we expect large
asymmetries at small $\sqrt{S}$, where $x$ is large.

\begin{figure}
\centerline{\epsfig{file=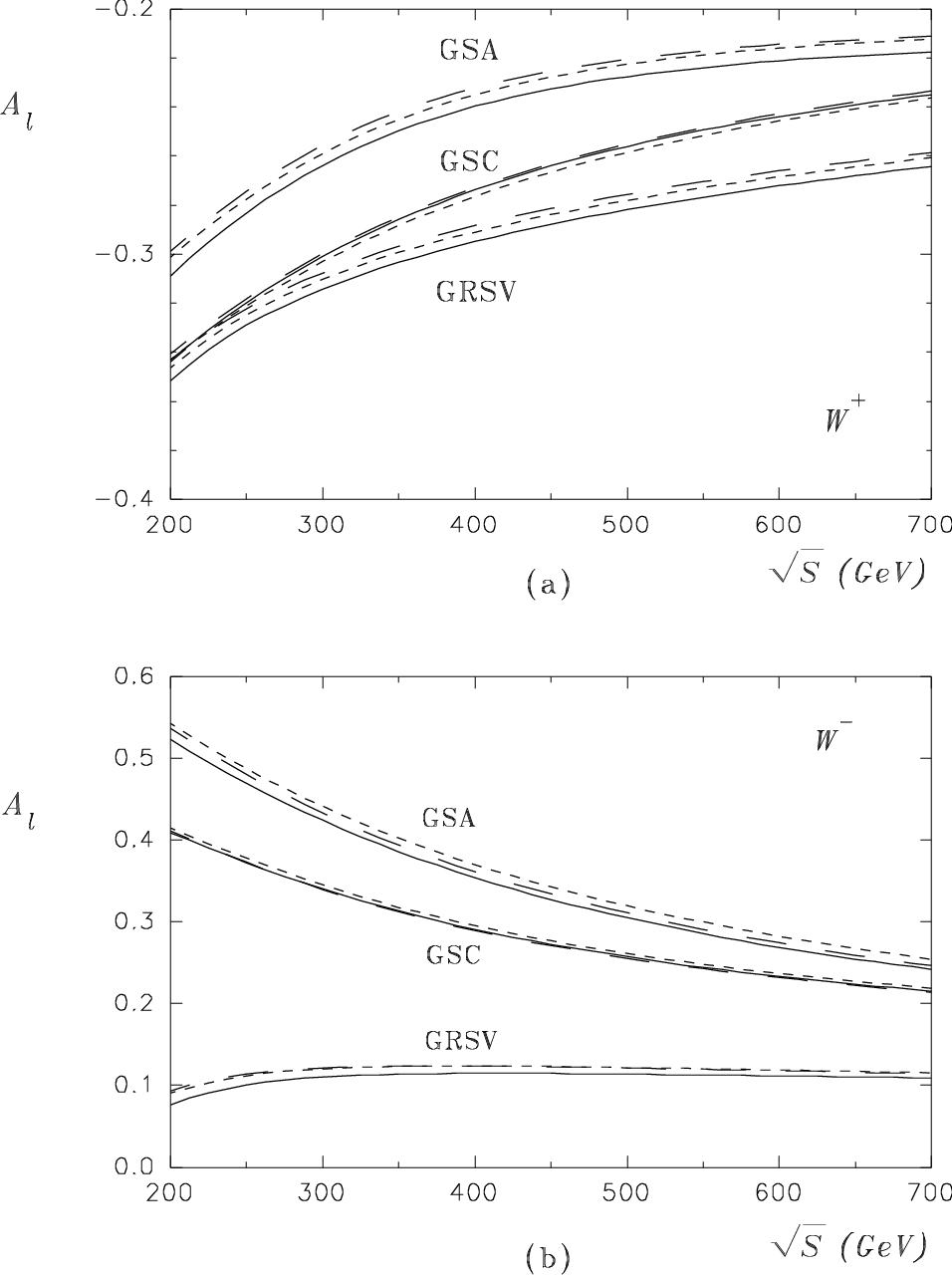,height=11cm,width=8cm}}
\vspace{10pt}
\caption{As Fig.\ \ref{Fig6}, except here the single-spin asymmetry,
$A_l$, is plotted.
}
\label{Fig7}
\end{figure}

Figures \ref{Fig7}(a) and \ref{Fig7}(b) present $A_l^{W^+}$ and $A_l^{W^-}$,
respectively. We see that the behaviour is exactly as expected. In all
cases, distinction between the various sets is straightforward; more
so for $A_l^{W^-}$, however. For $A_l^{W^+}$, knowledge of the HOC's
are
particularly important for this separation, especially at lower $\sqrt{S}$.
For $A_l^{W^-}$, we probe $\Dl \B{u}$ in a very direct fashion.
The asymmetries are large throughout. They are roughly constant for the
GRSV set, but increase in magnitude in the GS case as we go to lower
energies, allowing a precise determination of $\Dl \B{u}$
over a wide range of $x$. Of course,
no single measurement should be expected to determine any specific 
parton distribution exactly. One must fit all the data, including the
DIS data.

\begin{figure}
\centerline{\epsfig{file=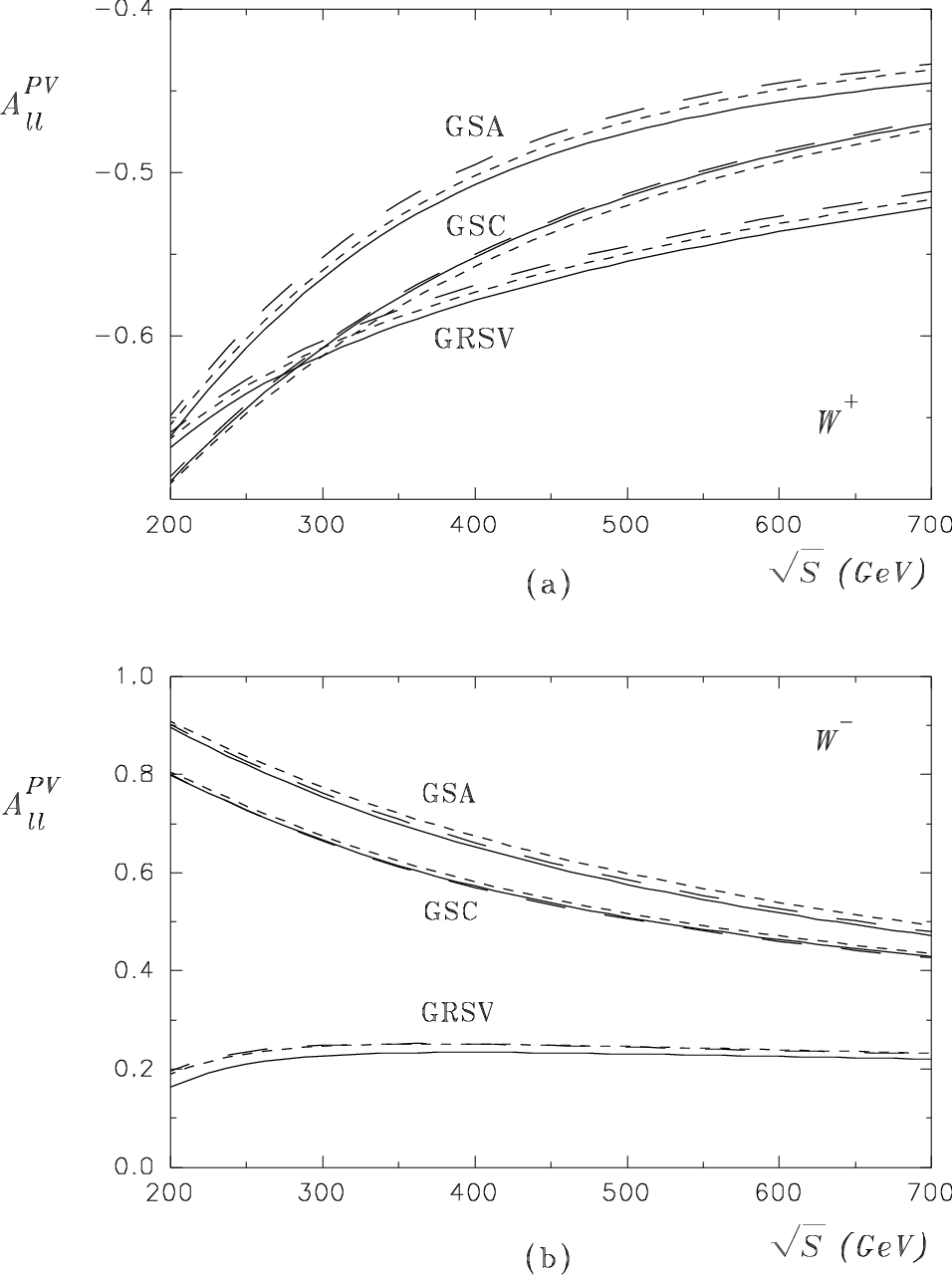,height=11cm,width=8cm}}
\vspace{10pt}
\caption{As Fig.\ \ref{Fig6}, except here the two-spin parity 
violating asymmetry,
$A_{ll}^{PV}$, is plotted.
}
\label{Fig8}
\end{figure}

Figures \ref{Fig8}(a) and \ref{Fig8}(b) 
give $A_{ll}^{PV}$ for $W^+$ and $W^-$ production,
respectively. We notice again roughly a factor of 2 enhancement over
$A_l$, except near $A_{ll}^{PV}=1$, where we are of course constrained
by $A_{ll}^{PV}\leq 1$. Already, precision $A_l$ measurements were
possible, so $A_{ll}^{PV}$, with the extra $\sqrt{2}$ in precision,
can pin down rather tightly the allowed sets of parton distributions.

$Z$ and $W^\pm$ production at RHIC have been previously examined in LO
as a tool for pinning down the polarized parton distributions in
\cite{Sof,SV}. In those studies ($Z$ and $W^\pm$) rapidity differential
asymmetries were considered, which are quite useful in pinning down
the $x$-dependence of the parton distributions. The general conclusions are
the same. Also, transverse momentum distributions for 
$Z$ and $W^\pm$ production at RHIC were studied using Monte Carlo
methods which sum up certain bremsstrahlung graphs in \cite{Schaf}. This
observable is known to be sensitive to the polarized gluon distribution.

Another interesting issue is the ratio $\B{u}/\B{d}$ at large $x$
and $\mu^2$, which is unrelated to spin physics. Knowledge of the
$\B{u}/\B{d}$ ratio at large $x$ and $\mu^2$ gives information on 
that ratio at even larger $x$ at lower $\mu^2$, where there is very
little experimental information. This was discussed at the beginning
of this section. DIS is insensitive to the sea quarks at large $x$, since
they are masked by the valence quark distributions.

From the experimentally measured violation \cite{viol} of the Gottfried
sum rule \cite{Gott}, we can conclude that $\B{u}(x)\neq \B{d}(x)$ for 
all $x$. Maximal violation of SU(2) flavor symmetry is usually taken
to occur at larger $x$, where RHIC is sensitive. A typical assumption
for $x$ (and $\mu^2$)
accessible to RHIC in $W^\pm$ production is $\B{u}/\B{d}
\simeq .5$ \cite{MRSG}. This is based on the Pauli exclusion principle
and explaining the Gottfried sum rule violation. The idea is that since
the relative number of valence to sea quarks is increasing with $x$, the
Pauli suppression effect will increase with $x$ such that the 
$\B{u}/\B{d}$ ratio decreases as there are more up valence quarks than
down valence quarks, leading to greater suppression of the up sea. 
More recently, low energy fixed target Drell-Yan experiments at
Fermilab have helped to disentangle $\B{d}$ and $\B{u}$ at lower
$\mu^2$. 

The quantity of experimental interest (again considering the muonic
decay channel) is
\beq
R_W \equiv \fr{\sg(W^-)}{\sg(W^+)} \simeq \fr{d\B{u}}{u\B{d}},
\eeq
in crude approximation. Since the $d$ and $u$ contributions are quite
well known, we directly probe $\B{u}/\B{d}$. With $\simeq$ 130,000
events, we can measure $R_W$ to high accuracy. At this level, we are
again sensitive to the effect of the HOC's.

\begin{figure}
\centerline{\epsfig{file=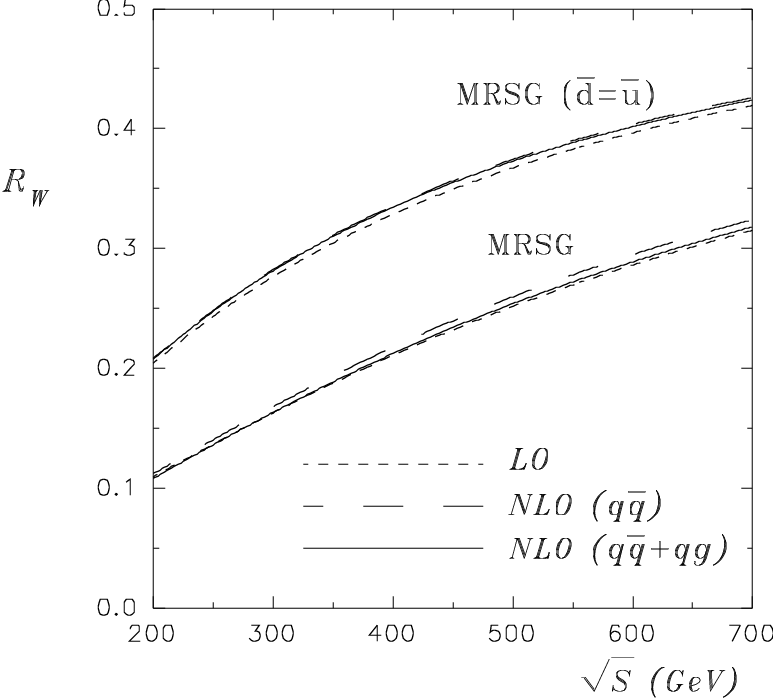,height=7cm,width=8cm}}
\vspace{10pt}
\caption{The ratio of the $W^-(\RA l^-\B{\nu}_l)$ to 
$W^+(\RA l^+\nu_l)$ cross sections, $R_W$, versus $\sqrt{S}$, in
$pp$ collisions. Predictions are made using both the MRSG set and the
MRSG set with the $\B{d}$ distribution set equal to the $\B{u}$ 
distribution. Lines as in Fig.\ \ref{Fig1}.
}
\label{Fig9}
\end{figure}

Fig.\ \ref{Fig9} shows $R_W$ for $200 \leq \sqrt{S} \leq 700\,\,\GeV$. Two
sets are considered. First, the MRSG set and second, the MRSG set with
$\B{d}$ set equal to $\B{u}$. We see a clear separation between the
two possibilities for all $\sqrt{S}$. This will therefore be a crucial
experiment for understanding the $\B{u}/\B{d}$ ratio.
This probe, at RHIC, has previously been studied in 
leading order and the same
conclusions were drawn (see, for instance, \cite{Sof,Kum2}).

%\vglue 1cm
%\newpage
%\begin{center}\begin{large}\begin{bf}
\section{THE FORWARD-BACKWARD DRELL-YAN ASYMMETRY}
%\end{bf}\end{large}\end{center}
%\vglue .3cm

As discussed in the introduction, the principal purpose of performing
a high precision measurement of the forward-backward Drell-Yan
asymmetry at Fermilab is to  precisely determine \stW. In order to 
accomplish this, it is necessary to take into account the QCD corrections.
We have seen that in the case of the spin dependent asymmetries in 
$pp$ collisions, the $qg$ subprocess often destabilizes the asymmetry.
As we will see, for the forward-backward asymmetry, the corrections
arise predominantly from the $q\B{q}$ subprocess. The explanation for 
this feature will be given in detail.

\begin{figure}
\centerline{\epsfig{file=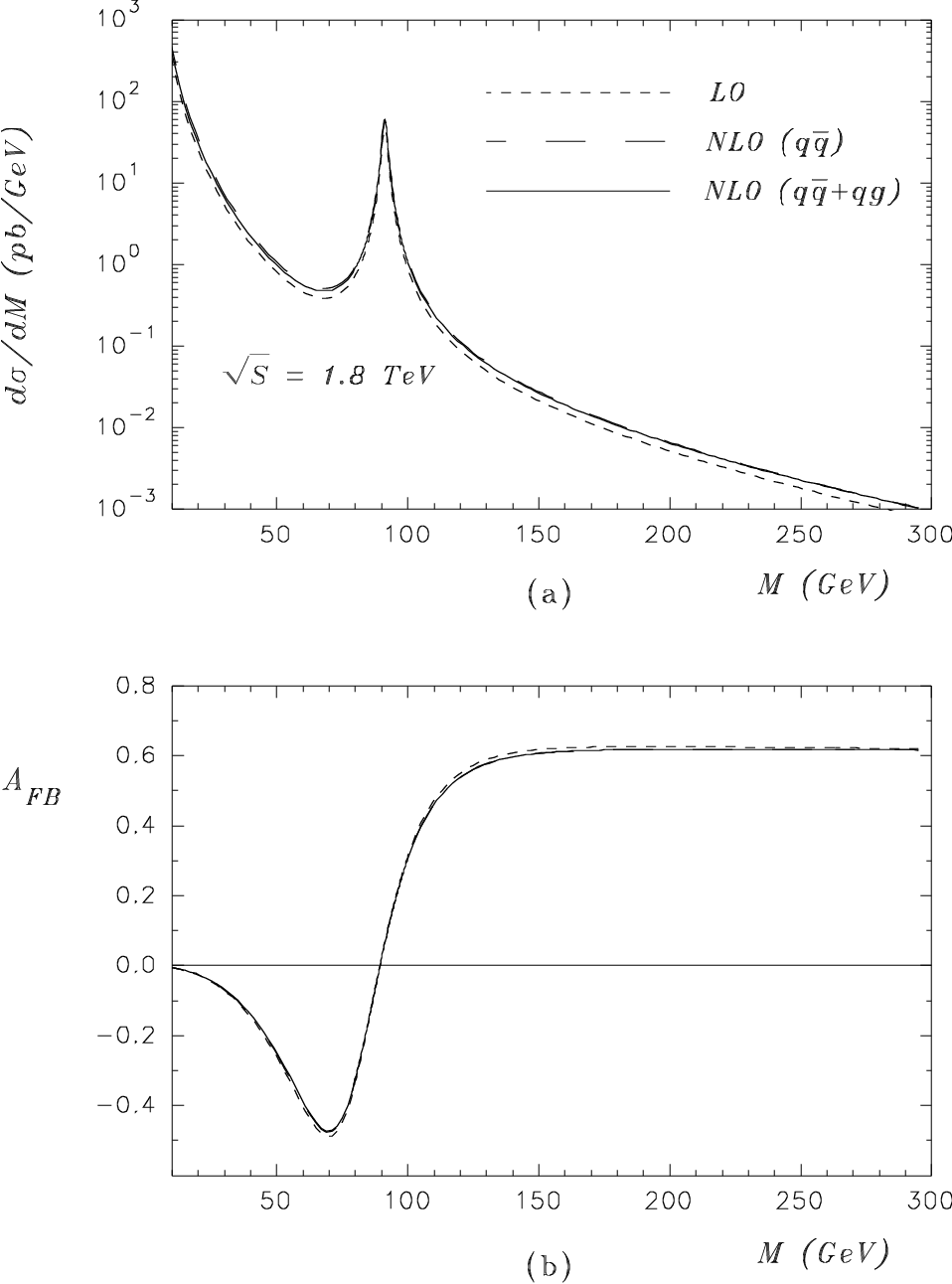,height=11cm,width=8cm}}
\vspace{10pt}
\caption{(a) The cross section, $d\sg/dM$, versus $M$, for $l^+l^-$
production in $p\B{p}$ collisions at $\sqrt{S}=1.8\,\,{\rm TeV}$; (b) 
corresponding forward-backward lepton asymmetry, $A_{FB}$,
obtained using $\sin^2\theta_W=.2315$. Lines as in Fig.\ \ref{Fig1}.
}
\label{Fig10}
\end{figure}

In Fig.\ \ref{Fig10}(a), we present $d\sg/dM$ for $l^+l^-$ production in 
$p\B{p}$ collisions at $\sqrt{S} = 1.8 \,\, {\rm TeV}$.
As usual, we include only one type of lepton-pair (i.e.\
$\mu^+\mu^-$ or $e^+e^-$).
Also, as in the previous numerical calculations, 
we use the MRSG \cite{MRSG} unpolarized
parton distributions in calculating physical cross sections. 
The peak in the
cross section is quite a bit larger than at RHIC due to the higher
energy (i.e.\ smaller $x$) and the valence-valence contributions. Here,
and throughout, we use $\mu=M$. The effect of varying $\mu$ will be
discussed later in this section.

Let us define the forward-backward (lepton) asymmetry as
\beq
A_{FB} = \fr{d\sg_{uu}^{p\B{p},F-B}/dM}{d\sg_{uu}^{p\B{p},F+B}/dM},
\eeq
where $\sg^{F\pm B}$ were defined in (\ref{fbdef}).
We expect large $A_{FB}$ in $p\B{p}$ collisions in mass regions where
$Z-\gm$ interference dominates since the asymmetry arises, at the
subprocess level, from axial -- vector interference so that the $Z$  
contribution
is pure axial and the asymmetry is unsuppressed. At the $Z$-pole, on
the other hand, the $Z$ vector couplings enter and lead to a suppression,
so the asymmetry will be somewhat smaller there. We can see this
explicitly by noting that the leading order subprocess level asymmetry
is proportional to
\beq
A_{FB}^{q\B{q}} \sim \fr{g_{al}g_{aq}g_{vl}g_{vq}}{(g_{al}^2+g_{vl}^2)
(g_{aq}^2+g_{vq}^2)}
\eeq
at the $Z$-pole. Hence there is an overall small factor,
\beq
g_{vl}=-\fr{1}{2} (1-4\sin^2 \theta_W),
\eeq
which is sensitive to \stW\  since \stW $\simeq .23$, so that a small
fractional change in \stW\  leads to a large fractional change in 
$A_{FB}$.

Fig.\ \ref{Fig10}(b) shows the effect of the
QCD corrections on $A_{FB}$. We see that, unlike  the
spin asymmetries (double-spin in particular), it is the $q\B{q}$ subprocess
which accounts for the dominant corrections. We understand this in 
the following way. From Fig.\ \ref{Fig10}(a), we see that the magnitude of the
$qg$ correction to $\sg^{F+B}$ is rather small compared to that of the
$q\B{q}$ subprocess. Thus, in order for the $qg$ subprocess to make an
appreciable contribution to $A_{FB}$, the corrections to $\sg^{F-B}$
would have to be quite different from those to $\sg^{F+B}$. Since one
has the same parton distributions in both cases, all the difference arises
from the differences in the subprocess cross sections. As we have explained
before, the dominant corrections come from the term $\sim \ln(1-w)$ in
(\ref{qgans}) which has the same form in both $\sg^{F+B}$ and $\sg^{F-B}$.
Consequently, the $qg$ corrections to $\sg^{F\pm B}$ amount to a 
multiplicative factor which basically cancels in the ratio, in the mass
region of interest. We also expect small corrections to $A_{FB}$ arising
from the $q\B{q}$ subprocess, for the same reasons. The question then is:
why are the $q\B{q}$ corrections to $A_{FB}$ observable  while the $qg$
corrections are not? The correction basically arises from the differences
in the hard bremsstrahlung contributions to $\sg^{F\pm B}$, 
$k_{uu}^{q\B{q}, F\pm B}$, given in (\ref{kqqb}). We note that the difference
vanishes in the limit $w\RA 1$, relevant for the large $M$ limit and for the
dominant corrections to $\sg^{F\pm B}$ at lower $M$. This explains the
vanishing of the corrections as $M$ increases. At intermediate and low $M$, 
however, there will be some contribution to $A_{FB}$ arising from smaller
$w$, in which case the term, $k_{uu}^{q\B{q}, F-B}=2(1+w)\ln w$, contributes
negatively (oppositely to the Born term) in a significant fashion, thus
accounting for the reduction in the magnitude of $A_{FB}$. Now, for the
$qg$ subprocess, the differences in the $k_{uu}^{qg, F\pm B}$ all vanish
in the limit $w\RA 1$ and are small at smaller $w$, thereby accounting
for the relative smallness of the $qg$ corrections to $A_{FB}$.

\begin{figure}
\centerline{\epsfig{file=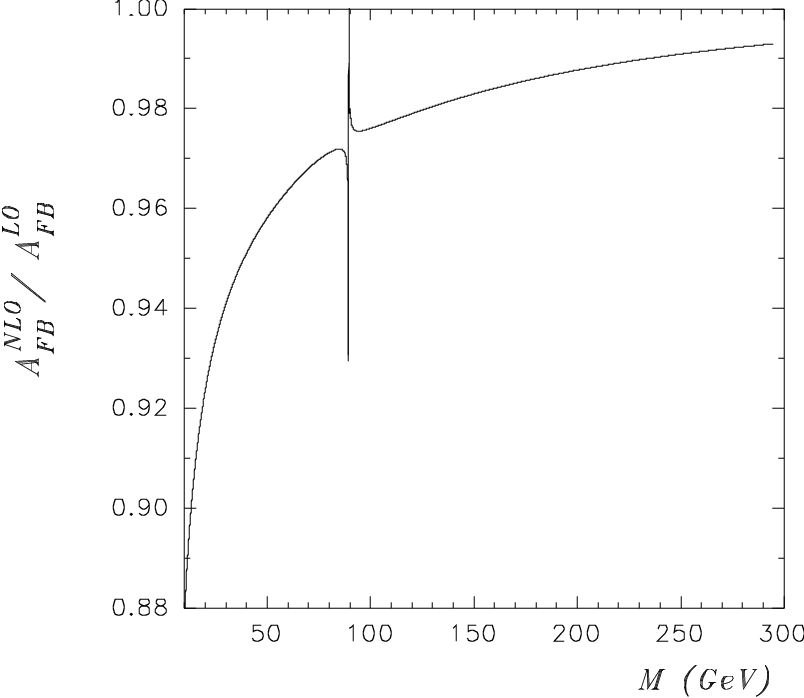,height=7cm,width=8cm}}
\vspace{10pt}
\caption{The ratio, $A_{FB}^{NLO}/A_{FB}^{LO}$, using the 
$A_{FB}^{(N)LO}$ of Fig.\ \ref{Fig10}(b).
}
\label{Fig11}
\end{figure}

In the end, the corrections to $A_{FB}$ almost exactly
amount to a multiplicative factor (less than 1) which
is given in Fig.\ \ref{Fig11}. We see that it increases with increasing mass,
approaching unity.  In the $Z$-pole region, it is $\sim .975$ . This result
is in good agreement with the finding of \cite{Baur}, considering 
that they use a different definition of $A_{FB}$ beyond LO, and 
a more involved approach, taking into account experimental cuts. Also,
in their approach, one could not understand properly the structure of
the QCD corrections, since the hard bremsstrahlung is handled via
Monte Carlo integration. The spikes at $M\simeq 89.3 - 89.5 \,\,
\GeV$ simply reflect the fact that $A_{FB}^{NLO}$ and $A_{FB}^{LO}$
vanish at {\em slightly} different points and intersect just above
the zero.

\begin{figure}
\centerline{\epsfig{file=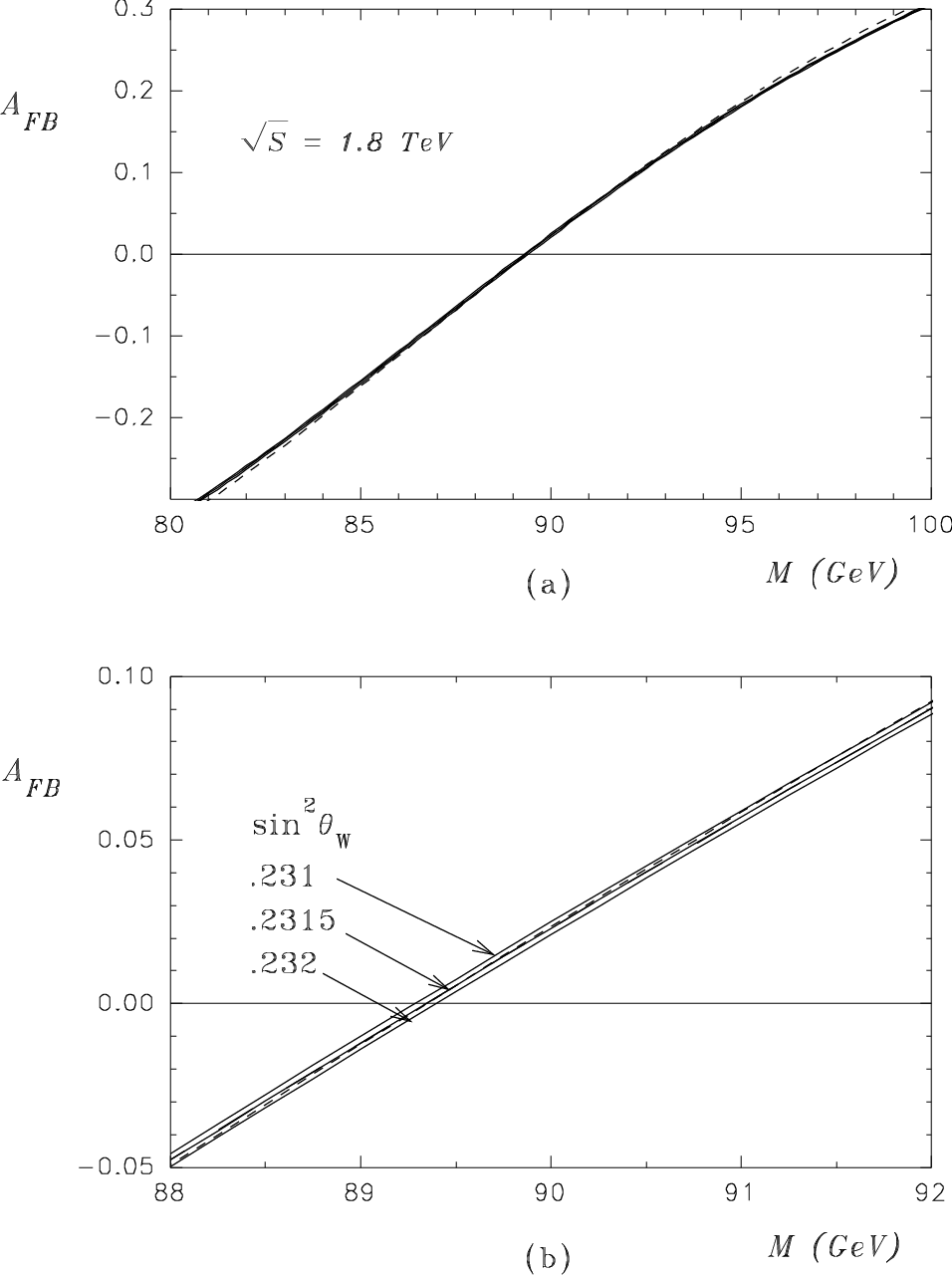,height=11cm,width=8cm}}
\vspace{10pt}
\caption{(a) The $A_{FB}$ of Fig.\ \ref{Fig10}(b), but in the mass region,
$80\leq M \leq 100\,\,\GeV$. The effect of varying \stW\ by 
$\pm .0005$, on $A_{FB}^{NLO}$, is shown. (b) Same as (a), but in the 
mass region, $88\leq M \leq 92\,\,\GeV$. 
}
\label{Fig12}
\end{figure}

Figures \ref{Fig12}(a) and \ref{Fig12}(b) show \afb\  in the $Z$-pole region
for \stW = .2315. The effect
of the QCD corrections is shown as is the effect of changing \stW\ 
by an amount of $\pm .0005$. The magnitudes of the effects are comparable,
but the character is distinctly different. Changing \stW\  amounts to a
shift in \afb\  rather than a multiplicative factor. As a result, changing
\stW\  shifts somewhat the zero of \afb. Since the QCD corrections do not
shift the zero appreciably, 
one sees that measuring precisely the zero in \afb\ 
may allow one to get a good handle on \stW\  without worrying about
neglected higher order QCD effects (and possibly other uncertainties
such as choice of parton distributions). How precisely this may be 
measured is left for a separate study.
Using the
GRV NLO ($\ovl{\rm MS}$)
set \cite{GRV} produced no appreciable change in the predictions.
The CTEQ3M set \cite{CTEQ}, on the other hand, gave slightly different 
predictions for the forward-backward asymmetry. Hence, closer agreement
between the various sets is required before precision determinations of 
\stW\  are possible.

The fact that the zero of \afb\  does not change appreciably under HOC's
is understandable since the zero essentially arises from the zero
in the subprocess cross section, which in turn depends on $M$ in a
way independent of parton distribution effects. Each subprocess, at 
NLO, is proportional to a Born term with a well-defined zero. So any shift
in the zero must arise predominantly
from the fact that each quark flavor passes
through zero at slightly different values of $M$. The NLO corrections
however, do not appreciably affect the weighting of the various quark
flavors relative to the LO weightings, since the same parton distributions
enter. Hence there is very little shift in the position of the zero.

We studied the effect of varying $\mu$ on $A_{FB}$ and it was found
to be negligible. This is not unexpected since the $\mu$-dependent part
of the corrections has the same form in $\sg^{F-B}$ and $\sg^{F+B}$.
Hence, the major uncertainties are in the parton distributions, the
neglected nonleading corrections, the QED corrections and possible
intrinsic transverse momentum effects. Normally, one would not consider
the latter effects at such high $\mu^2$, but as we are dealing with a very
high precision measurement, they should not be taken for granted as 
being negligible. There is also a pure QED contribution to $A_{FB}$
at order $\al^3$, which was originally studied in connection with
$e^+e^-\RA \mu^+\mu^-$ \cite{Berends}.

With ${\cal L}=110\,\, {\rm pb}^{-1}$, \afb\  for $e^+e^-$ pairs
in the $Z$-pole region was
measured with a statistical error of roughly $\pm 20$\% \cite{afb}. 
The maximum possible ${\cal L}$ for Run II after
several years running is 100 fb$^{-1}$ \cite{TeV2000}. We will take
${\cal L}_{\rm max} = $ 70 fb$^{-1}$ as being a more realistic
(if not optimistic) value in determining the best possible measurement
of \afb. Then, statistically, we expect to be able to measure \afb\ 
to (at best) $\pm .8$\%. Hence, our statistical error goes down by
a factor of roughly 25. The same statement applies to \stW. The
statistical error on \stW\  of $\pm .003$ which is obtained from
110 pb$^{-1}$ is reduced to $\pm .00012$ with  70 fb$^{-1}$. If one
takes into account two detectors and both muons and electrons, in a best
case scenario, we could multiply our number of events by a factor of 4
and get $\pm .00006$ as an error on \stW. For a more realistic   
30 fb$^{-1}$, we get an error of $\pm .00009 \simeq \pm .0001$. 
If we do not combine the data from both detectors, muons and electrons,
this goes up to $\pm .0002$. Either way, the error is very small.

Of course, this error analysis is very naiive and all the systematic
errors must be put under tight control for a realistic ultrahigh precision
measurement as discussed above. A more detailed analysis is beyond the
scope of this paper. Nonetheless, if any of the above scenarios could
be realized, this would  be the best \stW\  determination
available. In fact, it would be better than the world average. Even with
lower luminosity running such as 10 fb$^{-1}$ and not combining muons and
electrons, the measurement is of the same precision as the SLAC measurement
and hence could help to resolve the SLAC-LEP discrepancy.

%\vglue 1cm
%\begin{center}\begin{large}\begin{bf}
\section{CONCLUSIONS}
%\end{bf}\end{large}\end{center}
%\vglue .3cm

Complete analytical results for mass differential Drell-Yan type
cross-sections relevant to all possible initial longitudinal
polarization states were 
calculated at the one-loop level in QCD. Interference between bosons
of arbitrary mass, width and couplings was considered. For all 
observables considered, the corresponding forward-backward cross
sections were determined. The results were presented in a form valid
for all consistent
$n$-dimensional regularization schemes.  A survey of constraints
on allowable factorization and regularization schemes was given. The
mechanism behind scale dependences was discussed in some detail and
in a general fashion.

NLO predictions for all longitudinal Drell-Yan type processes
at RHIC ($W^\pm$, $Z$ and $\gm^*$) were made using polarized parton 
distributions which fit the recent DIS data.  The HOC's
increased the cross sections substantially and  had a major impact on the
asymmetries, while preserving the features of the LO asymmetries. The
exact sign and magnitude of the HOC's depended on the details of the
polarized parton distributions used, 
especially the sea and gluon distributions.
Faced with either low rates or small asymmetries, $\gm^*$ production
did not appear very interesting at face value,
for longitudinal polarization.  On the other hand, if the
agreement between the various parton distributions at small $x$ is 
accidental, rather than constrained, $\gm^*$ production at RHIC
could demonstrate that possibility by yielding an unpredicted asymmetry. 
The $Z$- asymmetries were all quite sensitive to the
sea quarks; the parity violating ones being the largest, with unexpected
sensitivity due to a coincidental cancellation between $u$ and $d$ valence
contributions. With large rates and asymmetries, $W^\pm$ production can 
directly measure the polarized
sea and valence distributions as well as the  unpolarized 
$\B{u}/\B{d}$ ratio. Lower energy running could directly measure
$\B{u}$ and $\B{d}$ at rather large $x$ (both polarized  and unpolarized).

NLO QCD predictions for the forward-backward lepton asymmetry at 
Fermilab were made. The QCD corrections amounted 
almost exactly to a multiplicative
factor on the asymmetry which was a function of the mass of the lepton
pair produced. This function was found to be less than unity throughout and
approached 1 with increasing invariant mass. In the $Z$-pole region,
the factor was roughly  $.975$\,.  It was thus observed that the 
zero in the asymmetry was quite stable under QCD corrections. The zero was
rather sensitive to \stW, however. This suggested an alternate method for
determining \stW\  (i.e.\ measuring precisely the zero).
The details were left for a separate study, however.
Based on expected luminosities for Fermilab's Run II and previous
Fermilab determinations which used a standard algorithm for 
extracting \stW\  from \afb,
high precision \stW\  measurements
were found to be possible at a level much better than (statistically),
or at worst competitive with,
the best measurements presently available and should be able to resolve
the SLAC-LEP discrepancy.

\vglue 1cm
\begin{center}\begin{large}\begin{bf}
\section*{ACKNOWLEDGEMENTS}
\end{bf}\end{large}\end{center}
\vglue .3cm

The author would like to thank W.J.\ Marciano for suggesting much of
the work done here and for countless valuable discussions, without
which, this project would not have been possible. I would also like to
thank F.E.\ Paige for insightful discussions and suggestions, which
stimulated my interest in some of the work done. I also acknowledge
very useful discussions with G.\ Bunce,
P.\ van Nieuwenhuizen, N.\ Saito,
M.\ Tannenbaum and T.L.\ Trueman as well as useful correspondence with
W.\ Vogelsang. 
Finally, I would like to thank
Z.\ Parsa and the ITP of UCSB, where part of this work was done,
for their hospitality during the workshop on future high energy 
colliders.
Supported by U.S.\ Department of Energy contract number 
DE-AC02-76CH00016.

%\newpage
%\vglue 1cm
%\begin{center}\begin{large}\begin{bf}
%APPENDIX
%\end{bf}\end{large}\end{center}
%\vglue .3cm

%\vglue 1cm
%\begin{center}\begin{large}\begin{bf}
%REFERENCES
%\end{bf}\end{large}\end{center}
%\vglue .3cm

%\begin{thebibliography}{99}

%\end{thebibliography}

%\newpage
\twocolumn

%\newpage
%\vglue 1cm
%\begin{center}\begin{large}\begin{bf}
%FIGURE CAPTIONS
%\end{bf}\end{large}\end{center}
%\vglue .3cm

%\begin{thecaptions}{99}

\mbox{}
\newpage

\mbox{}
\newpage

\mbox{}
\newpage
%
%\end{thecaptions}

%\newpage
%\vglue 1cm
%\begin{center}\begin{large}\begin{bf}
%TABLES
%\end{bf}\end{large}\end{center}
%\vglue .3cm

%\newpage

\begin{table}
\caption{The $\Dl_k T_{ij}$ which define the various factorization schemes
of interest; here $k = u,l$. }
\begin{center}
\begin{tabular}{ccccc} 
\mbox{}  & $\ovl{\rm MS}$ & $\ovl{\rm MS}_{\ve}$ & $\ovl{\rm MS}_{HC}$
\mbox{}  & $\ovl{\rm MS}_p$ \\ \hline
 $\Dl_k T_{qq}$ & 0 & $\Dl_k P_{qq}^{\ve}$ & $\Dl_l P_{qq}^{\ve}
 - \Dl_u P_{qq}^{\ve}$ &
  $\Dl_l P_{qq}^{\ve} - \Dl_u P_{qq}^{\ve}$ \\  
 $\Dl_k T_{qg}$ & 0 & $\Dl_k P_{qg}^{\ve}$ & 0 &
  $\Dl_l P_{qg}^{\ve}$ \\ 
 $\Dl_k T_{gq}$ & 0 & $\Dl_k P_{gq}^{\ve}$ & 0 &
  $\Dl_l P_{gq}^{\ve}$ \\  
 $\Dl_k T_{gg}$ & 0 & $\Dl_k P_{gg}^{\ve}$ & 0 &
  $\Dl_l P_{gg}^{\ve,<}$ \\ 
\end{tabular} 
\end{center}
\label{Tab1}
\end{table}

\end{document}